\documentclass[a4paper,11pt]{article}

\usepackage{amsmath}
\usepackage{authblk}
\usepackage[usenames,dvipsnames,svgnames,table]{xcolor}
\usepackage{graphicx}
\usepackage{subcaption}
\usepackage{lineno}
\usepackage{mathrsfs}
\usepackage{longtable}
\usepackage{multirow}
\usepackage{hyperref}
\usepackage[a4paper,margin=1in]{geometry}
\usepackage[utf8]{inputenc}
\usepackage{cite}
\usepackage{rotfloat}

\begin{document}
\title{Results of Dark Matter Search using the Full PandaX-II Exposure}
\author[2,3]{Qiuhong Wang}
\author[1]{Abdusalam Abdukerim}
\author[1]{Wei Chen}
\author[1,4]{Xun Chen\thanks{corresponding author, chenxun@sjtu.edu.cn}}
\author[5]{Yunhua Chen}
\author[6]{Chen Cheng}
\author[7]{Xiangyi Cui}
\author[8]{Yingjie Fan}
\author[9]{Deqing Fang}
\author[9]{Changbo Fu}
\author[10]{Mengting Fu}
\author[11,12]{Lisheng Geng}
\author[1]{Karl Giboni}
\author[1]{Linhui Gu}
\author[5]{Xuyuan Guo}
\author[1]{Ke Han}
\author[1]{Changda He}
\author[1]{Di Huang}
\author[5]{Yan Huang}
\author[13]{Yanlin Huang}
\author[1]{Zhou Huang}
\author[14]{Xiangdong Ji}
\author[15]{Yonglin Ju}
\author[7]{Shuaijie Li}
\author[15]{Huaxuan Liu}
\author[1,7,4]{Jianglai Liu\thanks{spokesperson, jianglai.liu@sjtu.edu.cn}}
\author[1]{Wenbo Ma}
\author[9,2]{Yugang Ma}
\author[10]{Yajun Mao}
\author[1,4]{Yue Meng}
\author[1]{Kaixiang Ni}
\author[5]{Jinhua Ning}
\author[1]{Xuyang Ning}
\author[16]{Xiangxiang Ren}
\author[5]{Changsong Shang}
\author[1]{Lin Si}
\author[11]{Guofang Shen}
\author[14]{Andi Tan}
\author[16]{Anqing Wang}
\author[2,17]{Hongwei Wang}
\author[16]{Meng Wang}
\author[10]{Siguang Wang}
\author[6]{Wei Wang}
\author[15]{Xiuli Wang}
\author[1,4]{Zhou Wang}
\author[6]{Mengmeng Wu}
\author[5]{Shiyong Wu}
\author[1]{Weihao Wu}
\author[1]{Jingkai Xia}
\author[14,18]{Mengjiao Xiao}
\author[7]{Pengwei Xie}
\author[1]{Binbin Yan}
\author[1]{Jijun Yang}
\author[1]{Yong Yang}
\author[8]{Chunxu Yu}
\author[16]{Jumin Yuan}
\author[1]{Ying Yuan}
\author[1]{Xinning Zeng}
\author[14]{Dan Zhang\thanks{corresponding author, dzhang16@umd.edu}}
\author[1,4]{Tao Zhang}
\author[1,4]{Li Zhao}
\author[13]{Qibin Zheng}
\author[5]{Jifang Zhou}
\author[1]{Ning Zhou}
\author[11]{Xiaopeng Zhou}

\affil[1]{INPAC and School of Physics and Astronomy, Shanghai Jiao Tong University, MOE Key Lab for Particle Physics, Astrophysics and Cosmology, Shanghai Key Laboratory for Particle Physics and Cosmology, Shanghai 200240, China}
\affil[2]{Shanghai Institute of Applied Physics, Chinese Academy of Sciences, Shanghai 201800, China}
\affil[3]{University of Chinese Academy of Sciences, Beijing 100049, China}
\affil[4]{Shanghai Jiao Tong University Sichuan Research Institute, Chengdu 610213, China}
\affil[5]{Yalong River Hydropower Development Company, Ltd., 288 Shuanglin Road, Chengdu 610051, China}
\affil[6]{School of Physics, Sun Yat-Sen University, Guangzhou 510275, China}
\affil[7]{Tsung-Dao Lee Institute, Shanghai 200240, China}
\affil[8]{School of Physics, Nankai University, Tianjin 300071, China}
\affil[9]{Key Laboratory of Nuclear Physics and Ion-beam Application (MOE), Institute of Modern Physics, Fudan University, Shanghai 200433, China}
\affil[10]{School of Physics, Peking University, Beijing 100871, China}
\affil[11]{School of Physics, Beihang University, Beijing 100191, China}
\affil[12]{International Research Center for Nuclei and Particles in the Cosmos \& Beijing Key Laboratory of Advanced Nuclear Materials and Physics, Beihang University, Beijing 100191, China}
\affil[13]{School of Medical Instrument and Food Engineering, University of Shanghai for Science and Technology, Shanghai 200093, China}
\affil[14]{Department of Physics, University of Maryland, College Park, Maryland 20742, USA}
\affil[15]{School of Mechanical Engineering, Shanghai Jiao Tong University, Shanghai 200240, China}
\affil[16]{School of Physics and Key Laboratory of Particle Physics and Particle Irradiation (MOE), Shandong University, Jinan 250100, China}
\affil[17]{Shanghai Advanced Research Institute, Chinese Academy of Sciences, Shanghai 201210, China}
\affil[18]{Center for High Energy Physics, Peking University, Beijing 100871, China}
\affil[ ]{(PandaX-II Collaboration)}

\maketitle
\abstract{We report the dark matter  search results obtained using
  the full 132 ton$\cdot$day exposure of the PandaX-II experiment,
  including all data from March 2016 to August 2018. No significant
  excess of events is identified above the expected background.
  Upper limits are set on the spin-independent dark matter-nucleon
  interactions. The lowest 90\% confidence level exclusion on the
  spin-independent cross section is $2.2\times 10^{-46}$~cm$^2$ at a
  WIMP mass of 30~GeV/$c^2$.}


PACS number(s): 95.35.+d, 29.40.–n

\section{Introduction}
\label{sec:intro}
The existence of dark matter (DM) has been supported by substantial
evidence from cosmology and astronomical
observations~\cite{Bertone:2004pz}. A large number of direct searches
for weakly interactive massive particles (WIMPs), a leading candidate
of particle DM, are ongoing worldwide~\cite{Liu:2017drf}.  Over the
last ten years, experiments employing dual-phase xenon time projection
chambers (TPCs) have produced the most stringent constraints on the
spin-independent interactions between WIMPs and nucleons with a mass
range from a few GeV/$c^2$ to 10 TeV/$c^2$~\cite{Tan:2016zwf,
  Akerib:2016vxi, Aprile:2017iyp, Cui:2017nnn, Aprile:2018dbl}.

The PandaX-II experiment~\cite{Tan:2016diz}, conducted at the China
Jinping Underground Laboratory (CJPL)~\cite{Yu-Cheng:2013iaa},
utilized a cylindrical dual-phase xenon TPC with a dodecagonal cross
section (distance to opposite side: 646~mm) confined by
polytetrafluoroethylene (PTFE) walls. The maximum drift distance is
600~mm in the vertical direction, defined by the distance from the
bottom cathode mesh to the top gate grid. A total of 580~kg of liquid
xenon is contained in the sensitive volume. Two arrays of Hamamatsu
R11410-20 photomultiplier tubes (PMTs) located at the top and bottom
of the TPC, respectively, view the sensitive volume. Recoil events
produce the prompt scintillation photons ($S1$) and delayed
electroluminescence photons ($S2$). To achieve good collection of
these photons, liquid xenon is purified with two circulation loops
with a total mass flow rate of approximately 560~kg/day through hot
getters to remove gaseous impurities. The relative sizes of $S1$ and
$S2$ provide a powerful way to discriminate the electron recoil (ER)
backgrounds from nuclear recoil (NR) signals.

The analog waveform of each PMT, linearly amplified by a factor of
approximately 10, is digitized by 100-MHz CAEN V1724 digitizers when
an event is triggered either by $S1$ or $S2$. The digitizers used
baseline suppression (BLS) firmware to suppress readouts for samples
below the configurable threshold.

Previous searches from PandaX-II have been reported in
Refs.~\cite{Tan:2016zwf} (33 ton$\cdot$day exposure)
and~\cite{Cui:2017nnn} (54 ton$\cdot$day exposure). In this work, we
report DM search results by combining all 132~ton$\cdot$day data
obtained in PandaX-II with a blind analysis carried out on data in a
fresh exposure. Major improvements in the data analysis are discussed
in detail in this paper. The remainder of this paper is organized as
follows. A simple description of the data sets is given in
Sec.~\ref{sec:pandax-ii}. The detailed data processing and improved
event selections are discussed in Sec.~\ref{sec:data_process_cut}. In
Sec.~\ref{sec:det_calib}, the data calibration, corrections, and
signal models are presented. The backgrounds are analyzed in
Sec.~\ref{sec:background}. The final candidates are reported in
Sec.~\ref{sec:candidate}, from which we derive the exclusion limits in
Sec.~\ref{sec:limit} using statistical analysis.

\section{Data sets in PandaX-II}
\label{sec:pandax-ii}
\begin{figure}[htb]
  \centering
\includegraphics[width=0.95\textwidth]{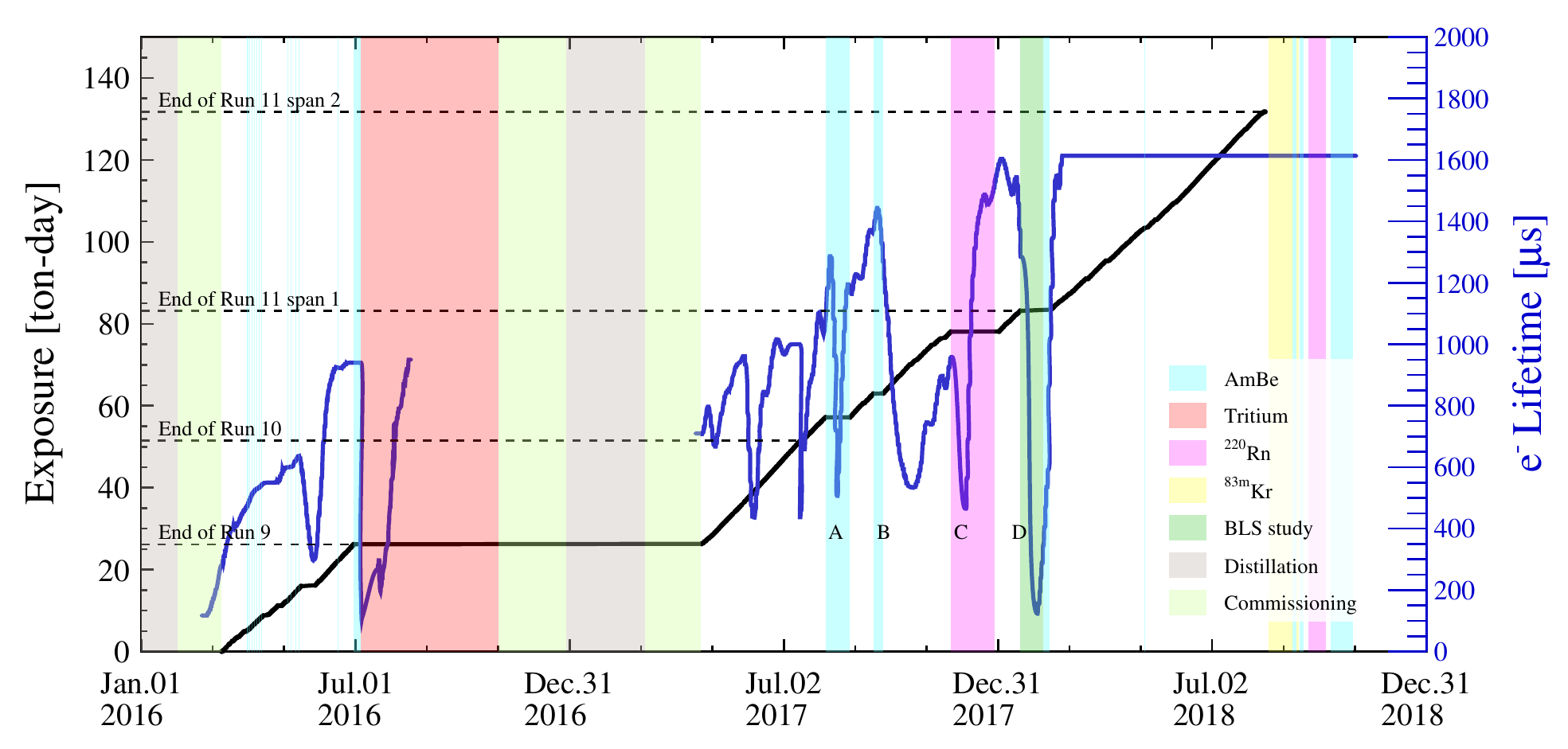}
\caption{The accumulation of DM exposure (black line) and the
  evolution of the electron lifetime (blue curve, right axis) in
  PandaX-II. The black dashed horizontal lines indicate the divisions
  of data sets. Various colored bands represent the different
  operation modes, including NR calibration with AmBe source (cyan),
  ER calibration with tritium (pale red) and $^{220}$Rn sources
  (pink), $^{83\rm{m}}$Kr calibration (yellow), baseline suppression
  study (green), xenon distillation (gray), and detector commissioning
  (light green).}
  \label{fig:exposure_elifetime}
\end{figure}

The operation history and the accumulation of DM exposure in PandaX-II
are summarized in Fig.~\ref{fig:exposure_elifetime}.  In total, three
major DM runs have been conducted in PandaX-II,
Run~9~\cite{Tan:2016zwf}, Run 10~\cite{Cui:2017nnn}, and Run 11.
Immediately after 79.6~days of data collection in Run 9, an ER
calibration with tritiated methane and a subsequent distillation
campaign were performed, after which Run 10 collected DM search data
for 77.1~days. Run 10 ended with a power failure, and Run 11 started
right after the recovery, collecting a total of 244.2~days of data
from July 17, 2017 to Aug. 16, 2018. The electron lifetime overlaid in
Fig.~\ref{fig:exposure_elifetime} indicates the change of detector
purity during the run. The first major drop in Run 11 (``A'' in
Fig.~\ref{fig:exposure_elifetime}) was due to an unexpected power
failure. The second drop ``B'', occured just after a neutron
calibration, during which we varied the recirculation pump speed to
study potential correlation with the background rate. The third drop
``C'', occurred during a calibration run with $^{220}$Rn
injection~\cite{Ma:2020kll}, which introduced impurities into the
detector. The last drop ``D'', was caused by a real air leak into the
detector due to a failed gate valve and, as a result, an increase in
the ER background rate was observed. In the analysis presented, Run 11
is therefore broken down into two spans, span 1 and span 2, separated
by ``D'', with live times of 96.3 and 147.9 days, respectively.  After
Run 11, the operation was dedicated to calibration and detector
systematic studies before the official shutdown of PandaX-II on June
29, 2019. In Run 9, the cathode and gate electrodes were set at
$-29$~kV and $-4.95$~kV, leading to an approximate drift field and
electron extraction field of $-400$~V/cm and 4.56~kV/cm (in liquid
xenon), respectively. In Runs 10 and 11, the cathode HV was lowered to
$-24$~kV, leading to a different drift field of $-317$~V/cm.

Calibration runs were interleaved with the DM data collection to study
detector responses. One set corresponding to the $^{241}$Am-Be (AmBe)
run was obtained at the end of Run 9. During and after Run 11, six
more sets of AmBe runs were taken. They are used to characterize the
NR responses. The low energy ER responses are characterized with
gaseous $\beta$ source injections runs, CH$_3$T (tritium) or
$^{220}$Rn, for Run 9 and Runs 10/11, respectively. Other types of
calibration runs include a $^{83\rm{m}}$Kr injection run for position
reconstruction studies and uniformity correction (Run 11), and
external $^{137}$Cs and $^{60}$Co source deployment to calibrate
detector response to higher energy gammas.

\section{Data processing, quality cuts and event reconstruction}
\label{sec:data_process_cut}

The basic data processing procedure from previous
analyses~\cite{Xiao:2015psa,Tan:2016diz} was followed in this
analysis. Only major improvements are highlighted here, including a)
inhibiting unstable PMTs for better consistency among data sets, b) an
improved gain calibration for low gain PMTs, c) refinements of data
quality cuts, and d) substantial improvements in the position
reconstruction algorithm. For the DM search in Run 11, we blinded the
data with $S1$ less than 45 PE (previous search window) to avoid
subjective choices until the background estimation was finalized.

\subsection{Unstable PMTs}
For consistency, particularly in terms of position reconstructions,
seven malfunctioning PMTs (five top, two bottom) among the 110
R11410-20 PMTs are fully inhibited in this analysis for all data
sets. Among them, three were turned turned off in
Ref.~\cite{Cui:2017nnn}, one due to severe afterpulsing, and two due
to failures in PMT bases. During Run 11, four more PMTs became
unstable; one was physically turned off due to the high discharge rate,
and the other three were inhibited by software: one due to
afterpulsing, and the other two due to abnormal gains and baseline
noises. The five inhibited top PMTs are indicated in
Fig.~\ref{fig:xycandidates} in Appendix~\ref{sec:app_b}.

\subsection{Low-gain PMTs}
\label{sec:low_gain_pmts}
PMT gains were calibrated twice a week using low-intensity blue
light-emitting diodes (LEDs) inside the detector by fitting the single
photoelectron (SPE) peak in the spectrum. After a vacuum failure
between Runs 9 and 10, which may have caused degradation in some
high-voltage feedthroughs, a number of PMTs could only run at lower
high voltages. Some had to be gradually lowered throughout Runs 10 and
11. These led to significantly reduced gain values (average gain
changed from 1.41$\times10^{6}$ in Run 9 to 0.96$\times10^{6}$ in
Run~11, before the linear amplifiers). For low gain channels
($\sim10^5$), the LED calibration had two problems. First, the
corresponding SPE peaks could not be distinguished from baseline
noises, leading to failed fits and jumps. Second, the LED calibration
could not catch up with the temporal change of the gains.

\begin{figure}
  \centering
  \includegraphics[width=0.45\textwidth]{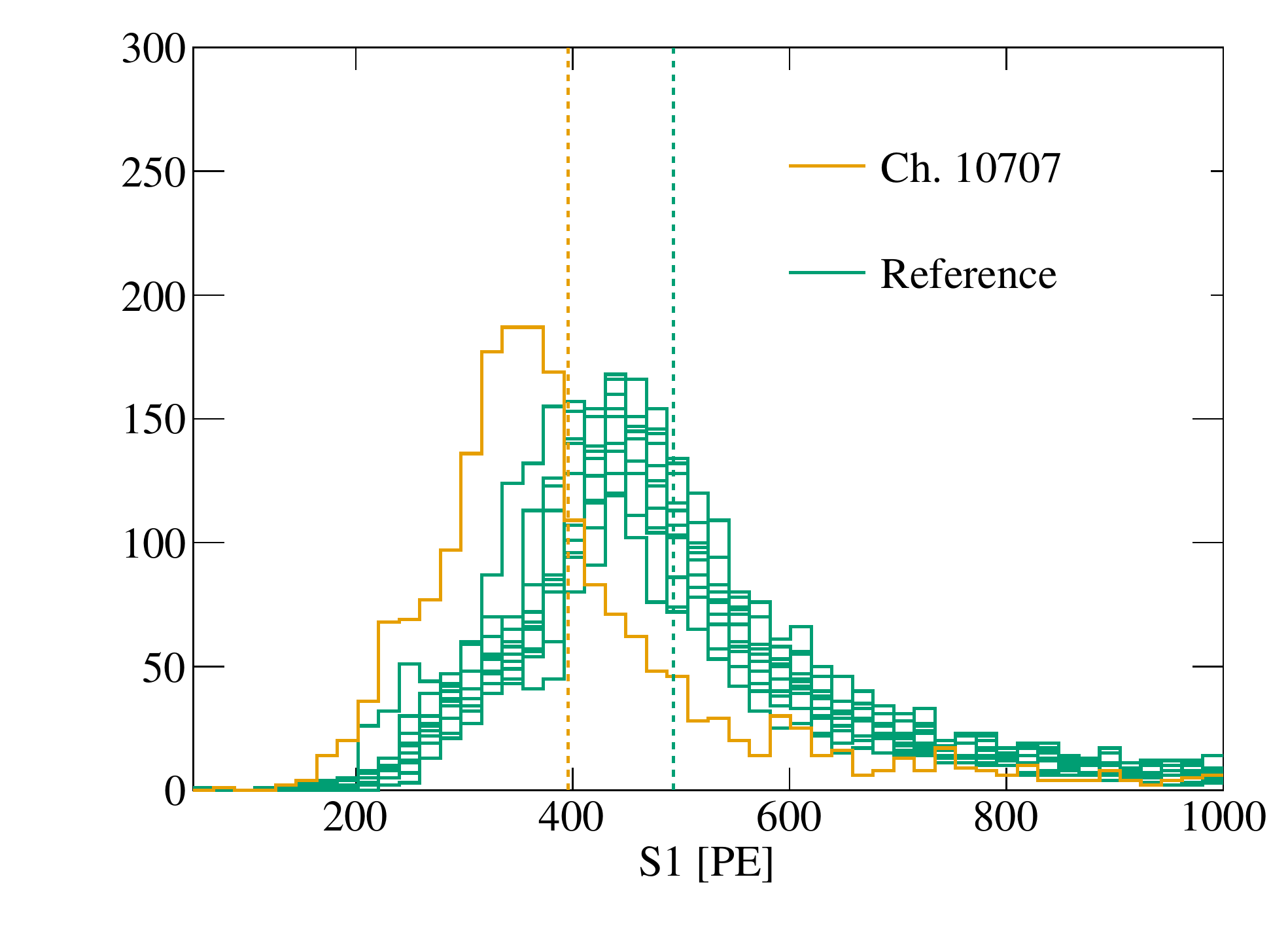}
  \includegraphics[width=0.45\textwidth]{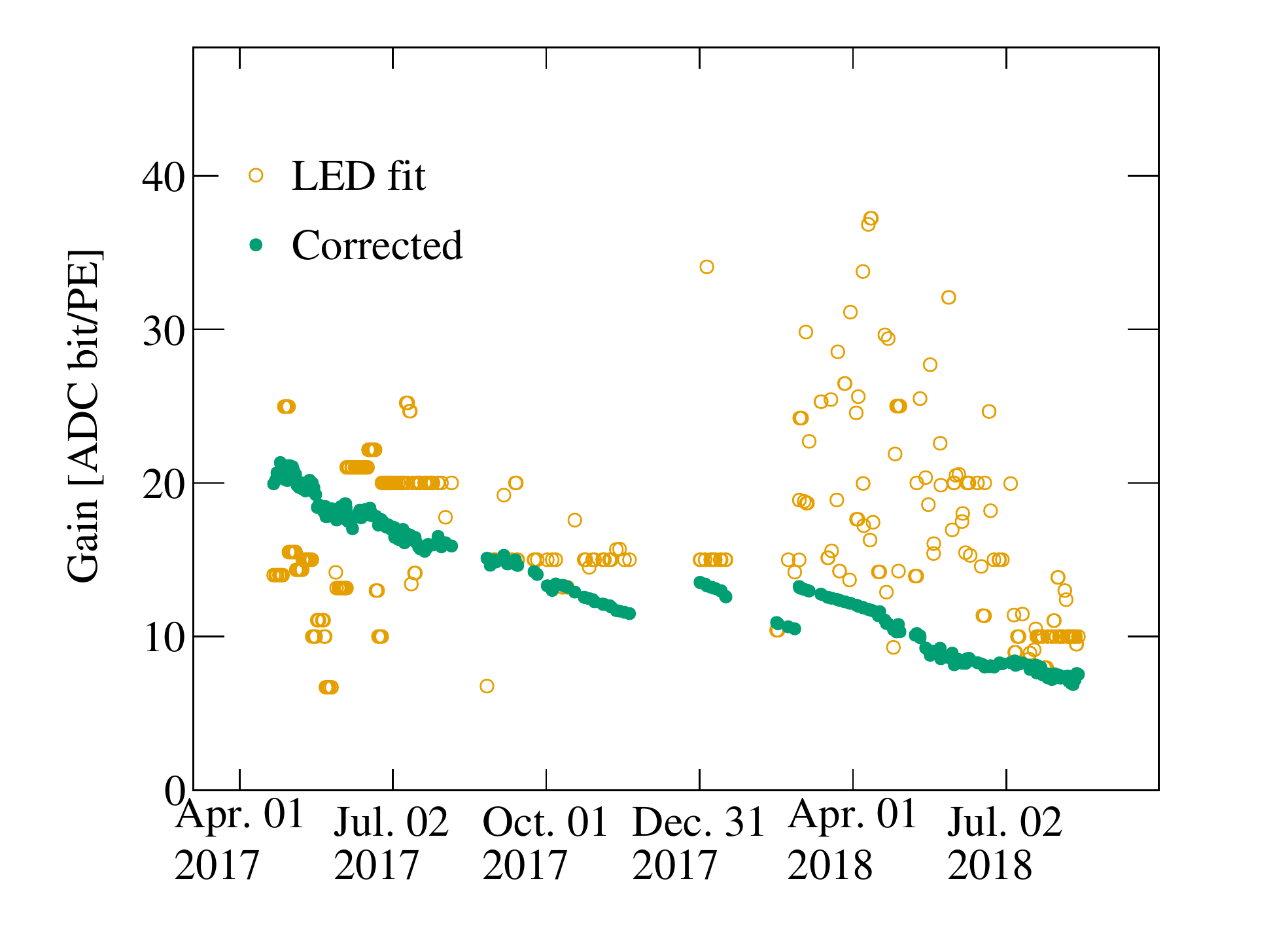}
  \caption{An example of gain correction to low gain PMTs. Left: The
    $S1$ charge distribution of $\alpha$ events in the low gain
    Ch.10707 (yellow line) and the other reference PMTs located at the
    same radius (green lines). The yellow and green dashed line mark
    the mean of the $S1$ detected by Ch.10707 and the reference PMTs,
    respectively.  The ratio of the means is used to correct the gain
    of Ch.10707. Right: The evolution of gain of Ch.10707. The yellow
    open circles represent gains obtained in the LED calibration and
    the green dots are the corrected gains. The gradual decrease is
    due to the continuous reduction of the supply voltage, and the
    residual jumps are due to the attempts to recover the high voltage
    during the run.}
  \label{fig:low_gain_correction}
\end{figure}

To mitigate these effects, we developed {\it in situ} gain estimates
by selecting alpha events and using the mean $S1$ charge in the normal
PMTs located at the same radius in the same array as a
reference. This procedure is illustrated in
Fig.~\ref{fig:low_gain_correction}. After this correction, the gain
evolution of all channels becomes stable.

\subsection{Data quality cuts}
The data selection cuts used in Ref.~\cite{Tan:2016diz} are also
inherited in this analysis. The criteria for some cuts have been
updated in this analysis due to the updated PMT configurations.  Two
more cuts are developed to suppress spurious events. 1) In PandaX-II,
the PMT cathode and metal housing were set at approximately
$-700$~V. Ionized electrons produced in the gaseous region between the
top PMT array and the anode (ground) would drift toward the anode and
some may get amplified close to the anode wires, producing
$S2$s. These events have a typical drift time of $\sim40 \mu$s, and
due to longer tracks and diffusions in weak drift fields, these $S2$s
have larger width in comparison with the normal events. A cut on the
$S2$ widths is developed and applied. 2) We observed that occasionally
some mini-discharges occurred in the detector, resulting in waveforms
containing ``trains'' of small pulses. A cut on the ``cleanliness'' of
the waveform is developed to remove such events. By analyzing the NR
and ER calibration data, the inefficiency of the two cuts for ``good''
single-scattering events is estimated to be less than $5\%$.

The loss of efficiency is more significant for smaller signals, and it
gradually plateaued. Similar to the previous analyses, the overall
selection efficiency can be parameterized into
\begin{equation}
\label{eq:eff}
\epsilon = \epsilon_1(S1)\epsilon_2(S2)\epsilon_{\rm{BDT}}\epsilon_{\rm{plateau}}\,,
\end{equation}
in which $\epsilon_1(S1)$ ($\epsilon_2(S2)$) are $S1$ ($S2$) data
quality cut efficiency normalized to unity toward high energy, which
is discussed in Sec.~\ref{sec:calibration_model}, and
$\epsilon_{\rm{BDT}}$ refers to the efficiency of the boosted decision
tree (BDT) cut to suppress accidental backgrounds (see
Sec.~\ref{sec:background}). The plateau efficiency from all quality
cuts is estimated to be $91\%$ for $S1 > 20$~PE and $S2_{\rm{raw}}$
(uncorrected for uniformity) $>500$~PE, by studying the NR events in
AmBe calibration runs within the $\pm3\sigma$ region of the NR band.

\subsection{Position reconstruction}
Only single scattered events, containing a single $S1$ and $S2$ pair,
are selected for final analysis. The separation between the two
signals determines the vertical position of the event, by assuming a
constant drift velocity. The maximum drift time is measured to be
350~$\mu$s in Run 9 and 360~$\mu$s in Runs 10 and 11 due to
differences in drift fields.

\begin{figure}[hbt]
  \centering
  \includegraphics[width=0.3\textwidth]{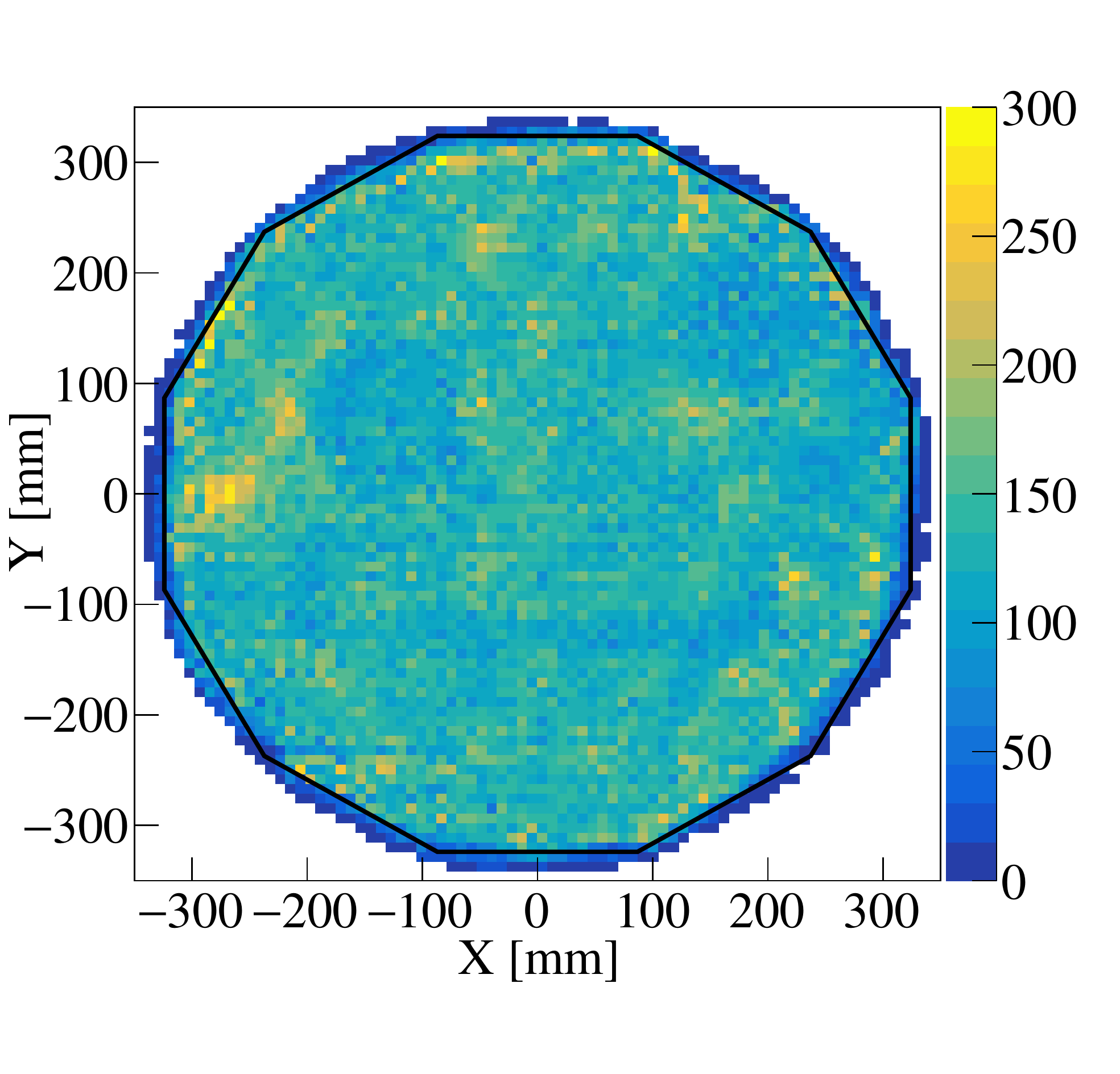}
  \includegraphics[width=0.3\textwidth]{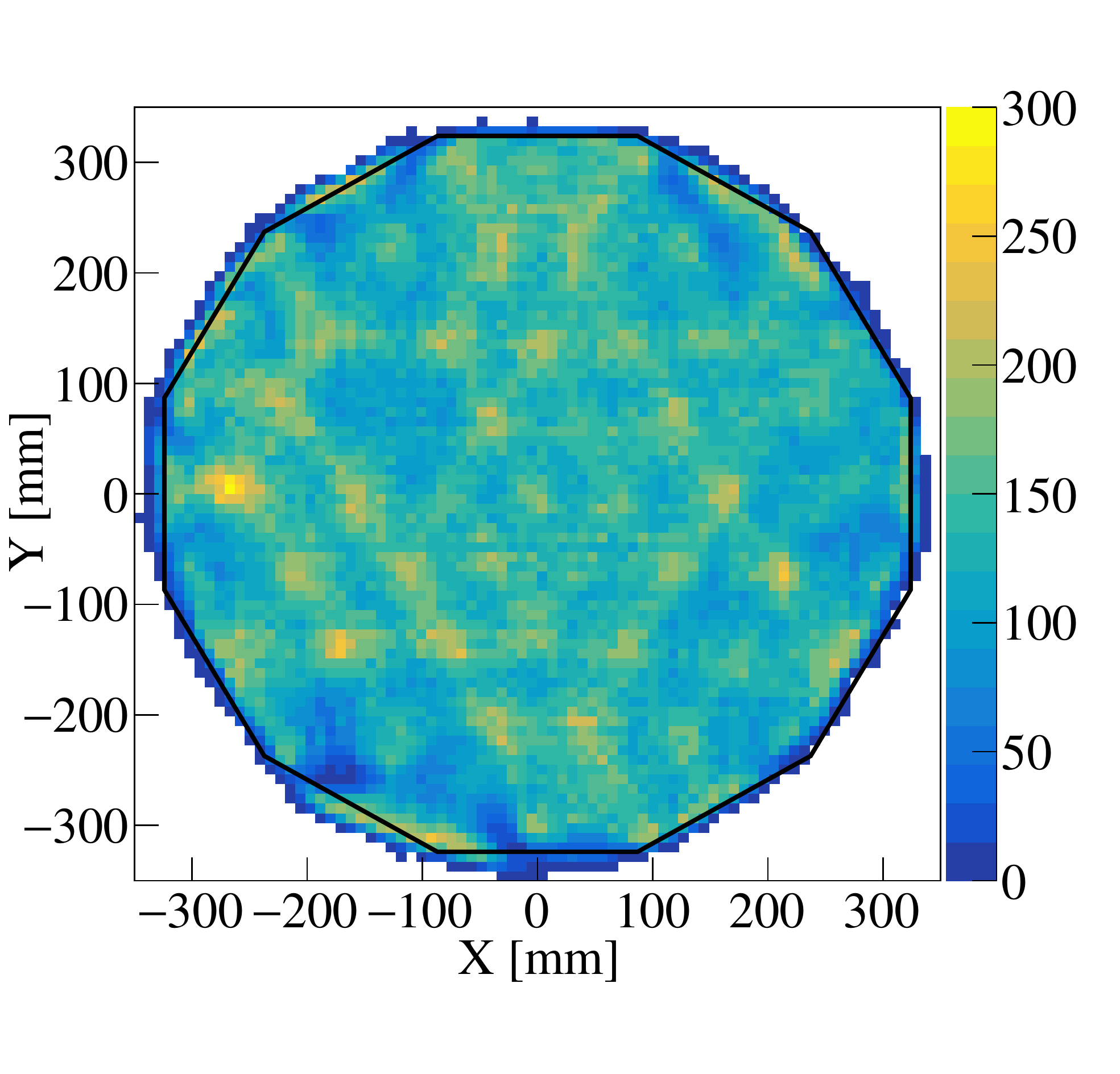}
  \includegraphics[width=0.3\textwidth]{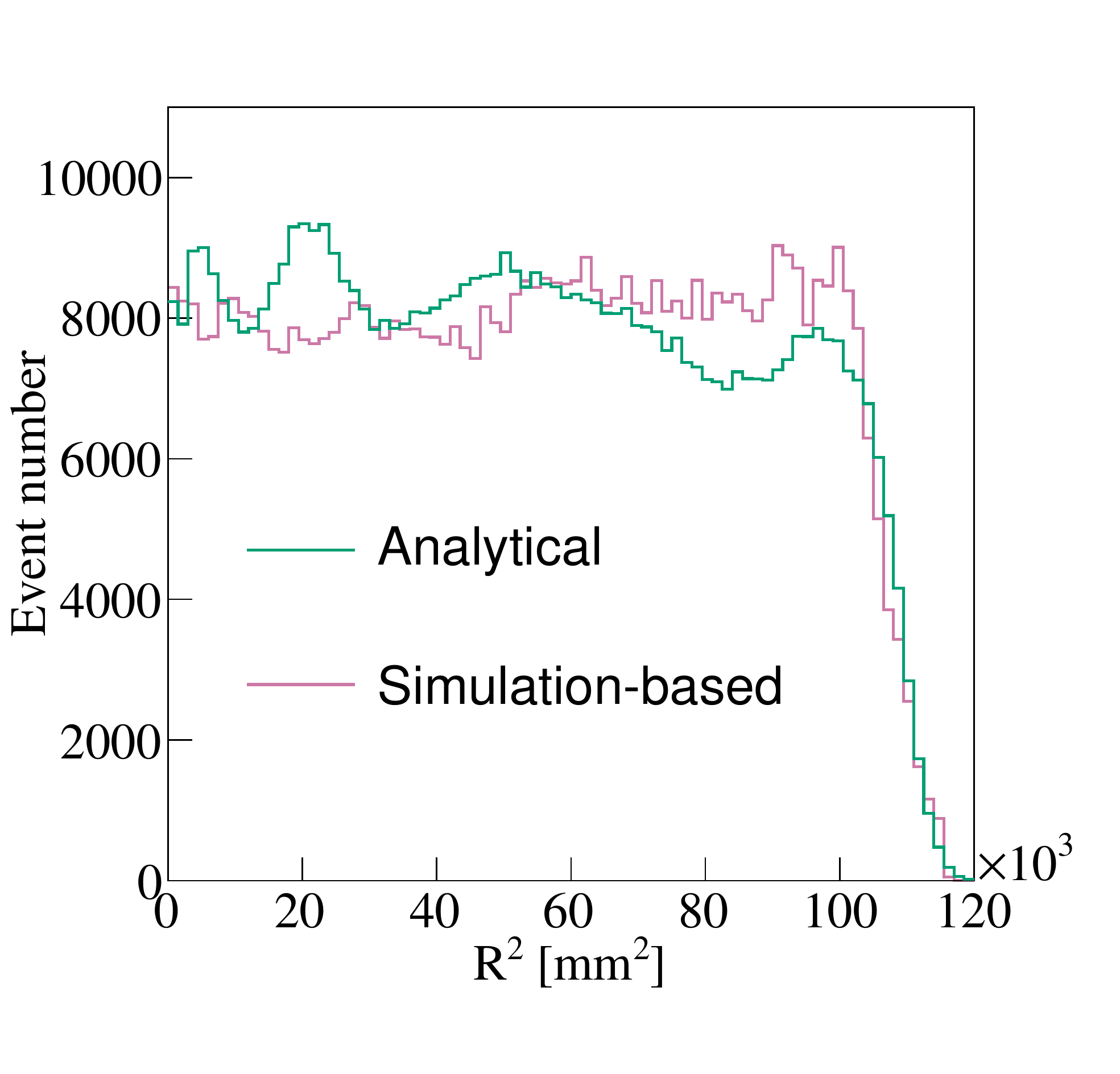}
  \caption{Horizontal distributions of $^{83\rm{m}}$Kr events
    reconstructed with simulation-based (new, left) and analytical
    (old, middle) algorithms. The simulation-based algorithm is more
    stable at the bottom-left corner where two PMTs are turned off.}
  \label{fig:kr83m_pos}
\end{figure}

The horizontal position is extracted from the $S2$ charge pattern on
the top PMT array, exploiting a data-driven photon acceptance function
(PAF), i.e., the proportion of $S2$ deposited onto each PMT for a
given horizontal position. In Ref.~\cite{Tan:2016zwf}, the PAF was
parameterized analytically which allowed position reconstruction via a
likelihood fit. However, it is found that reconstructed positions have
local distortions (clustering toward the center of the PMTs). It is
also not sufficiently stable with PMTs turned off. In this analysis,
we develop an improved PAF method utilizing the $^{83\rm{m}}$Kr
calibration data obtained in 2018, with ER peaks of 41.6~keV
distributed throughout the detector. The procedure is as follows.
\begin{enumerate}
\item The average charge pattern of $S2$ on the top PMT array is
  extracted for each given reconstructed position pixel, using the
  analytical PAF as the starting point.
\item Geant4~\cite{Agostinelli:2002hh, Allison:2006ve} optical
  simulations with realistic PandaX-II geometry and optical properties
  are carried out, assuming $S2$ photons are produced as a point
  photon source in the gas gap. The vertical photon production point
  $z_{S2}$ is adjusted for each horizontal position to find a match
  in charge pattern between the data and the simulations. 
\item Once the optimal $z_{S2}(x,y)$ is found, a new PAF is produced,
  based entirely on the tuned Geant4 simulations. The $^{83\rm{m}}$Kr
  positions are reconstructed again using the new PAF, after which a
  new data-driven charge pattern vs. position is produced.
\end{enumerate}
This procedure is carried out iteratively until the reconstruction
becomes stable. It outperforms the previous method significantly,
especially after the malfunctioned PMTs are inhibited, as shown in
Fig~\ref{fig:kr83m_pos}. For each event, the difference between the
horizontal positions reconstructed by the new and old method is
required to be smaller than 40~mm, serving as a data quality cut. The
uncertainty in horizontal position is estimated to be 5~mm based on
the $^{83\rm{m}}$Kr data, which propagated into an uncertainty in the
fiducial volume (FV).

\section{Calibration and signal model}
\label{sec:det_calib}
Various calibration runs were performed throughout the PandaX-II
operation to measure detector responses, such as the signal
uniformity, the single electron gain (SEG), the average photon
detection efficiency (PDE), and the electron extraction efficiency
(EEE), and to model the low-energy DM signal and background events. We
discuss them in turn.

\subsection{Detector uniformity calibration}
\label{sec:non_uniformity}
Mono-energetic events uniformly distributed in liquid xenon are
selected to calibrate the non-uniform distribution of detected
signals. The correction to $S1$ is a smooth three-dimensional map
based on the internal background peaks, since there is no simple
analytical parameterization. The correction of $S2$s is separated into
two parts, first an exponential attenuation in the vertical direction
due to electron losses during the drift, parameterized by the electron
lifetime (see Fig.~\ref{fig:exposure_elifetime}), then a
two-dimensional smooth map based on internal background peaks.
Ideally, {\it in situ} background peaks keep track with potential
temporal changes in the detector; thus, they are the best choice for
uniformity correction, but statistics are also an important
consideration. In the three runs, the corrections are obtained
differently, as summarized in Table~\ref{tab:uniformity}.

\begin{table}[hbt]
  \centering
  \begin{tabular}{cccc}
    \hline
    Item & Run 9 & Run 10 & Run 11 \\ \hline
    $S1$ & $^{131\rm{m}}$Xe & $^{83\rm{m}}$Kr & $^{83\rm{m}}$Kr \\
    $S2$ electron lifetime & $^{131\rm{m}}$Xe  & $^{131\rm{m}}$Xe & internal $\alpha$\\
    $S2$ horizontal & $^{131\rm{m}}$Xe and tritium & $^{131\rm{m}}$Xe & $^{83\rm{m}}$Kr\\
    \hline
  \end{tabular}
  \caption{Uniformity calibration approach used in the three runs.}
  \label{tab:uniformity}
\end{table}

\begin{figure}[hbt]
  \centering
  \begin{subfigure}{.32\textwidth}
    \includegraphics[width=1.0\textwidth]{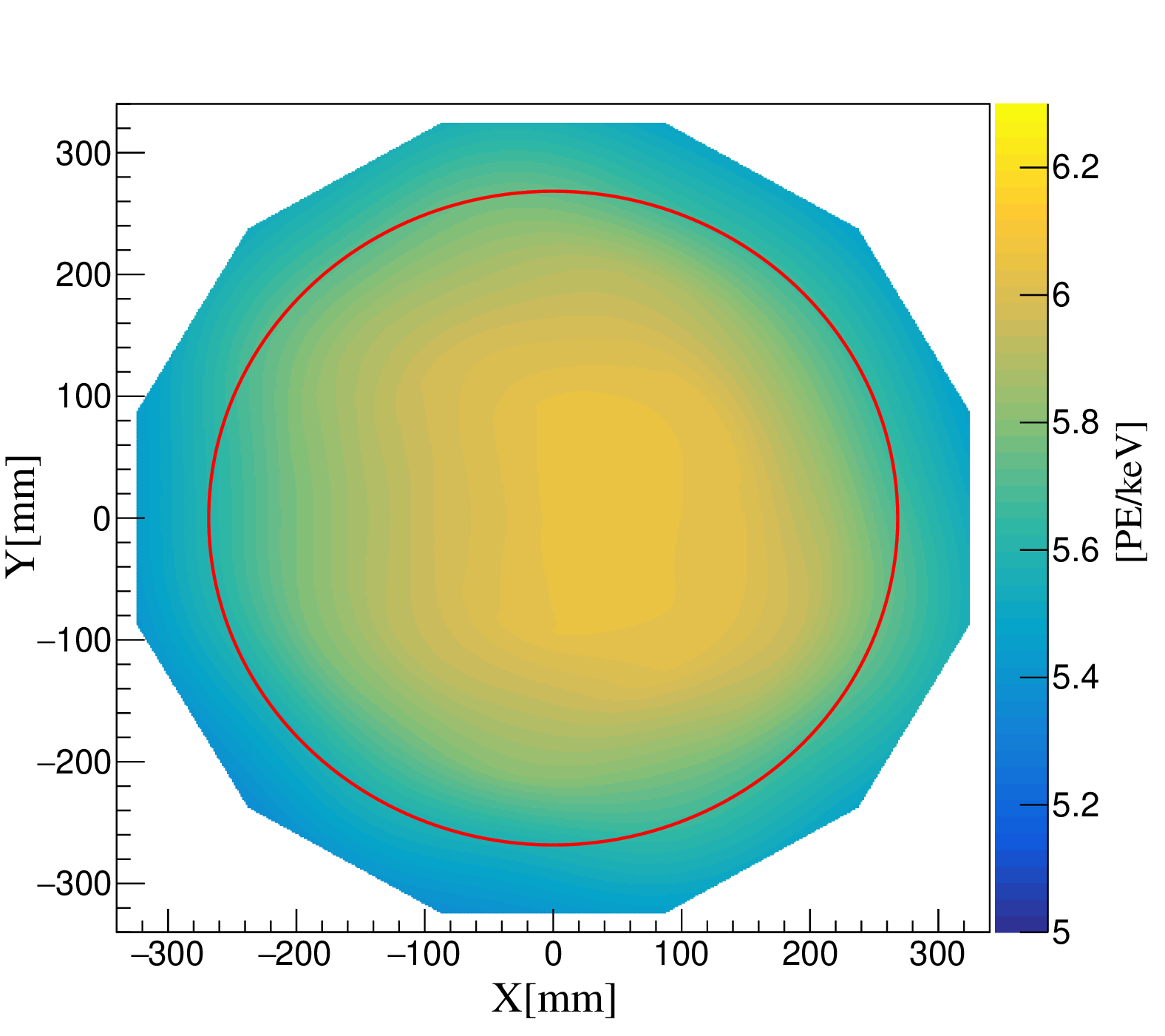}
    \caption{$S1$, horizontal y vs. x}
   \end{subfigure}
  \begin{subfigure}{.32\textwidth}
    \includegraphics[width=1.0\textwidth]{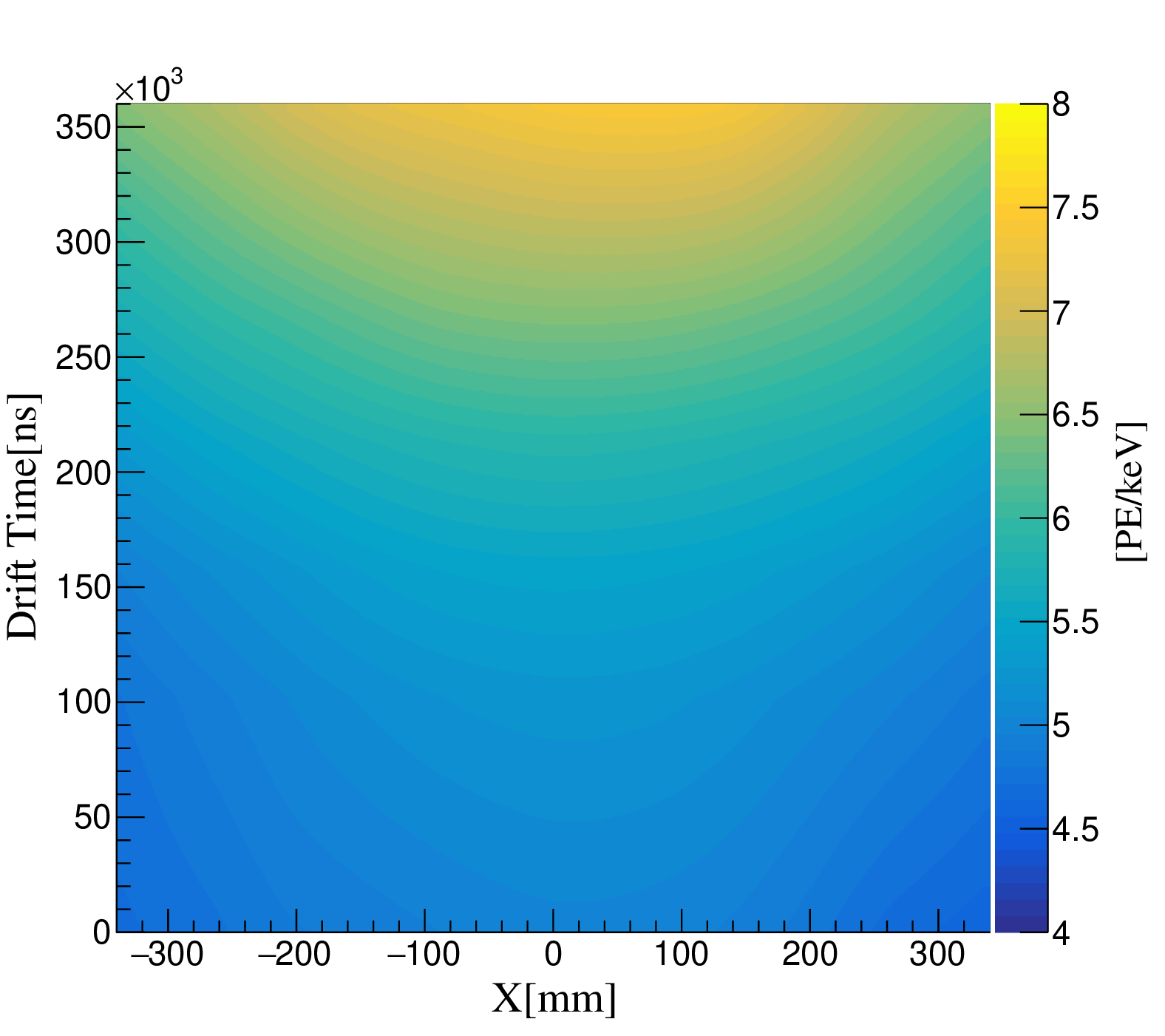}
    \caption{$S1$, drift time vs. x}
   \end{subfigure}
  \begin{subfigure}{.32\textwidth}
    \includegraphics[width=1.0\textwidth]{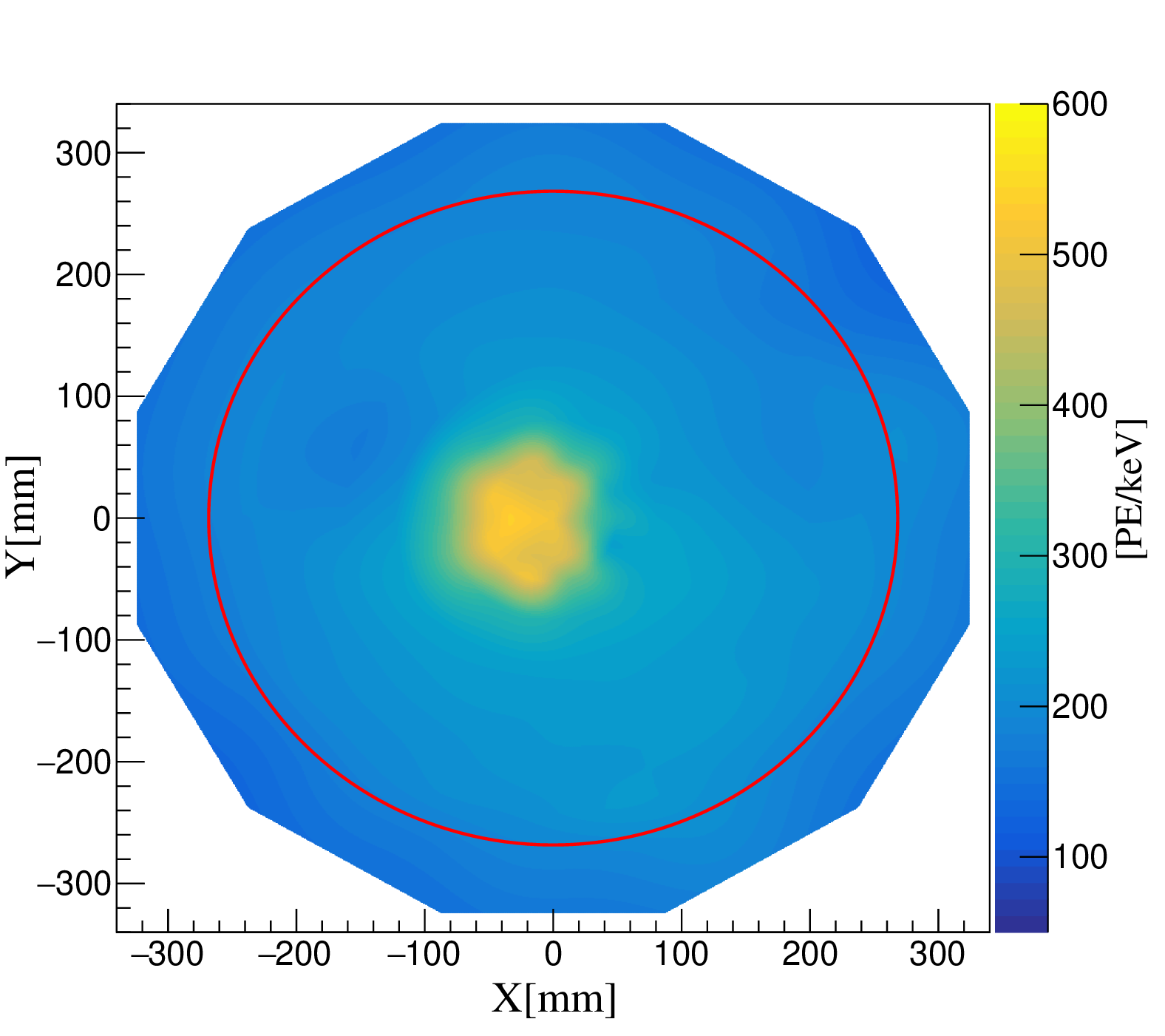}
    \caption{$S2$, horizontal y vs. x}
   \end{subfigure}
   \caption{The non-uniformity correction map obtained from the
     $^{83\rm{m}}$Kr data for $S1$ and $S2$.}
  \label{fig:kr_maps_run11}
\end{figure}

In Run 9, due to the long exposure of xenon on the surface, the
detector has a rather high rate of 164-keV gamma events from the
neutron-activated $^{131\rm{m}}$Xe, based on which both $S1$ and $S2$
maps are produced. In Run 10, the $S2$ horizontal correction and the
electron lifetime are obtained from $^{131\rm{m}}$Xe, but the
three-dimensional $S1$ correction is based on the $^{83\rm{m}}$Kr data
in 2019 due to its excellent statistics. In Run 11, as the
$^{131\rm{m}}$Xe rate becomes insufficient, a $^{83\rm{m}}$Kr map is
applied to the $S1$ and $S2$ horizontal corrections. The electron
lifetime, on the other hand, is obtained {\it in situ} using
$^{222}$Rn and $^{218}$Po alpha events with $S2_b$, the portion of
$S2$ detected by the bottom PMT array~\footnote{On the other hand, in
  Ref.~\cite{Ni:2019kms}, the electron lifetime was obtained based on
  the high energy gamma data to optimize the resolution in that
  regime, but may already contain the saturation effects.}. The
$^{83\rm{m}}$Kr maps used in Run 11 for $S1$ (two projections) and
$S2$ are shown in Fig.~\ref{fig:kr_maps_run11}. The maximum variations
in the FV are $[-19.0\%, 30.2\%]$ in $S1$ and $[-31.9\%, 16.4\%]$ in
$S2$.

Note that $S2$ photons are substantially clustered on the top PMT
array; therefore, they are subject to saturation effects. In this
analysis, both $S2$ and $S2_b$ are corrected using their corresponding
maps according to Table~\ref{tab:uniformity}. It is found that that
the $S2$ map in Run 9 (when most PMTs were operated under normal gain)
is biased due to saturation; thus, the $S2$ horizontal map is further
corrected based on the mean value of the $S2$s in the tritium
calibration data.

\subsection{Measurement of BLS nonlinearity}
\label{sec:readout_efficiency}
The BLS threshold for each digitizer channel was set at an amplitude
of 2.75 mV above the baseline. For comparison, the SPE for a gain of
10$^6$ corresponds to a mean amplitude of 4.4~mV in the
digitizer. Although the gains vary from PMT to PMT, fixed thresholds
are needed to avoid excessive data size due to baseline noises. The
channel-wise BLS inefficiency is negligible for Run 9, since all PMTs
were operating under the normal gain, but becomes more significant
during Runs 10 and 11 due to the low-gain PMTs
(Sec.~\ref{sec:low_gain_pmts}). Consequently, the detected $S1_d$ and
$S2_d$ are suppressed from the actual $S1$ and $S2$. As long as $S1_d$
and $S2_d$ fall into selection windows, BLS does not cause an event
loss but rather a nonlinearity in $S1$ and $S2$, stronger for small
signals and approaching unity for large ones.

\begin{figure}[htb]
  \centering
  \includegraphics[width=.45\textwidth]{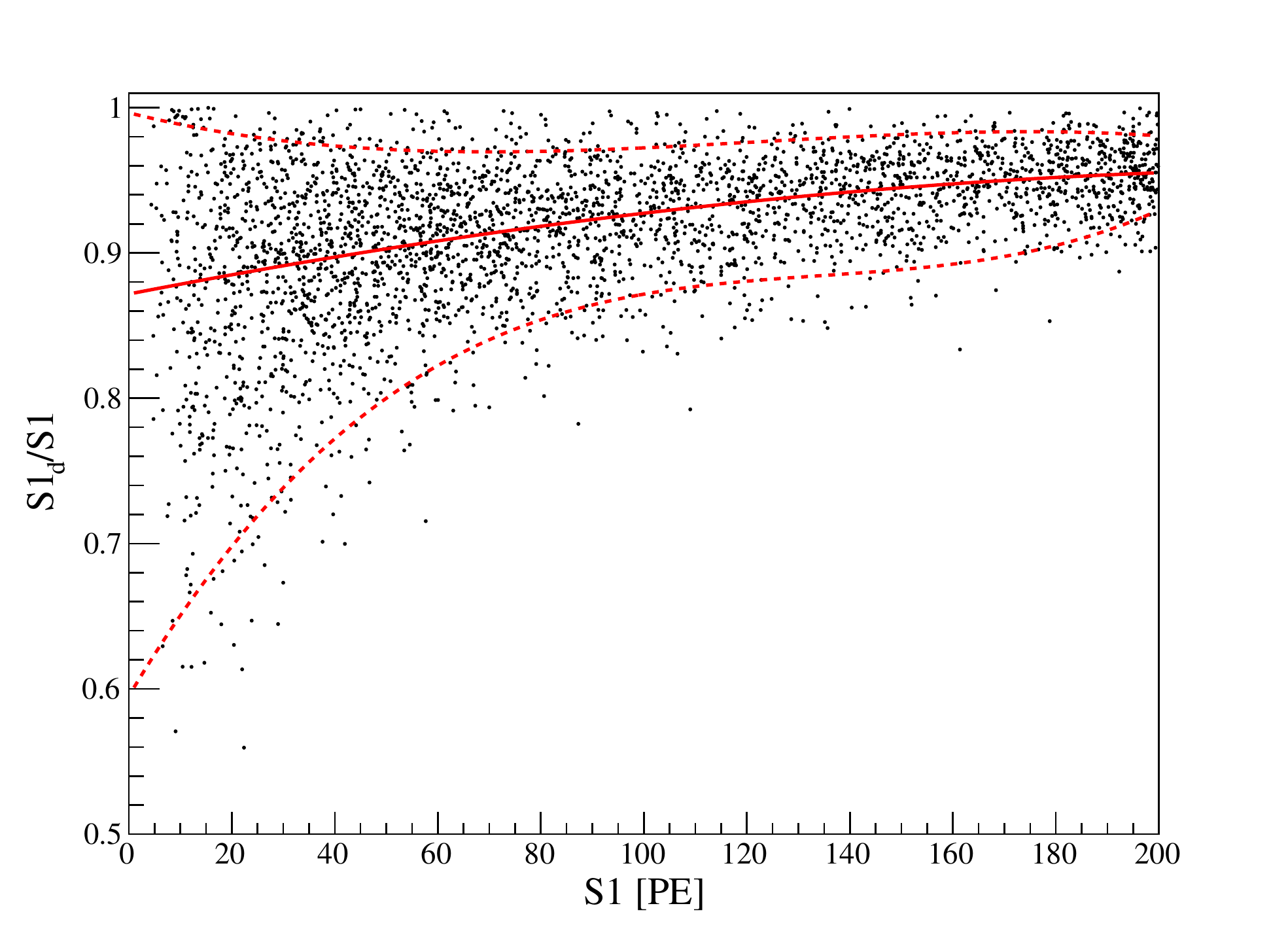}
  \includegraphics[width=.45\textwidth]{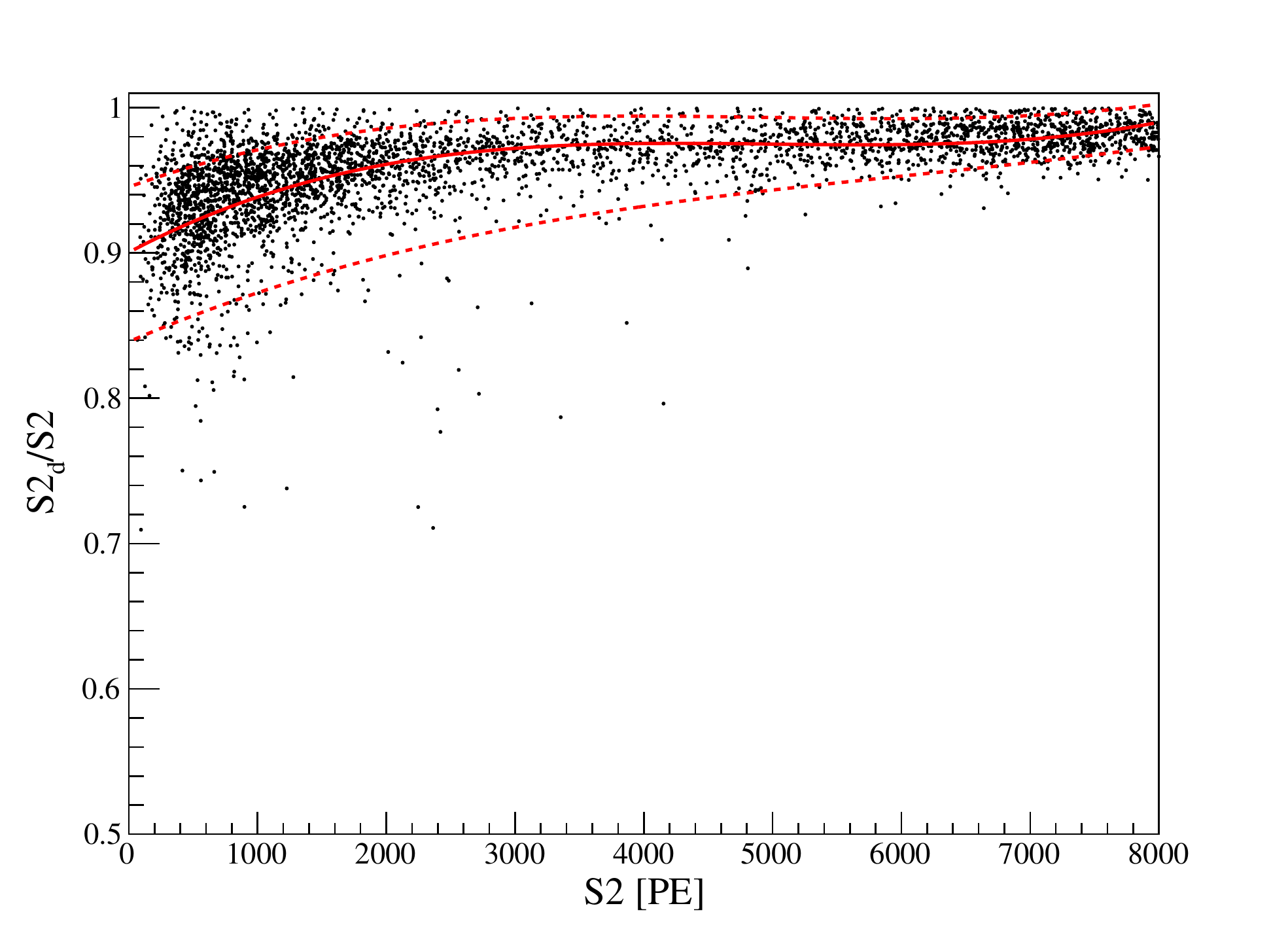}
  \caption{The distribution of BLS nonlinearities $f_1$ ($f_2$) versus
    $S1$ ($S2$) using neutron calibration data with the BLS firmware
    disabled. The solid and dashed lines are the median and 90\%
    quantile fits.}
  \label{fig:zle_efficiency}
\end{figure}

The nonlinearities depend subtly on the shapes and actual
distributions of $S1$ and $S2$ on individual PMTs. Therefore, instead
of adopting the single-channel BLS efficiency from the LED calibration
as in Ref.~\cite{Cui:2017nnn}, in Run 11 we performed direct
measurement using neutron calibration data with low energy events
distributed throughout the detector. During this special data
acquisition, the BLS firmware was disabled, so that all waveform data
were saved (thresholdless), and the standard $S1$ and $S2$
identifications were performed on the data~\footnote{The software
  threshold for pulse identification is very low with negligible
  inefficiency.}. We then applied the BLS algorithm on the data as
that in the firmware and obtained $S1_d$ and $S2_d$, from which the
BLS nonlinearities $f_1=\frac{S1_d}{S1}$ and $f_2=\frac{S2_d}{S2}$
were determined in an event-by-event manner. The distributions of
$f_1$ and $f_2$ are shown in Fig.~\ref{fig:zle_efficiency} with clear
spreads due to fluctuations in the data. They are modeled into smooth
probability density functions (PDFs) when later converting $S1$ and
$S2$ into $S1_d$ and $S2_d$ in our signal and background models. In
the remainder of this paper, $S1$ and $S2$ refer to the detected
$S1_d$ and $S2_d$ for simplicity, unless otherwise specified.

\subsection{Calibration for PDE, EEE, and SEG}
With updates in the lower level analysis mentioned above, we extract
detector parameters for Run 9 and Run 10. For each event, the energy
is reconstructed as
\begin{equation}
\label{eq:energy_comb}
  E_{\rm{rec}} = 0.0137\,{\rm{keV}}\times
  \left(\frac{S1}{\rm{PDE}}+\frac{S2}{\rm{EEE}\times\rm{SEG}}\right),
\end{equation}
where $S1$ and $S2$ have been corrected for uniformity in all the
runs, and BLS non-linearity in Runs 10 and 11. Note that in all three
run sets, $S2$ saturation is found for energy higher than 200 keV. For
conservativeness, for energy higher than 30~keV, we calculate $S2$ as
$\alpha \times S2_b$, and $\alpha$ is 3.0 and 3.18 for Run 9 and Runs
10/11, determined using data in the low energy DM search region. The
SEG is determined by selecting the smallest $S2$s, and the spectrum is
fitted with a combination of a threshold function and a double
Gaussian encoding the single and double electron signals. The best
fit parameters of PDE and EEE are determined by performing a parameter
scan when fitting the energy spectra of known ER peaks from the
calibration data, requiring a global minimization of $\chi^2$ between
the reconstructed and expected energies.

\begin{figure}[htb]
  \centering
\includegraphics[width=0.7\textwidth]{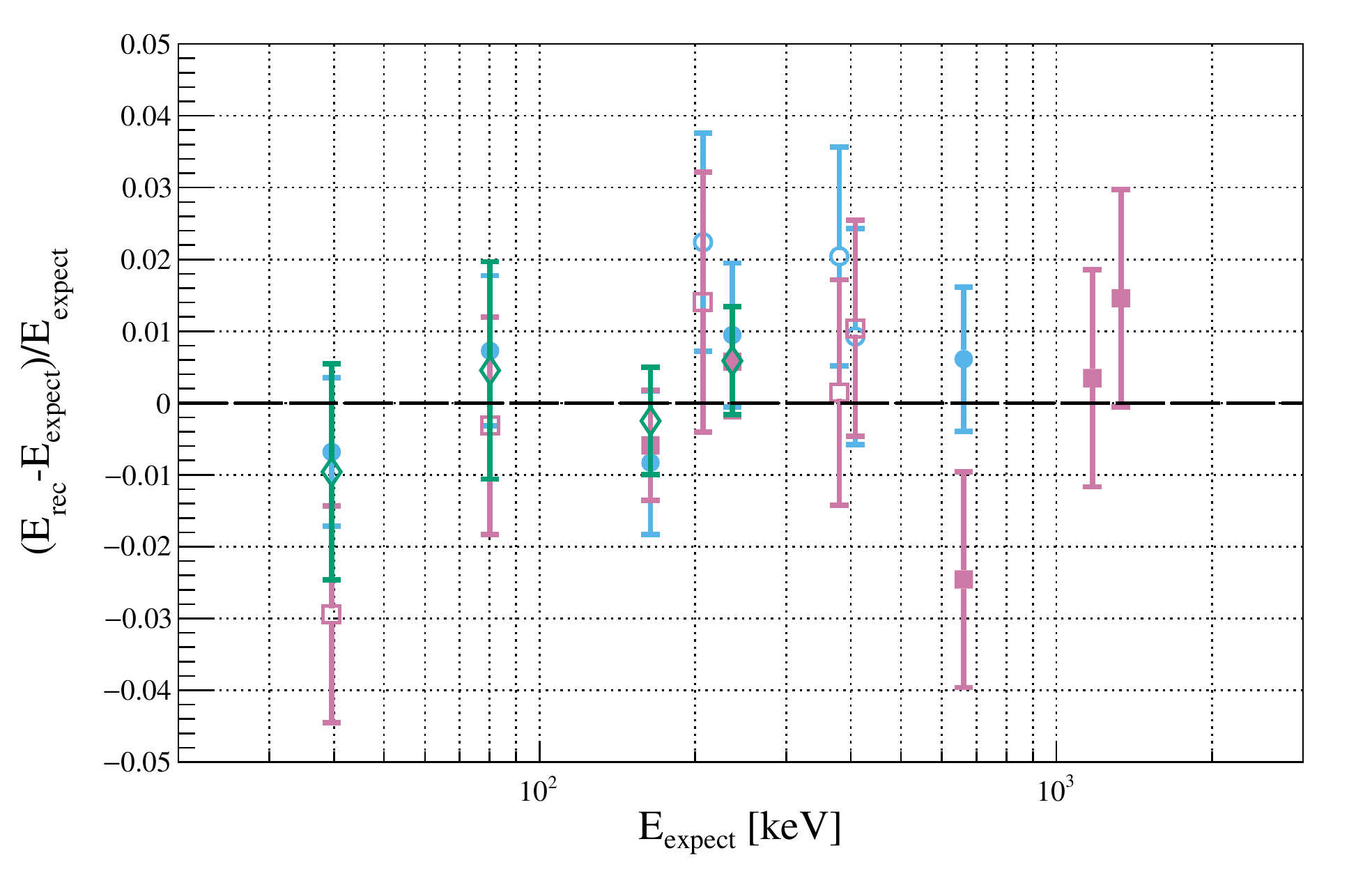}
\caption{Fractional difference between the reconstructed energy
  $E_{\rm{rec}}$ and expected energy $E_{\rm expect}$ for
  characteristic gamma peaks in Run 9 (blue circles), Run 10 (magenta
  squares), and Run 11 (green diamonds). Peaks include {39.6 keV (n,
    $^{129}$Xe), 80.2 keV (n, $^{131}$Xe), 164 keV ($^{131\rm{m}}$Xe),
    236 keV ($^{129\rm{m}}$Xe), 202.8, 375, and 408 keV ($^{127}$Xe),
    662 keV ($^{137}$Cs ) and 1173 and 1332 keV($^{60}$Co)}.
  Uncertainties are dominated by systematic components.  Closed
  symbols represent points used in the fits, and open symbols are
  those test peaks. See text for details.}
  \label{fig:energy_diviation}
\end{figure}

In Run 9, we select the prompt de-excitation gamma rays from the
neutron calibration, 39.6~keV from $^{129}$Xe, and 80.2~keV from
$^{131}$Xe, both corrected for the small shifts caused by the mixture
of NR energy. ER peaks due to the same neutron illumination, 164~keV
($^{131\rm{m}}$Xe) and 236~keV ($^{129\rm{m}}$Xe), are also
selected. For higher energy gamma peaks, we only select the 662~keV
peak from $^{137}$Cs to avoid potential bias in energy due to the
saturation of $S2$. In Run 10, to avoid BLS nonlinearities at lower
energies, higher energy peaks, including 164~keV, 236~keV and 662~keV,
together with gammas of 1173~keV and 1332~keV from $^{60}$Co, are
selected. The systematic uncertainty of each peak is initially set to
be $1\%$, guided by the sensitivity of the 164 keV to data cuts,
uniformity corrections, fit range, temporal drifts, etc. Additional
uncertainties are assigned to peaks that entail more uncertainty due
to saturation effects, mixture of NR, etc.  The quality of the fits is
illustrated in Fig.~\ref{fig:energy_diviation} where the relative
differences between the reconstructed and the expected energies are
plotted. Peaks not used in fitting serve as critical checks, shown as
the open symbols in the figure. The overall agreement is better than
$3\%$. The uncertainties of PDE and EEE are estimated based on
parameter contours bounded by $\Delta\chi^2=1$.

The resulting parameters in different run sets are summarized in
Tab.~\ref{tab:pde_eee_seg}. For Run 11, since the field configurations
stays the same as in Run 10, the PDE and EEE are obtained by scaling the Run 10
values according to the average $S1$ and $S2$ from the 164-keV peak in
the detector.
\begin{table}[htb]
\centering
\begin{tabular}{cccc}
\hline
Run & PDE (\%) & EEE (\%) & SEG (PE/$e^-$) \\
\hline
9 & $11.5\pm0.2$ & $46.3\pm1.4$ & $24.4\pm0.4$ \\
10 & $12.1\pm0.5$ & $50.8\pm2.1$ & $23.7\pm0.8$ \\
11 & $12.0\pm0.5$ & $47.5\pm2.0$ & $23.5\pm0.8$ \\
\hline
\end{tabular}
\caption{Summary of PDE, EEE and SEG in three DM search data runs in PandaX-II.}
\label{tab:pde_eee_seg}
\end{table}

\subsection{Calibration of low-energy ER and NR responses}
\label{sec:calibration_model}
For the ER calibration, as in Ref.~\cite{Cui:2017nnn}, the tritiated
methane data are used for Run 9, but reanalyzed using an updated PMT
configuration and uniformity correction. Different from
Ref.~\cite{Cui:2017nnn}, the two $^{220}$Rn data sets in Run
11~\cite{Ni:2019kms} are combined and used for both Runs 10/11. For
the NR calibration, in Run 11, 19, 158 low energy single-scatter NR
events in the FV are identified, allowing a more accurate modeling of
the NR responses. These data are used as the NR calibration for both
Run 10 and Run 11.

\begin{figure}[htb]
  \centering
  \includegraphics[width=.45\textwidth]{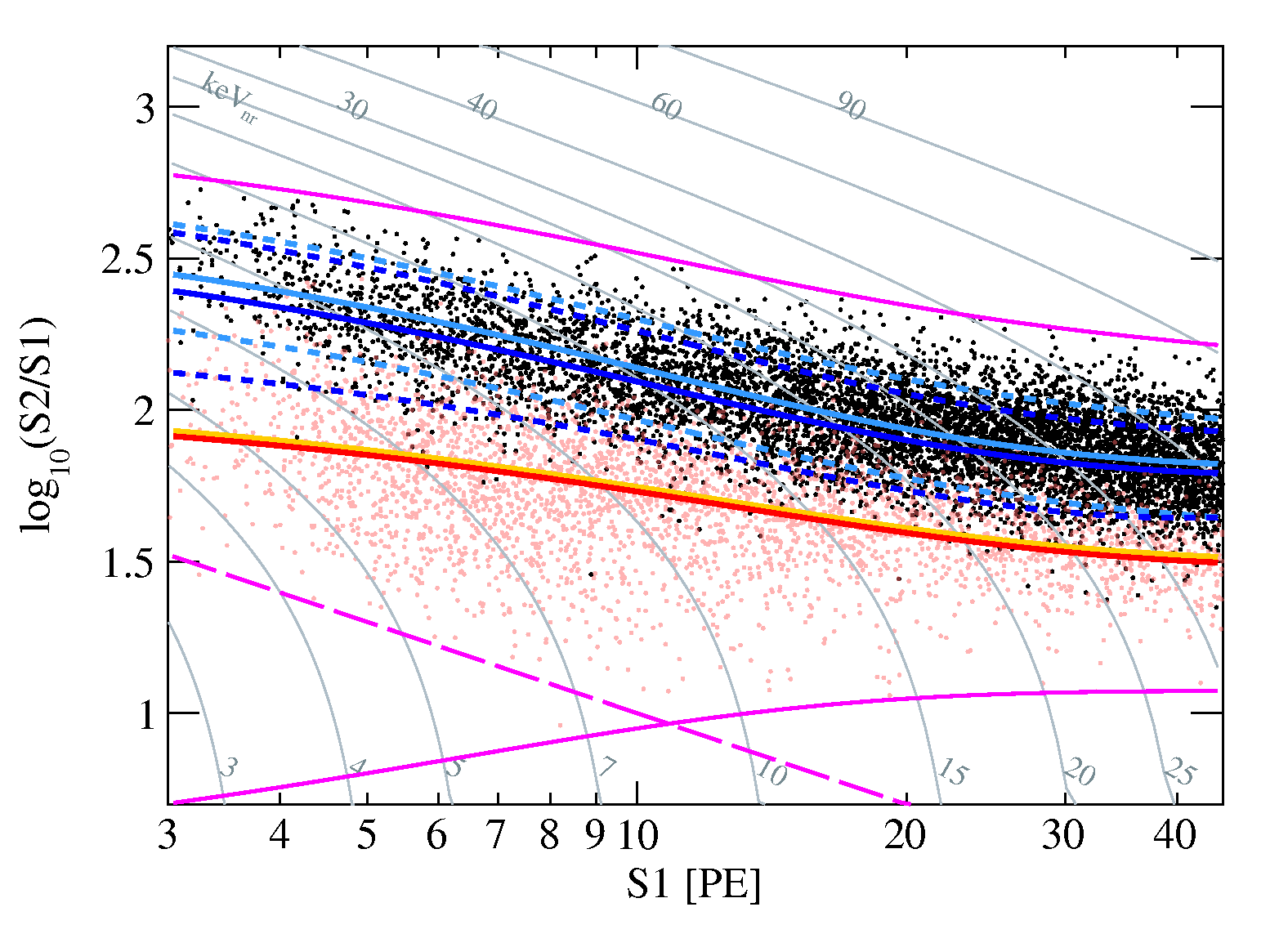}
  \includegraphics[width=.45\textwidth]{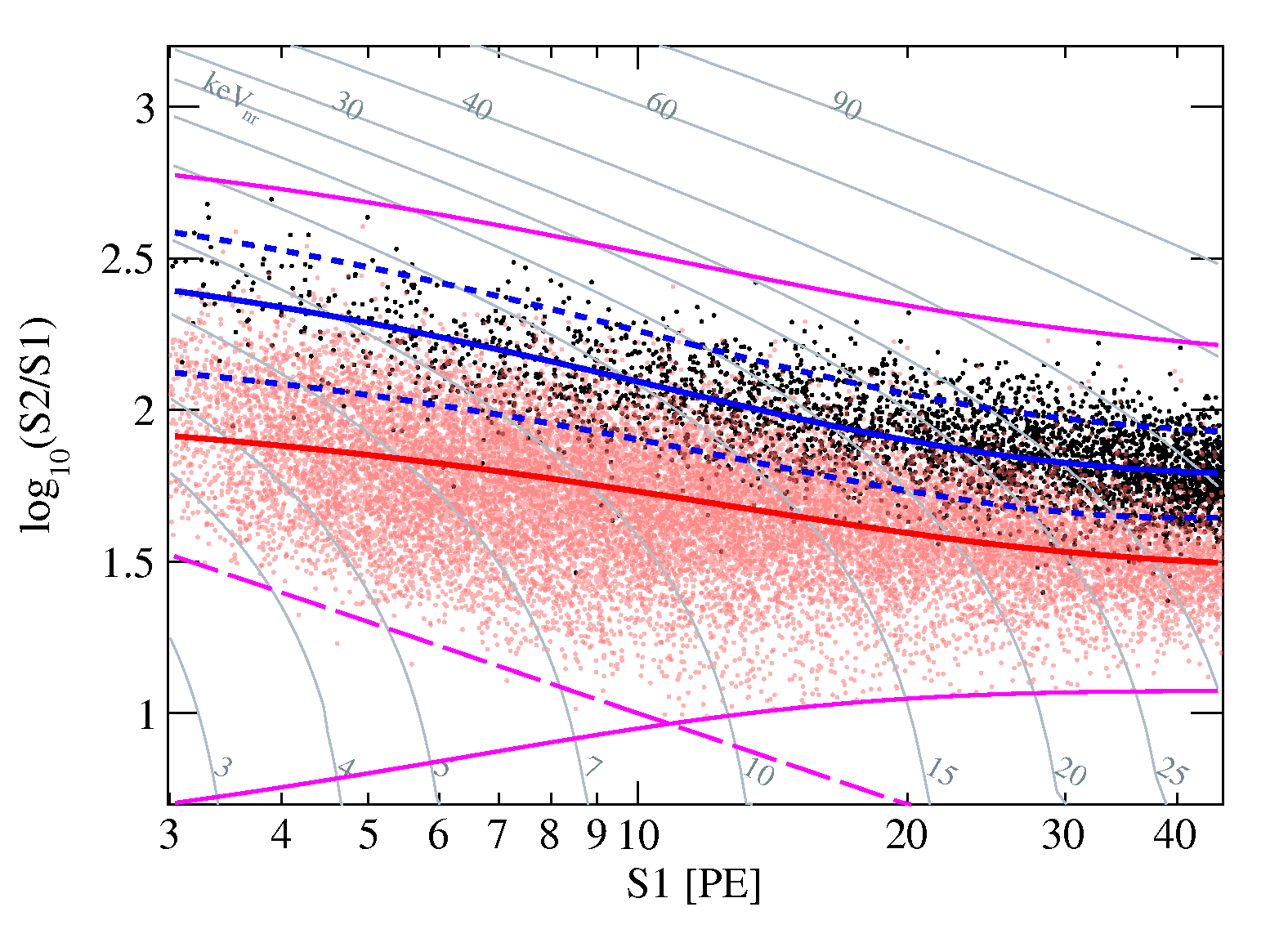}
  \caption{The distributions of calibration data in Run 9 and
    Runs~10/11 in $\log_{10}(S2/S1)$ vs. $S1$: ER (black), NR
    (red). The light and dark blue lines represent the fitted ER
    medians for Run 9 and Runs 10/11, respectively, and the dashed
    blue lines are the corresponding $90\%$ quantiles. The orange and
    red lines represent the fitted NR medians for Run 9 and Runs
    10/11, respectively. The impact of differences in PDE/EEE/SEG is
    confirmed to be negligible.  }
  \label{fig:calibration_data}
\end{figure}

The distributions of $\log_{10}(S2/S1)$ vs. $S1$ for ER and NR
calibration events in Run 9 and Runs 10/11 are shown separately in
Figs.~\ref{fig:calibration_data}. As expected, a shift in the ER
distribution is observed due to different drift fields, but not in the
NR distribution~\cite{Aprile:2006kx}. The discrimination power of the
detector to reject ER backgrounds can be evaluated by the leakage ratio
$r$, defined as the number of ER signals leaking below the NR band
median, which is measured in Fig.~\ref{fig:calibration_data} to be
$53/7089=0.75\pm0.10\%$ in Run 9, and $28/3463=0.81\pm0.15\%$ in Runs
10/11.
 
\begin{figure}[htb!]
  \centering  
  \begin{subfigure}{0.3\textwidth}
    \includegraphics[width=1\textwidth]{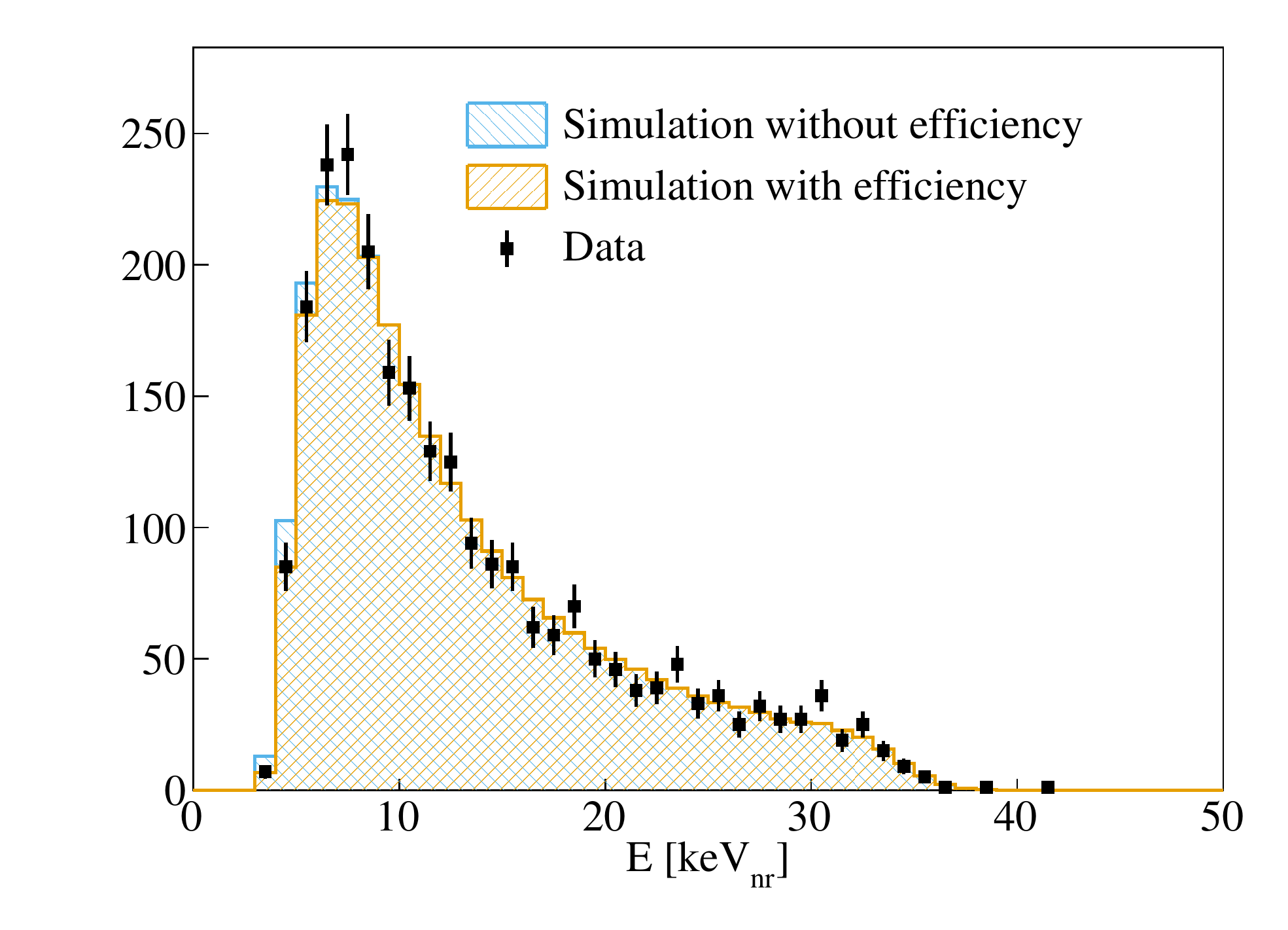}
    \caption{Run 9 AmBe energy}
  \end{subfigure}
  \begin{subfigure}{0.3\textwidth}
    \includegraphics[width=1\textwidth]{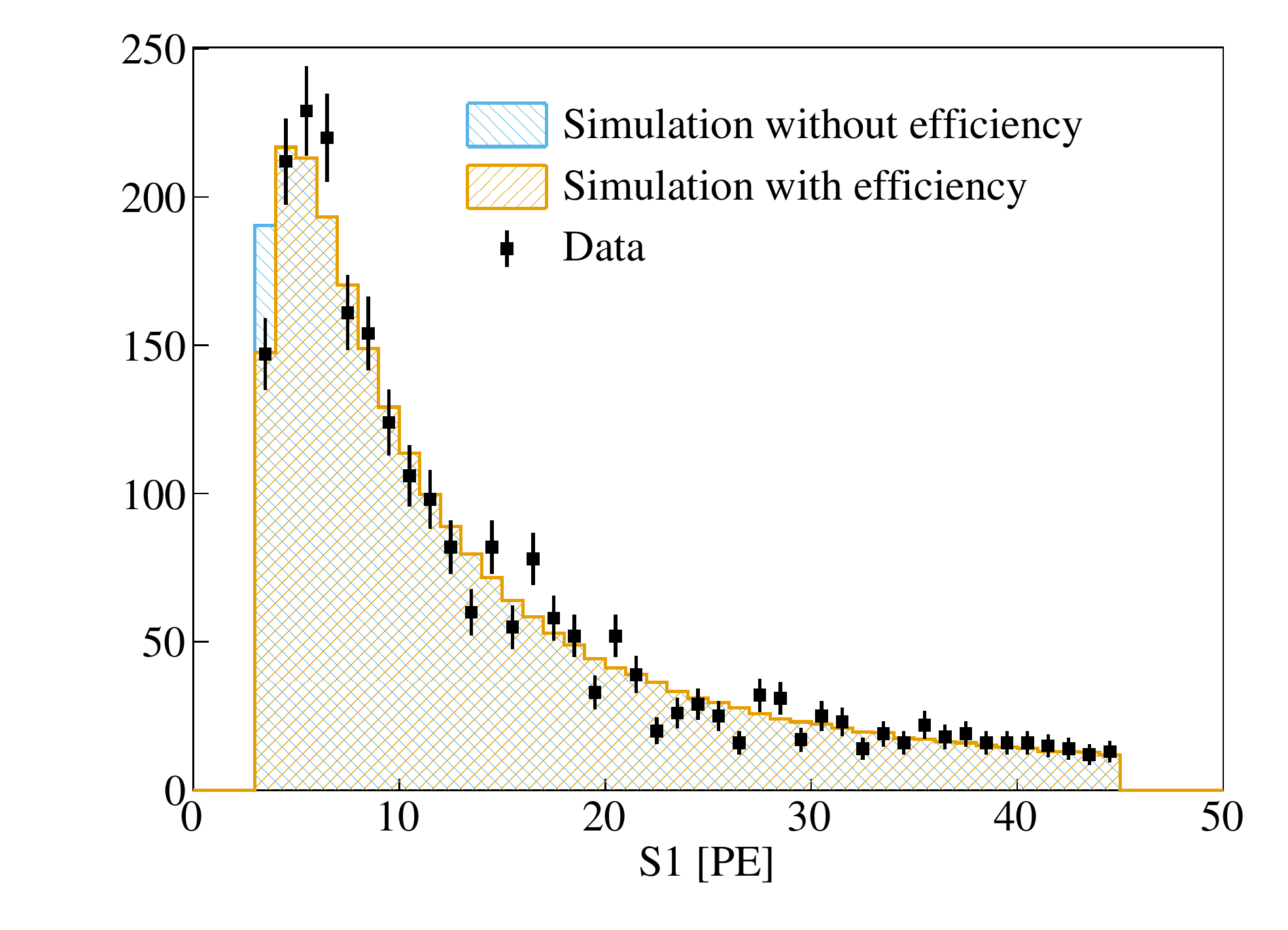}
    \caption{Run 9 AmBe $S1$}
  \end{subfigure}
  \begin{subfigure}{0.3\textwidth}
    \includegraphics[width=1\textwidth]{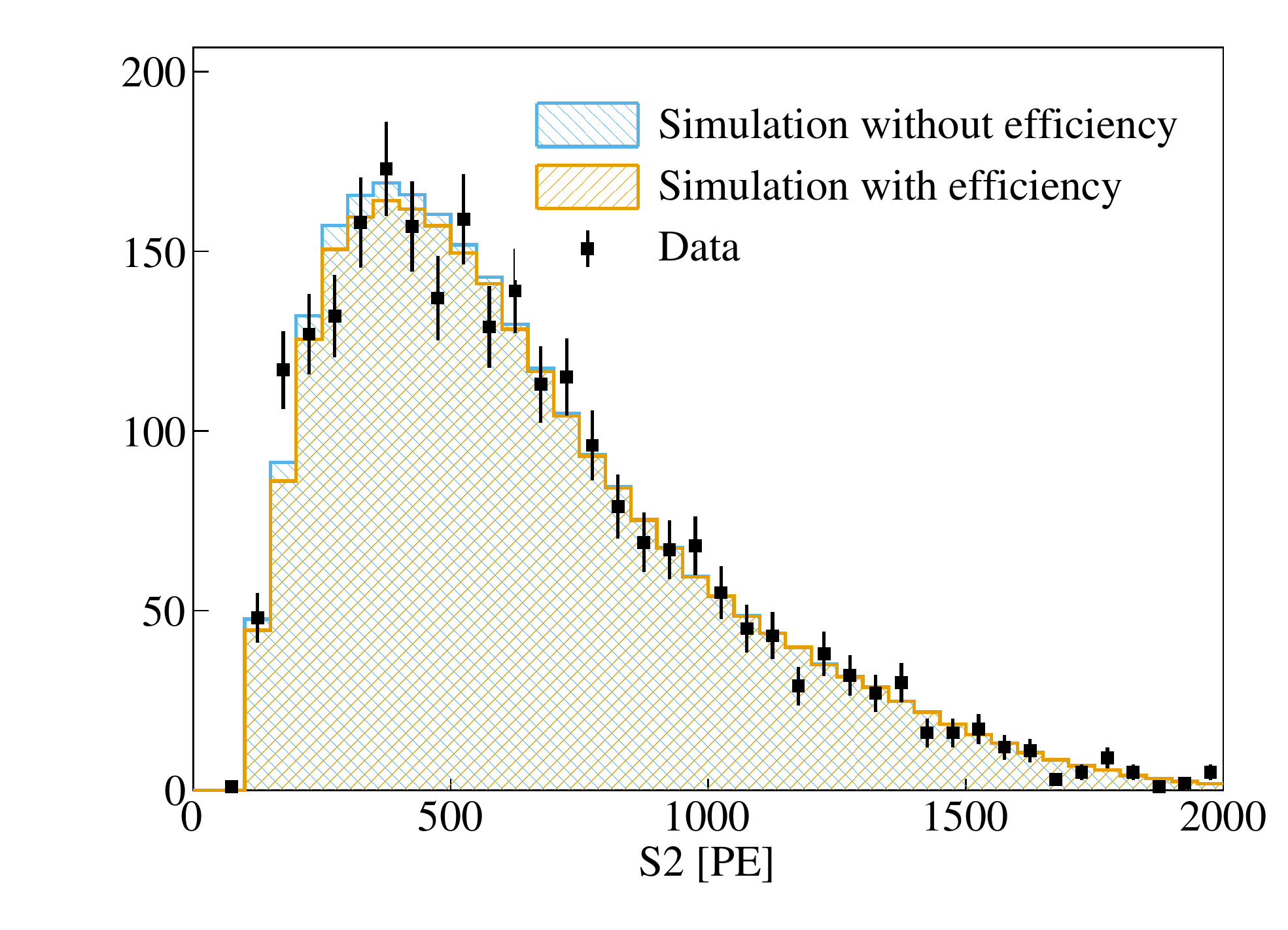}
    \caption{Run 9 AmBe $S2$}
  \end{subfigure}
  \begin{subfigure}{0.3\textwidth}
    \includegraphics[width=1\textwidth]{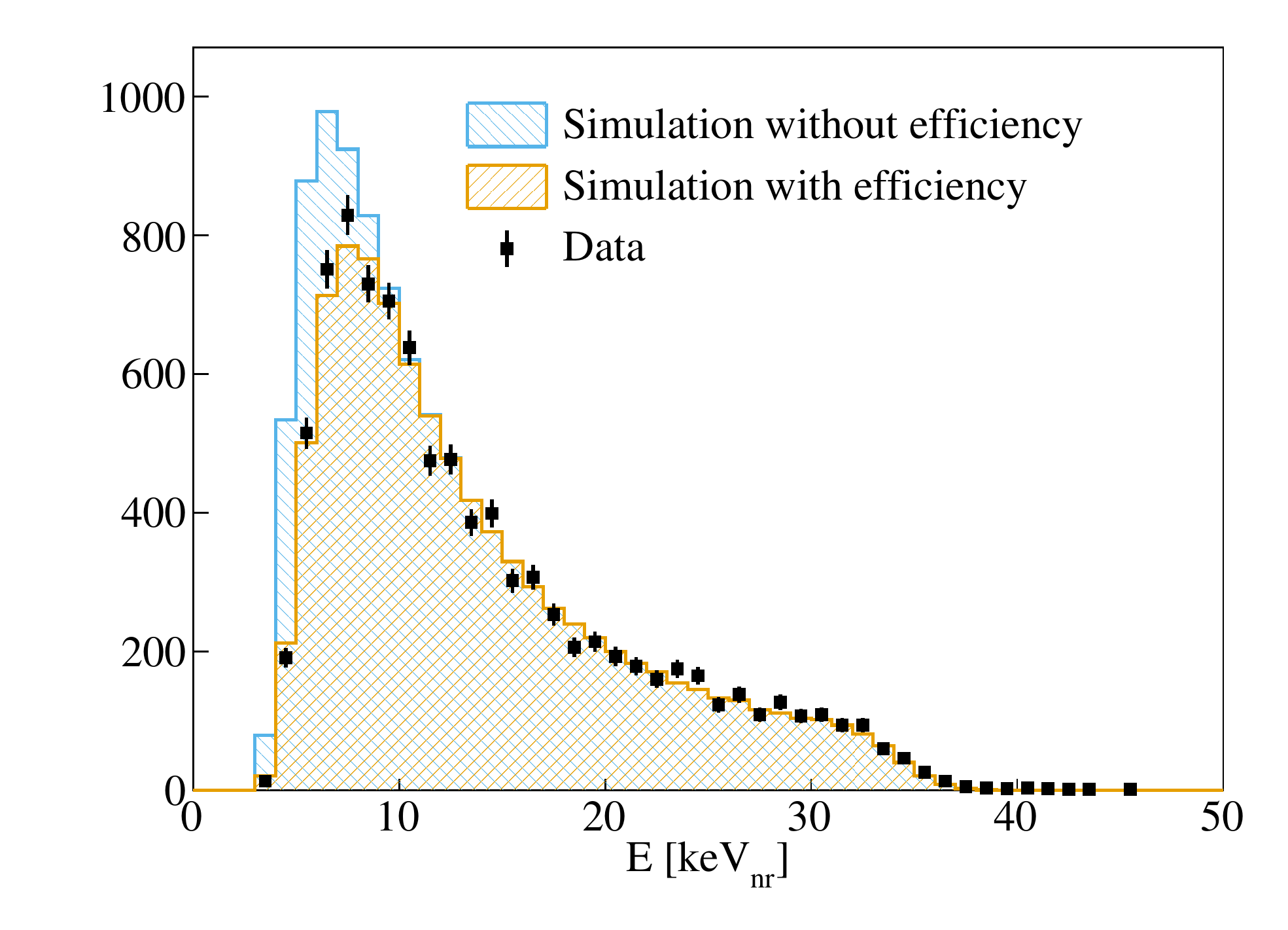}
    \caption{Runs 10/11 AmBe energy}
  \end{subfigure}
  \begin{subfigure}{0.3\textwidth}
    \includegraphics[width=1\textwidth]{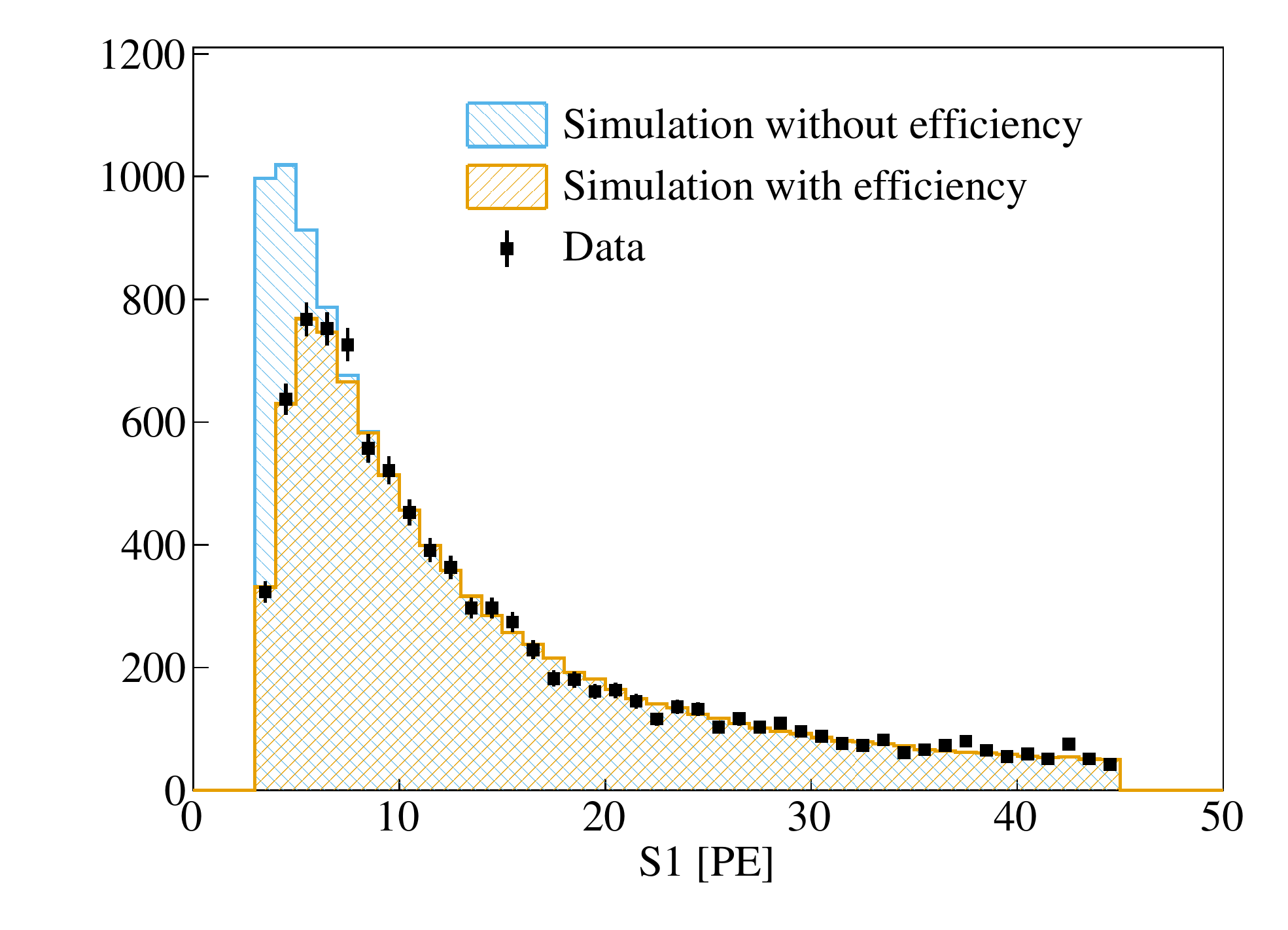}
    \caption{Runs 10/11 AmBe $S1$}
  \end{subfigure}
  \begin{subfigure}{0.3\textwidth}
    \includegraphics[width=1\textwidth]{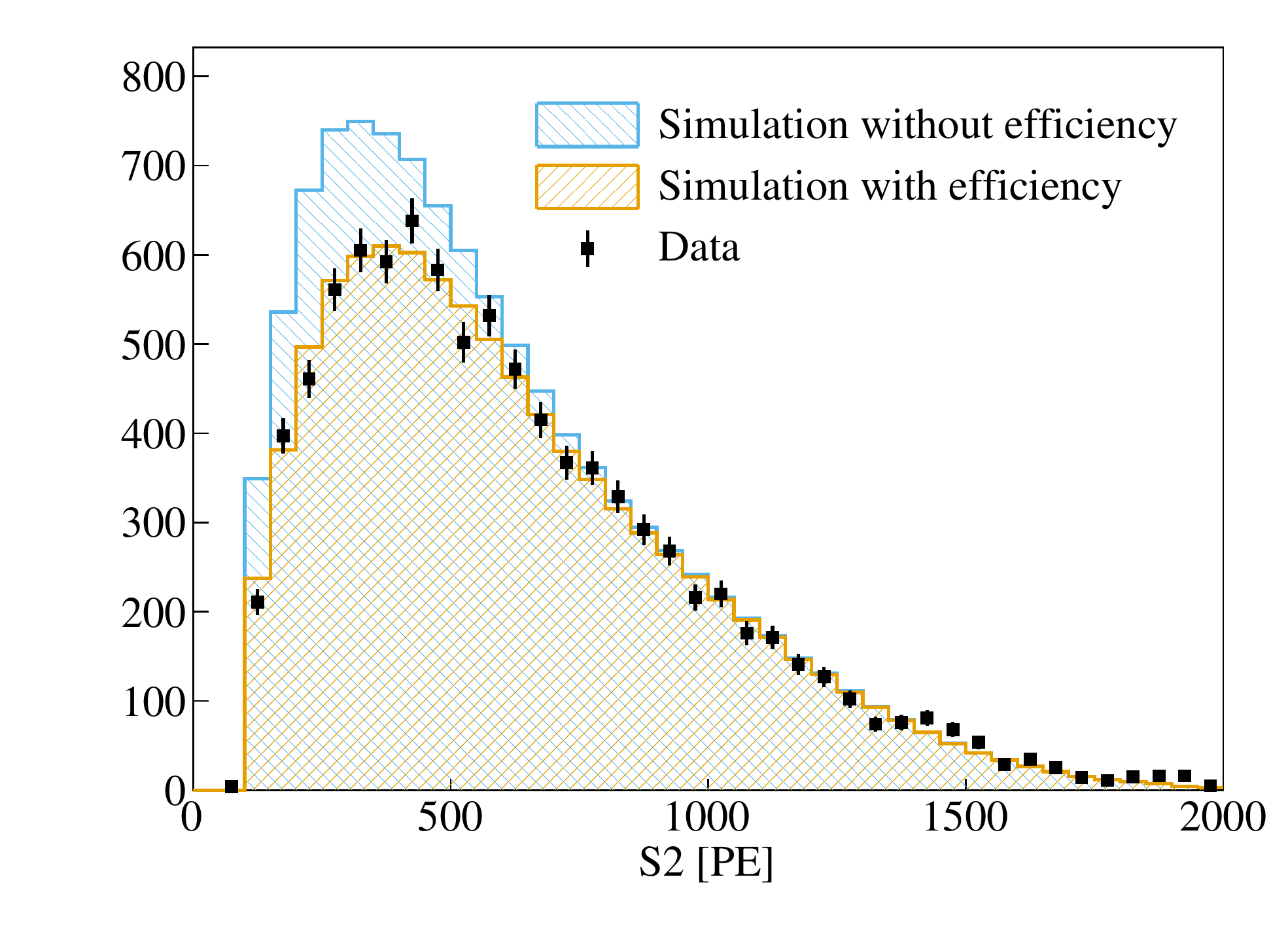}
    \caption{Runs 10/11 AmBe $S2$}
  \end{subfigure}
  \begin{subfigure}{0.3\textwidth}
    \includegraphics[width=1\textwidth]{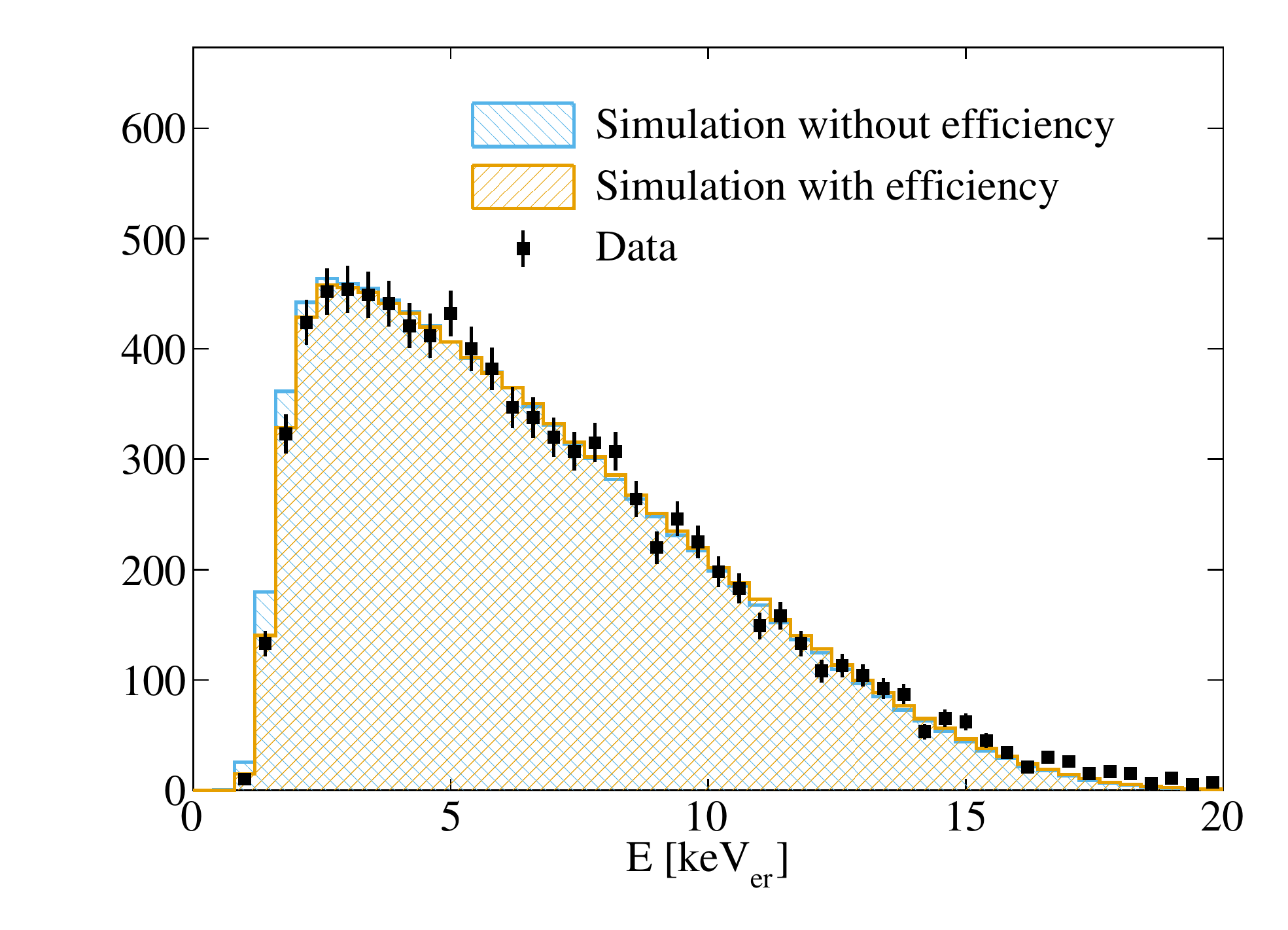}
    \caption{Run 9 CH$^{3}$T energy}
  \end{subfigure}  
  \begin{subfigure}{0.3\textwidth}
    \includegraphics[width=1\textwidth]{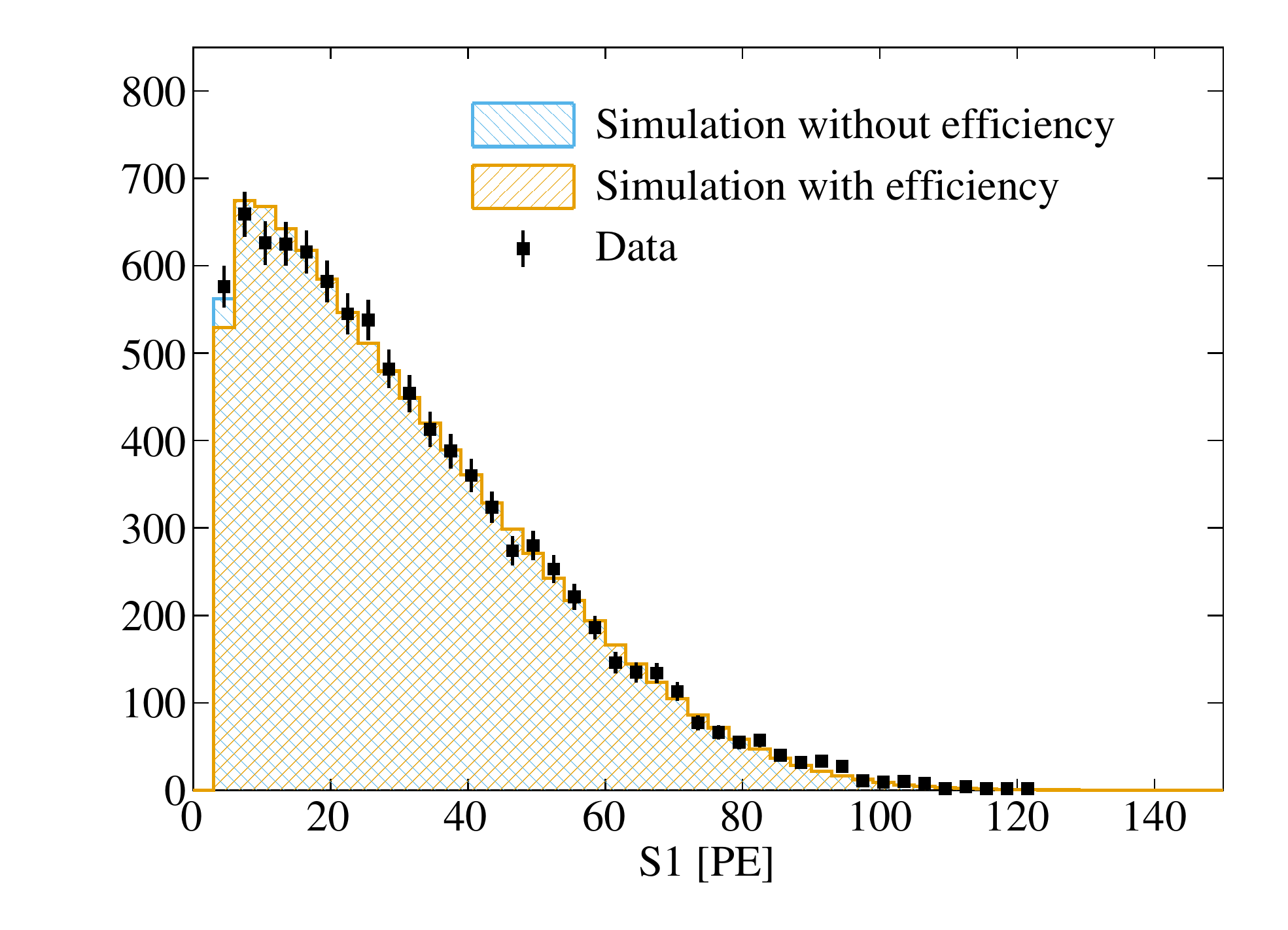}
    \caption{Run 9 CH$^{3}$T $S1$}
  \end{subfigure}  
  \begin{subfigure}{0.3\textwidth}
    \includegraphics[width=1\textwidth]{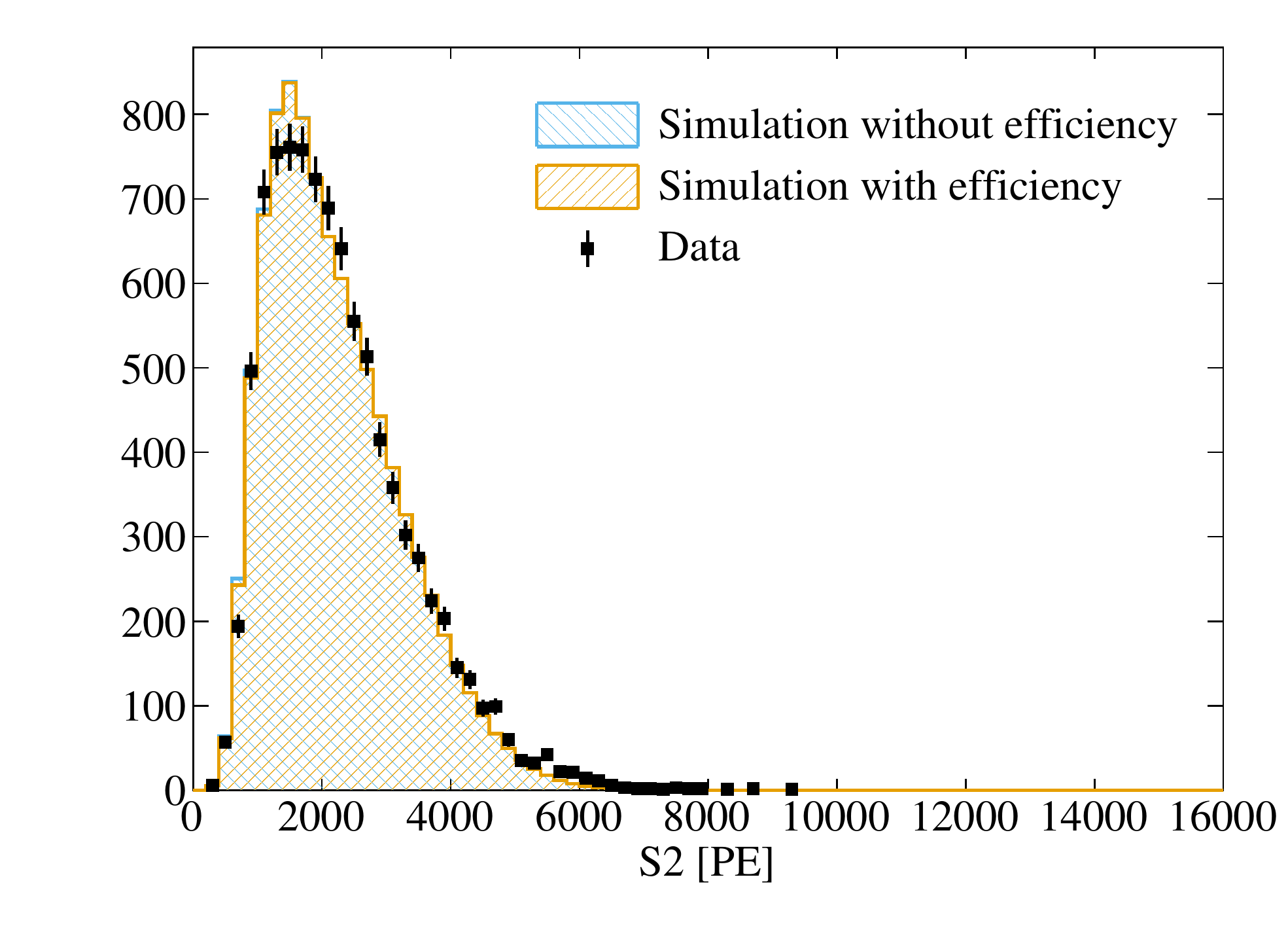}
    \caption{Run 9 CH$^{3}$T $S2$}
  \end{subfigure}  
  \begin{subfigure}{0.3\textwidth}
    \includegraphics[width=1\textwidth]{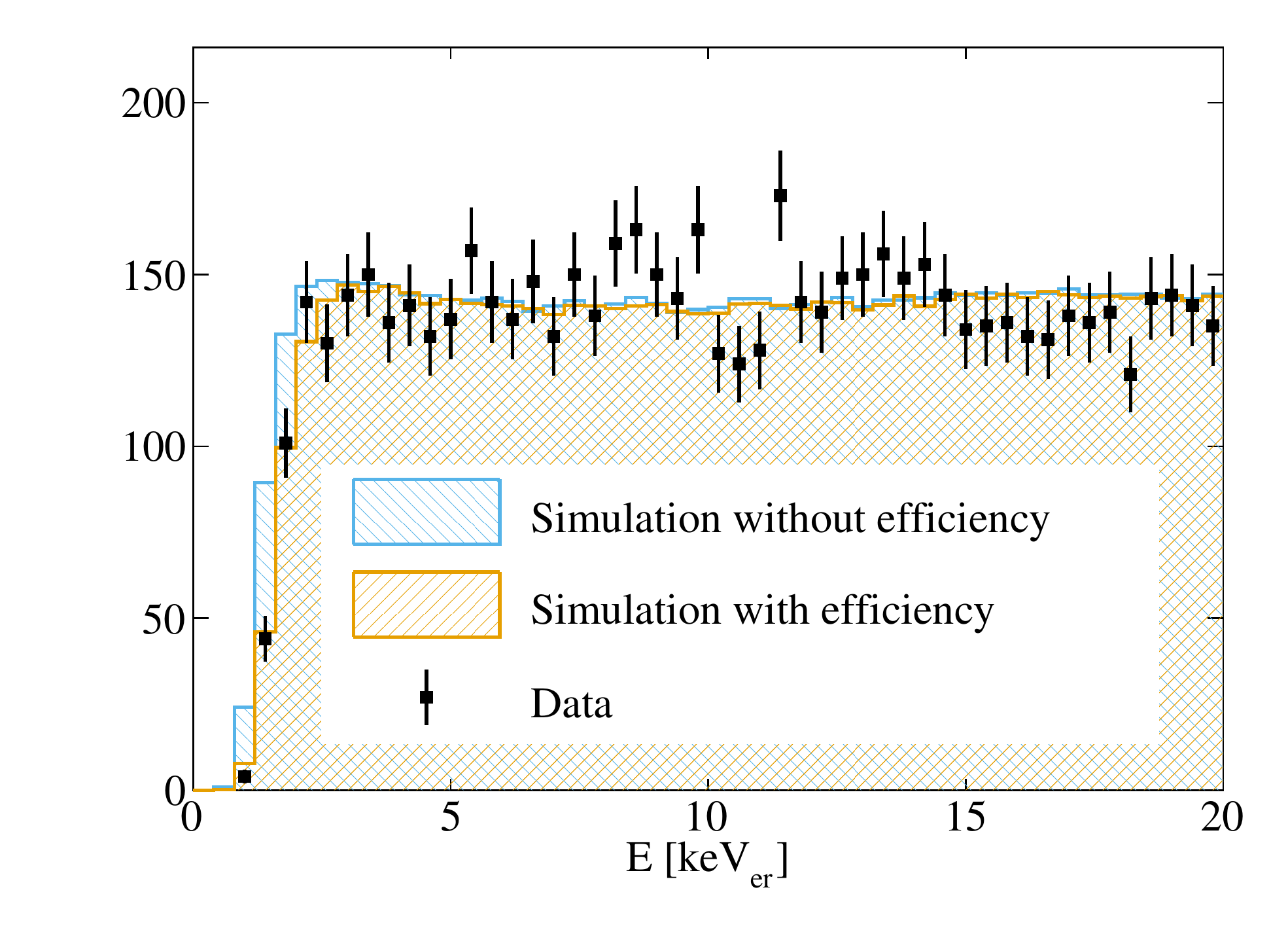}
    \caption{Runs 10/11 $^{220}$Rn energy}
  \end{subfigure}  
  \begin{subfigure}{0.3\textwidth}
    \includegraphics[width=1\textwidth]{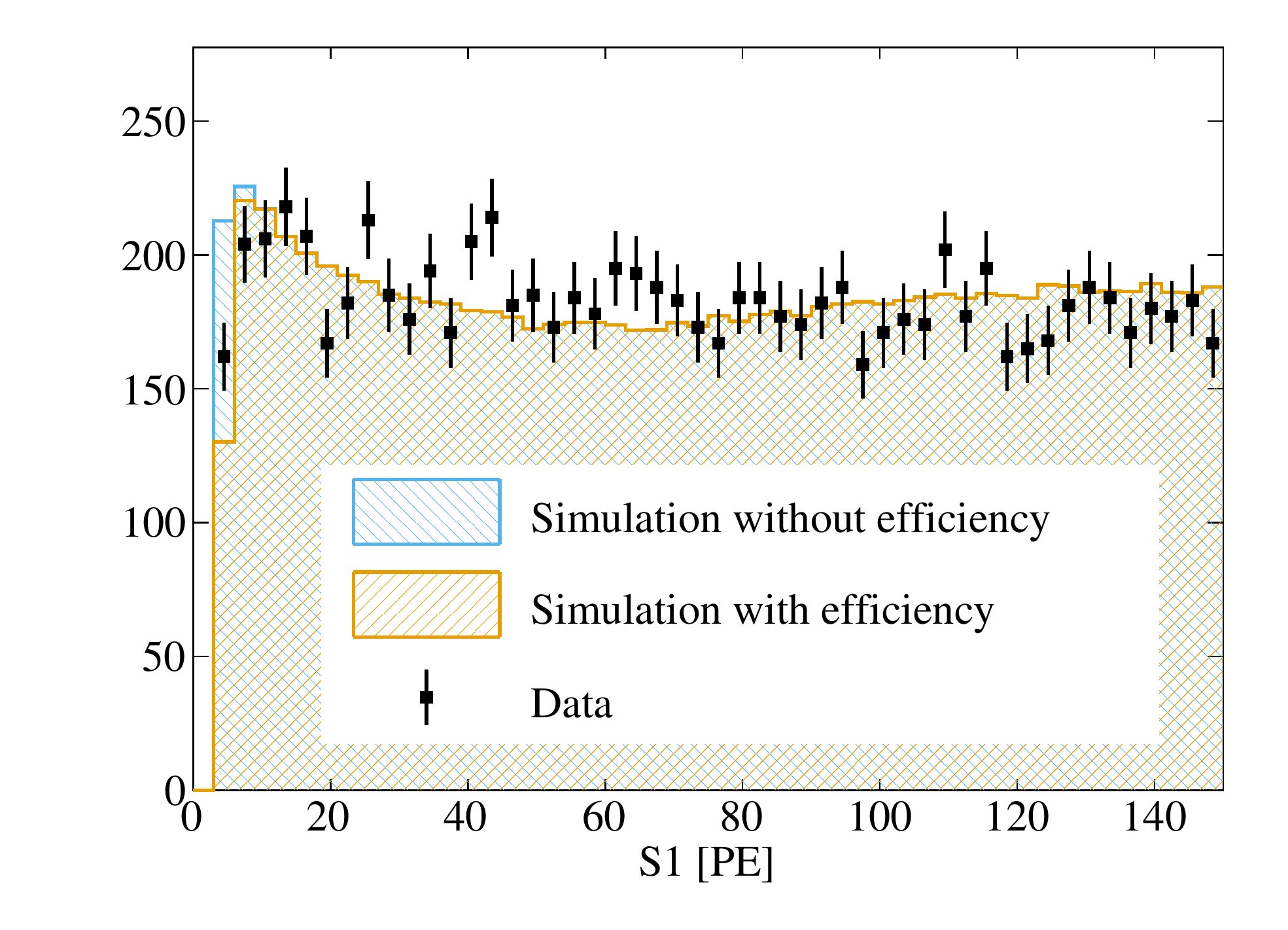}
    \caption{Runs 10/11 $^{220}$Rn $S1$}
  \end{subfigure}  
  \begin{subfigure}{0.3\textwidth}
    \includegraphics[width=1\textwidth]{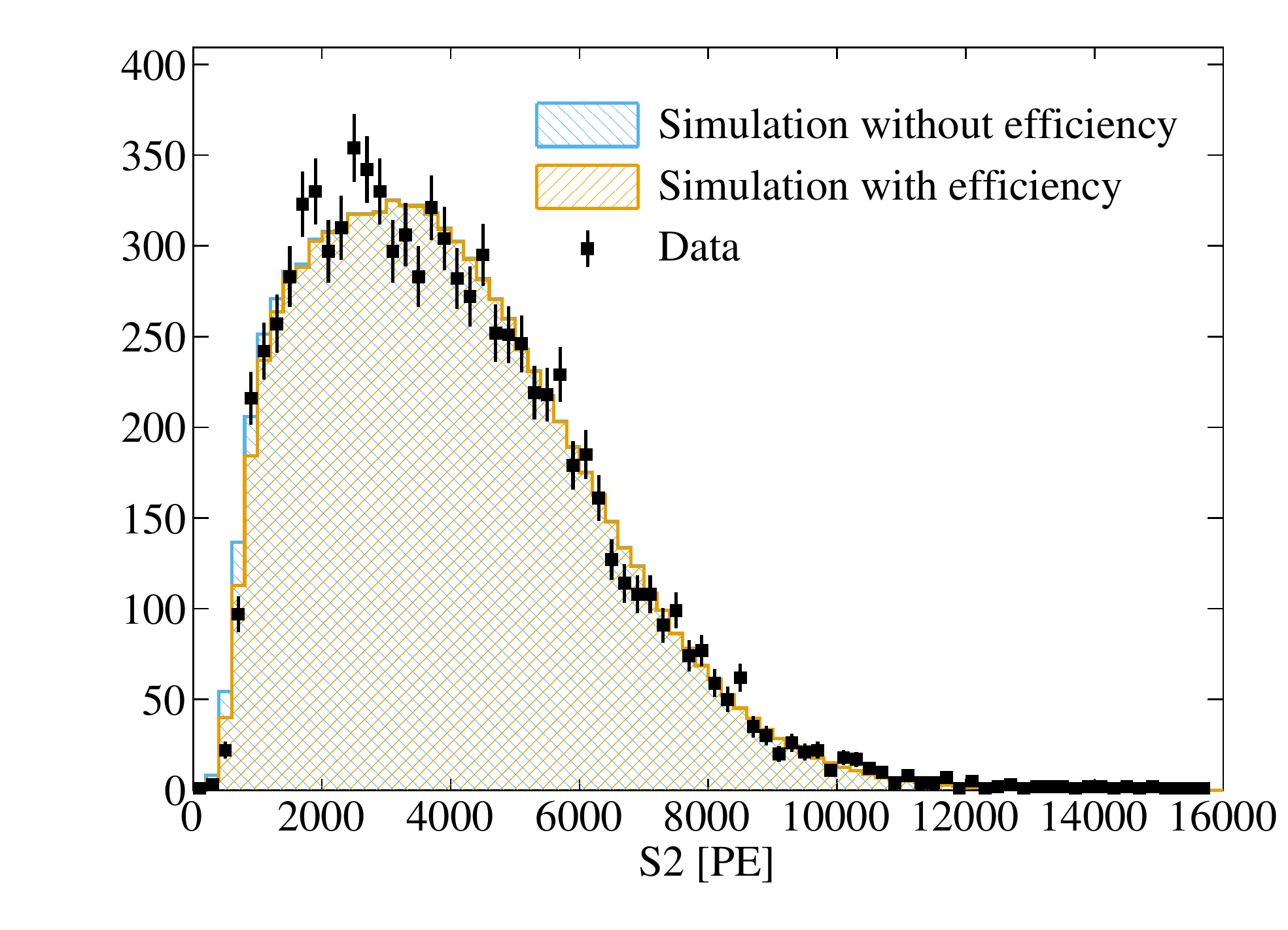}
    \caption{Runs 10/11 $^{220}$Rn $S2$}
  \end{subfigure}  
  \caption{The comparison of model simulation and calibration data in the projection of deposited energy, $S1$ and $S2$, in Run 9 and Runs 10/11.}
  \label{fig:ER_NR_comparison}
\end{figure}

Our ER and NR response model follows the construction of the
so-called NEST2.0~\cite{szydagis_m_2018_1314669}. In this analysis,
the initial excitation-to-ionization ratio, $N_{\rm{ex}}/N_{\rm i}$,
is taken from NEST2.0. On the other hand, the charge yield and light
yield (per unit energy) are initially fitted from the centroids of our
data as
\begin{equation}
\mathrm{CY}_0 = \frac{S2}{\mathrm{EEE}\times \mathrm{SEG}}/E_{\mathrm{rec}}\,,  \mathrm{LY}_0 = \frac{S1}{\mathrm{PDE}}/E_{\mathrm{rec}}\,,
\end{equation}
where the so-called electron-equivalent energy is reconstructed based
on Eqn.~\ref{eq:energy_comb}. For an NR event, the NR energy is
estimated by further dividing out the so-called Lindhard
factor~\cite{Lenardo:2014cva}. Note that $E_{\rm rec}$ contains energy
smearing introduced by the fluctuations in $S1$ and $S2$. Therefore
NEST2.0-based simulations are carried out in which CY and LY are
adjusted iteratively until a good fit in the two-dimensional
distributions in the data and simulation is reached. Our model also
takes into account the detector parameters extracted above and the
spatial non-uniformity (Sec.~\ref{sec:non_uniformity}), the double
photoelectron emission measured {\it in situ}
($\sim${21.5\%})~\cite{Cui:2017nnn}, and the SPE resolution to
properly include the fluctuations. In the simulation, $S1$ is randomly
distributed onto individual PMTs so that the three-PMT coincidence
selection cut could be simulated. The BLS nonlinearity is included in
$S1$ and $S2$. To fit the entire data distribution, the fluctuation in
the recombination rate is tuned against the calibration as well. The
comparisons between our best model simulation and calibration data in
$S1$, $S2$, and $E_{\rm rec}$ are shown in
Fig.~\ref{fig:ER_NR_comparison}, in which good agreement is found.

\begin{figure}[htb]
  \centering
  \includegraphics[width=0.6\textwidth]{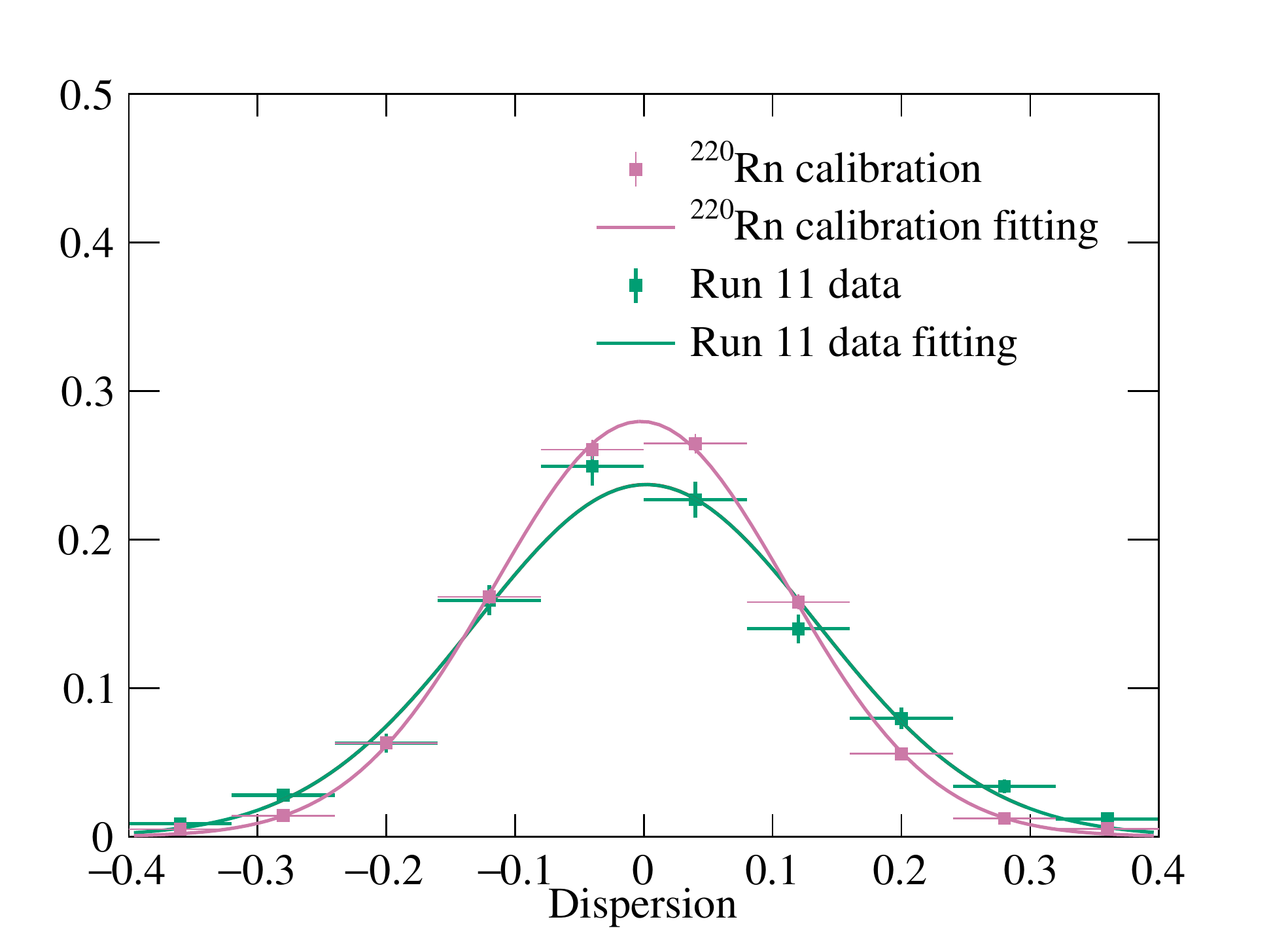}
  \caption{Comparison of the distribution of events away from the
    median of ER band with $S1\in (45,200)$~PE between the $^{220}$Rn
    calibration data (magenta dots) and Run 11 DM search data (green
    dots). The fitted Gaussian functions are overlaid. }
  \label{fig:erwidth}
\end{figure}

We also compare the aforementioned ER model with the ER event
distributions in Runs 10 and 11, by selecting the events within
$S1\in(45, 200)$~PE (outside DM search window; see
Sec.~\ref{sec:candidate}). Although the band centroids agree well, the
observed width in the data is larger than that from the calibration
(Fig.~\ref{fig:erwidth}), presumably due to the accumulated
fluctuations over time. Therefore, we increase the fluctuations in the
ER model for Runs 10 and 11 accordingly, leading to larger leakage
ratios $r$ (see Tab.~\ref{tab:background_budget} in
Sec.~\ref{sec:background}).

\begin{figure}[htb]
  \centering
  \begin{subfigure}{0.49\textwidth}
    \includegraphics[width=1.0\textwidth]{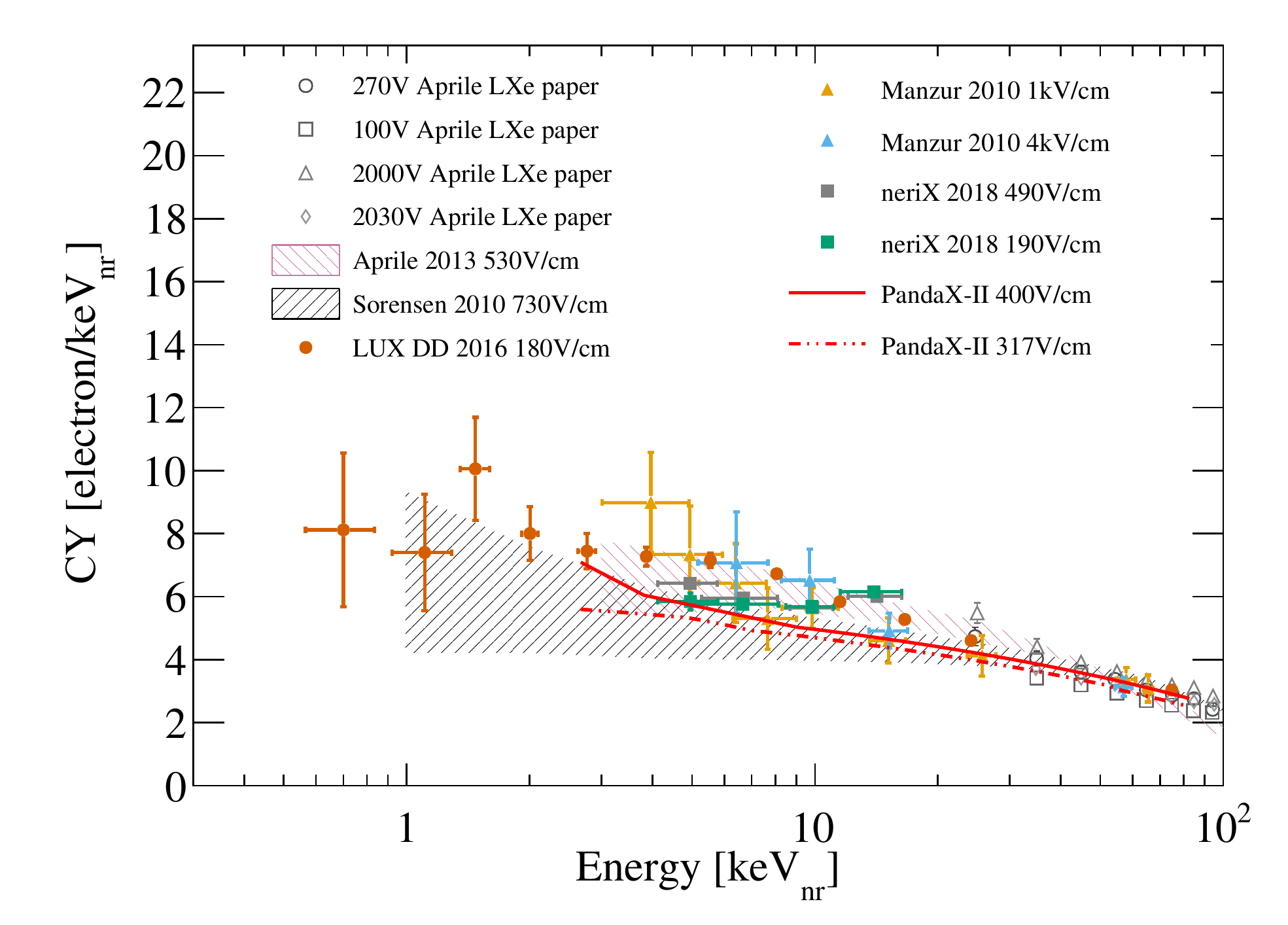}
    \caption{CY of NR}
    \label{fig:cy_nr_worlddata}
  \end{subfigure}
  \begin{subfigure}{0.49\textwidth}
    \includegraphics[width=1.0\textwidth]{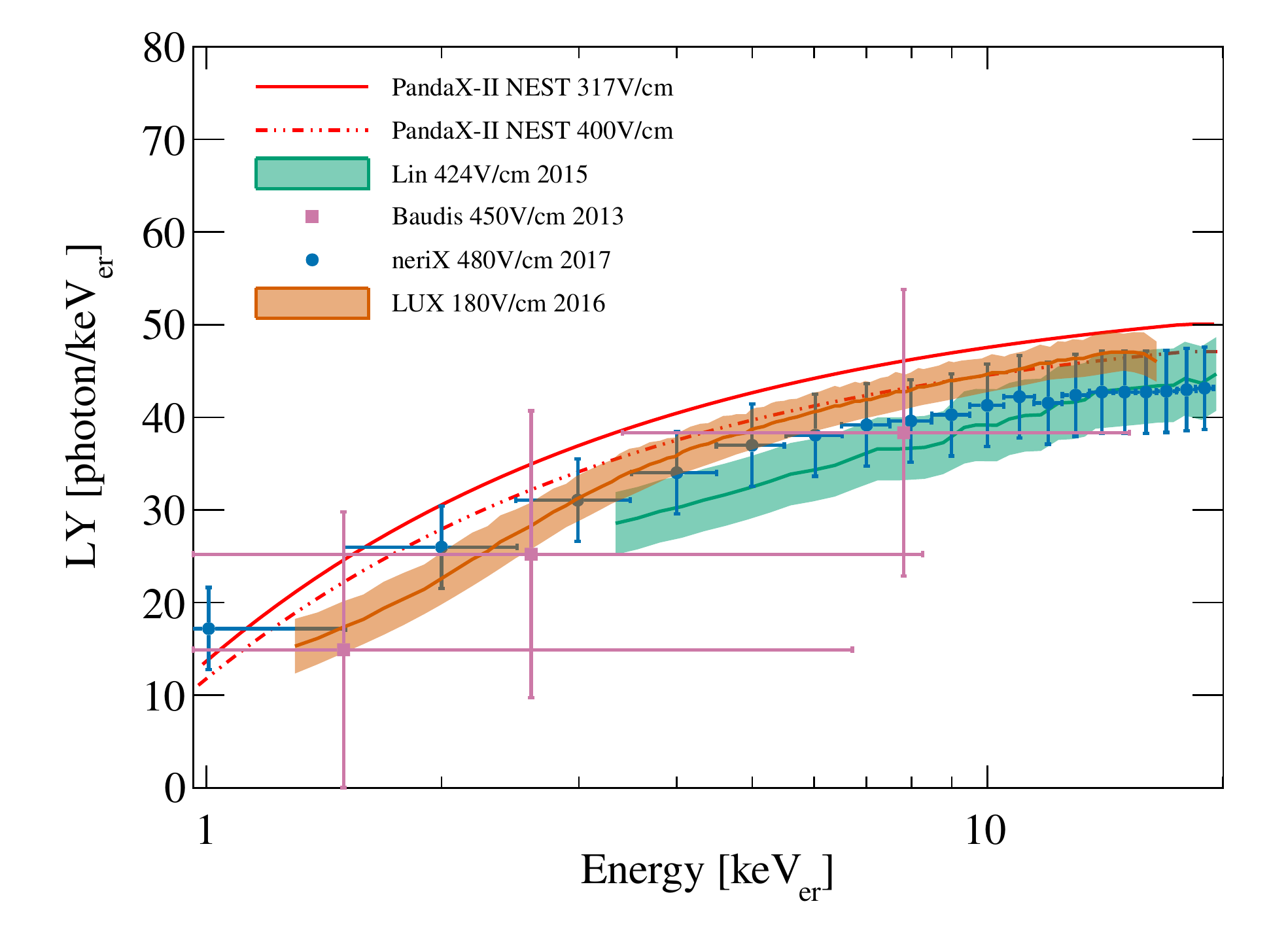}
    \caption{LY of ER}
    \label{fig:ly_er_worlddata}
  \end{subfigure}
  \caption{Charge yield of NR (a) and light yield of ER (b) from
    PandaX-II calibration data compared with results from the worldwide
    data (ER:
    Refs.~\cite{Baudis:2013cca,Lin:2015jta,Akerib:2015wdi,Goetzke:2016lfg},
    NR:
    Refs.~\cite{Aprile:2006kx,Manzur:2009hp,Sorensen:2010hq,Aprile:2013teh,Akerib:2016mzi,Aprile:2018jvg},
    see legend).}
  \label{fig:cy_nr_er_worlddata}
\end{figure}

The best fit LY (for ER) and CY (for NR) in PandaX-II are plotted in
Fig.~\ref{fig:cy_nr_er_worlddata}, together with the values from other
xenon-based experiments. Our NR model is in agreement with the
worldwide data within uncertainties. On the other hand, our ER model
is consistent with that of Ref.~\cite{Baudis:2013cca}, but has certain
dissimilarities with those of Refs.~\cite{Lin:2015jta,
  Akerib:2015wdi,Goetzke:2016lfg}.  Nevertheless, since our model
describes the calibration data, it is a self-consistent model for
producing the signal and background distributions.

The $S1$ and $S2$ simulation noted above do not include the data
quality cut efficiencies. Therefore, anchoring the distributions at
the high end, $\epsilon_1$ and $\epsilon_2$ in Eqn.~\ref{eq:eff} can
be performed by a comparison between the simulation and data. This is
also illustrated in Fig.~\ref{fig:ER_NR_comparison}. Because
$\epsilon_1$ agrees well for NR and ER events, we adopted
$\epsilon_1 =1/\left({1+\exp(\frac{S1-3.1}{0.075})}\right)$ (Run 9)
and $\epsilon_1 = 1/(\left({1+\exp(\frac{S1-4.0}{0.8})}\right)$(Runs
10/11). It is found that no $S2$ efficiency is needed, presumably due
to the fact that our analysis cut is for $S2_{\rm raw}>100$~PE, but
the $S2$ trigger threshold is approximately 50~PE~\cite{Wu:2017cjl}.

\section{Backgrounds in Dark Matter Search Data}
\label{sec:background}
After all data selections, four main backgrounds remain in the data:
the ER, neutron, accidental, and surface background. Improvements of
the estimates are addressed in turn.

The ER backgrounds come from varies sources, including gammas from
radioactive decay in the detector materials, the radioactive xenon
isotope of $^{127}$Xe (Runs 9 and 10), the beta decay of tritium (Runs
10 and 11), radioactive krypton and radon identified in the detector,
the solar neutrinos, and the double beta decay of $^{136}$Xe.  A
summary of the background level is presented in
Tab.~\ref{tab:er_background}.
\begin{table}[htb]
  \centering
  \begin{tabular}{cccccc}
    \hline\hline
    \multicolumn{2}{c}{{Item}} & Run 9 & Run 10 & Run 11, span 1 & Run 11, span 2\\
    \hline
     &$^{85}$Kr     & $1.19\pm0.2$ & $0.18\pm0.05$ & $0.20\pm0.06$ & $0.40\pm0.07$ \\
    {Flat ER}&$^{222}$Rn  & $0.19\pm0.10$ & $0.17\pm0.02$ & $0.19\pm0.02$& $0.19\pm0.02$ \\
    components&$^{220}$Rn  & $0.01\pm0.01$ & $0.01\pm0.01$ & $0.01\pm0.01$& $0.01\pm0.01$\\
    {(mDRU)}&ER (material) & $0.20\pm0.10$ & $0.20\pm0.10$ & $0.20\pm0.10$&$0.20\pm0.10$\\
    &Solar $\nu$  & $0.01$ & $0.01$ & $0.01$ &$0.01$\\
    &$^{136}$Xe & $0.0022$ & $0.0022$ & $0.0022$ &$0.0022$\\
\hline
\multicolumn{2}{c}{Total flat ER (mDRU)} &  $1.61\pm0.24$ & $0.57\pm0.11$
& $0.73\pm 0.08$ & $1.03\pm0.08$\\
    \hline
    \multicolumn{2}{c}{$^{127}$Xe (mDRU)}  & $0.14\pm0.03$ & $0.0069\pm0.0017$ &  \multicolumn{2}{c}{$<0.0001$}\\
    \hline
    \multicolumn{2}{c}{$^3$H (mDRU) }         &   0   & \multicolumn{3}{c}{$0.11$} \\
    \hline
    \multicolumn{2}{c}{Neutron (mDRU) } &
   \multicolumn{4}{c}{$ 0.0022\pm 0.0011$} \\
   \hline
   \multicolumn{2}{c}{Accidental (event/day) }  & \multicolumn{4}{c}{$ 0.014\pm 0.004$} \\
   \hline
    \multicolumn{2}{c}{Surface (event/day)} & \multicolumn{2}{c}{$0.041\pm0.008$} &\multicolumn{2}{c}{$0.063\pm 0.0013$}\\
    \hline\hline
  \end{tabular}
  \caption{Backgrounds in the dark matter search runs inside the
    FV. Among the ER backgrounds, the radioactivity level of $^{3}$H
    is the best fit in Ref.~\cite{Zeng:2020axion}. Others are
    estimated independently. ER and neutron backgrounds are estimated
    in $0$-$25$~keV. Accidental and surface backgrounds are estimated
    in the search window of S1 within $3$-$45$~PE and S2 within
    $100$(raw)-$10000$~PE. $1$~mDRU = $1\times10^{-3}$~evt/keV/day/kg.
    The total flat ER backgrounds of Runs 9 and 10 are sums of the
    components, and that of Run 11 is estimated with data in the
    region of $20$-$25$~keV (see text for details).}
  \label{tab:er_background}
\end{table}

The estimates of the ER backgrounds from the detector material, solar
neutrinos and $^{136}$Xe used in a previous analysis~\cite{Tan:2016diz}
are inherited.

The $^{127}$Xe background is also inherited from
Ref.~\cite{Cui:2017nnn}. Due to the relatively short half-life (35.5-day),
no $^{127}$Xe events are identified in Run 11.

A significant tritium background is introduced in the CH$_{3}$T
calibration in 2016 after Run 9, and it could not be effectively removed
by hot getters. The distillation campaign thereafter reduced the
tritium level by a factor of roughly 100. The residual tritium rate is
found to be stable at $0.030 \pm 0.004 \mu$Bq/kg, based on
unconstrained fits to data in different runs within $(0, 10]$~keV.

The level of the $^{220}$Rn background is estimated by the
$^{212}$Bi-$^{212}$Po and $^{220}$Rn-$^{216}$Po coincidence
events. The updated $^{220}$Rn level is $0.37\pm0.20$~$\mu$Bq/kg in
Runs 10 and 11. For $^{222}$Rn, the dominating background in the DM
region is the $\beta$-decay of the daughter $^{214}$Pb. In
Ref.~\cite{Ni:2019kms}, it is found that ions from the $^{222}$Rn
decay chains would drift toward the cathode, causing a decay rate
depletion for decay daughters in the FV. Based on this study, a more
robust method for estimating $^{214}$Pb background is developed by
combining the $\alpha$ rate from $^{222}$Rn and $^{218}$Po and
$\beta$-$\alpha$ coincidence of
$^{214}$Bi-$^{214}$Po~\cite{Ni:2019kms}. The resulting $^{214}$Pb
level is found to be rather stable in the three data sets, at
approximately 10~$\mu$Bq/kg.

Krypton causes one of the most critical background in
PandaX-II. During Run 11, the concentration of Kr background is
estimated to be $7.7\pm2.2$ ($15.2\pm2.5$) ppt before (after) the
leakage, using delayed $\beta$-$\gamma$ coincidence of $^{85}$Kr decay
($0.5\%$ branching ratio) and assuming a $^{85}$Kr abundance of
$2\times10^{-11}$. Accordingly, we separate Run 11 data into span 1
and span 2. In Tab.~\ref{tab:er_background}, we list the average Kr
background estimated by the $\beta$-$\gamma$ coincidence and the
corresponding statistical uncertainties. The increase of the Kr
background in Run 11 span 2 is also confirmed by an increase of the
event rate in $20$-$25$ keV, where the ER background rate rises from
$0.73\pm0.08$ to $1.03\pm0.08$ mDRU in Run 11, consistent with
Tab.~\ref{tab:er_background}. The total flat ER backgrounds summarized
in Tab.~\ref{tab:er_background} are used as inputs in the final
statistical fitting.

The neutron background from detector materials is evaluated based on a
new method discussed in Ref.~\cite{Wang:2019opt}, using the
high-energy gammas to constrain the low-energy single-scattering
neutrons. The total number of neutron events is estimated to be
$3.0\pm1.5$ in the full exposure of PandaX-II DM search, within the
final signal window and with all cuts applied.

\begin{figure}[htb]
  \centering
  \includegraphics[width=0.6\textwidth]{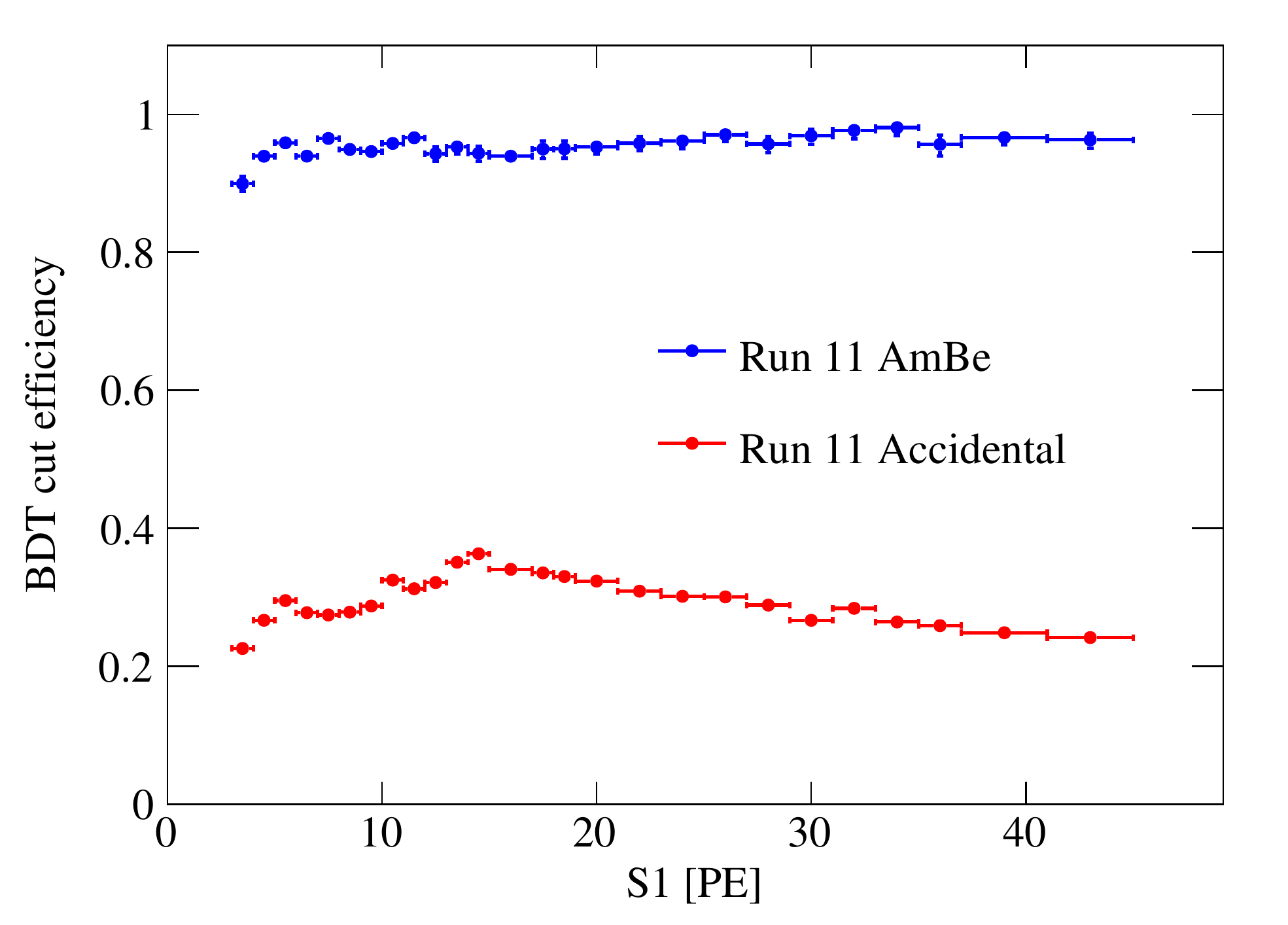}
  \caption{BDT efficiency $\epsilon_{\rm{BDT}}$ and the suppression
    ratio of accidental background vs. $S1$ for Run 11.}
  \label{fig:BDT} 
\end{figure}

The accidental background, produced by the random coincidence of
isolated $S1$ and $S2$, is calculated with refined treatment to
isolated $S1$s.  The isolated $S1$s are searched in the pre-trigger
window of high-energy single-scattering events, resulting in a much
larger data sample for better spectral measurement. The estimated rate
of isolated $S1$ is consistent with that in the previous analyses, in
which isolated $S1$s were searched in $S1$-only events before the
trigger~\cite{Tan:2016diz}, or in random trigger
data~\cite{Cui:2017nnn}, both with limited statistics. The updated
rate of isolated $S1$s is estimated to be 1.5~Hz in Run 9, 0.5~Hz in
Run 10, and 0.7~Hz in Run 11. The identification of isolated $S2$s
follows previous treatments with a stable rate of 0.012~Hz in all
runs. Isolated $S1$s and $S2$s are randomly paired and go through the
same data quality cuts with a 15\%-20\% survival ratio in the three
runs, leading to an unbiased estimate of the rate and spectrum of this
background. As in Refs.~\cite{Tan:2016diz,Tan:2016zwf}, a
BDT~\cite{Roe:2004na} is adopted using the AmBe and accidental samples
as the training data for the signal and background, respectively. The
BDT cuts suppress the accidental background to approximately 26\%
while maintaining a high efficiency ($\epsilon_{\mathrm{BDT}}$) for NR
events, which is illustrated in Fig.~\ref{fig:BDT} for Run 11. The
residual accidental background in the FV is 2.1 (Run 9), 1.0 (Run 10),
and 2.5 (Run 11), as displayed in Tab.~\ref{tab:evtnum}. Those for
Runs 9 and 10 are reduced from the previous analysis, mostly due to a
larger sample of isolated $S1$s and better-trained BDT cuts. The
average uncertainties are estimated to be $30\%$ by the variance of
the rate of isolated $S1$s throughout the runs.

\begin{figure}[hbt]
  \centering
  \begin{subfigure}{0.32\textwidth}
    \includegraphics[width=1.0\textwidth]{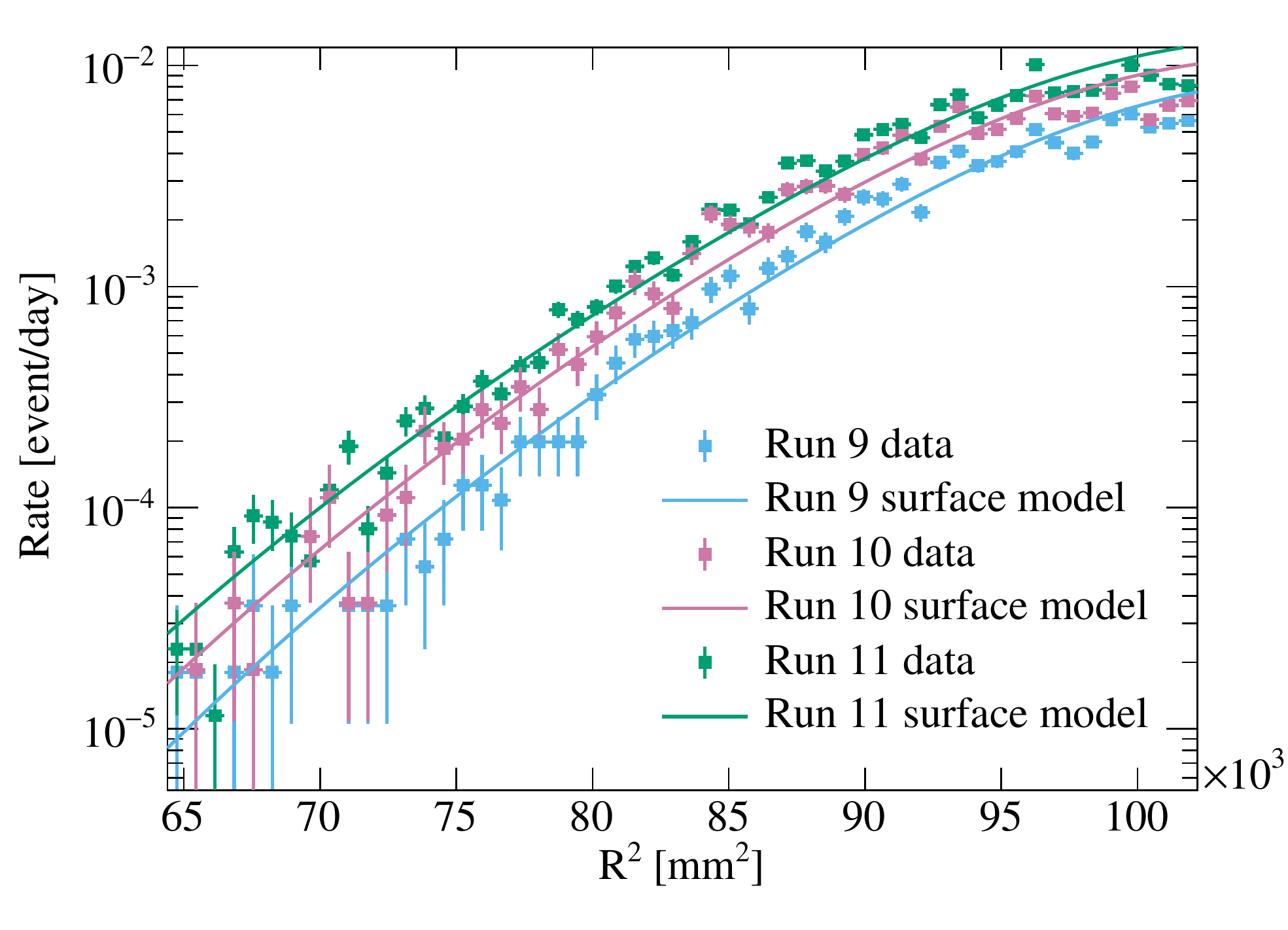}
    \caption{$S1\in (50, 100)$PE }
  \end{subfigure}
  \begin{subfigure}{0.32\textwidth}
    \includegraphics[width=1.0\textwidth]{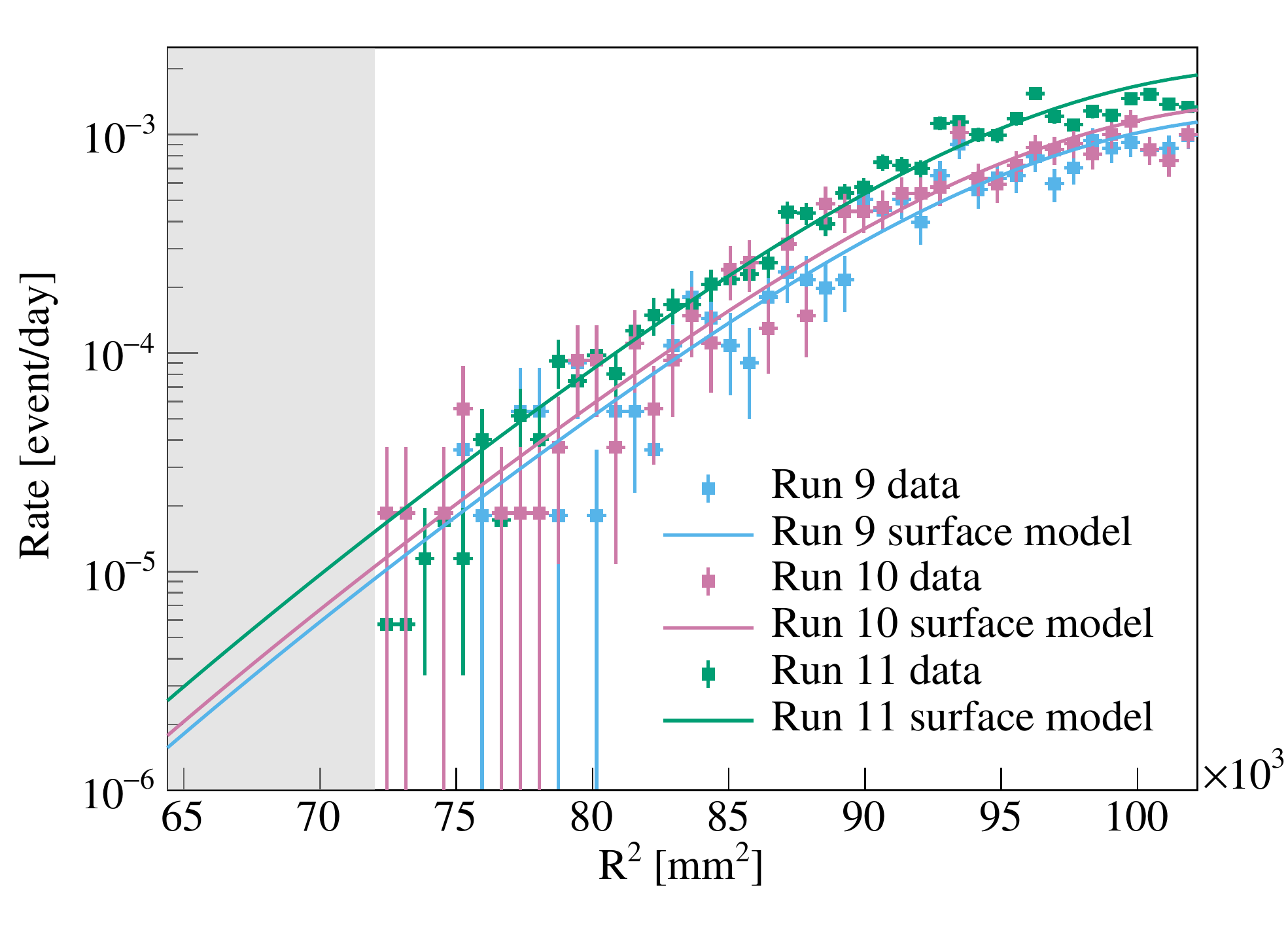}
    \caption{$S1\in (3, 45)$PE }
  \end{subfigure}
  \begin{subfigure}{0.32\textwidth}
    \includegraphics[width=1.0\textwidth]{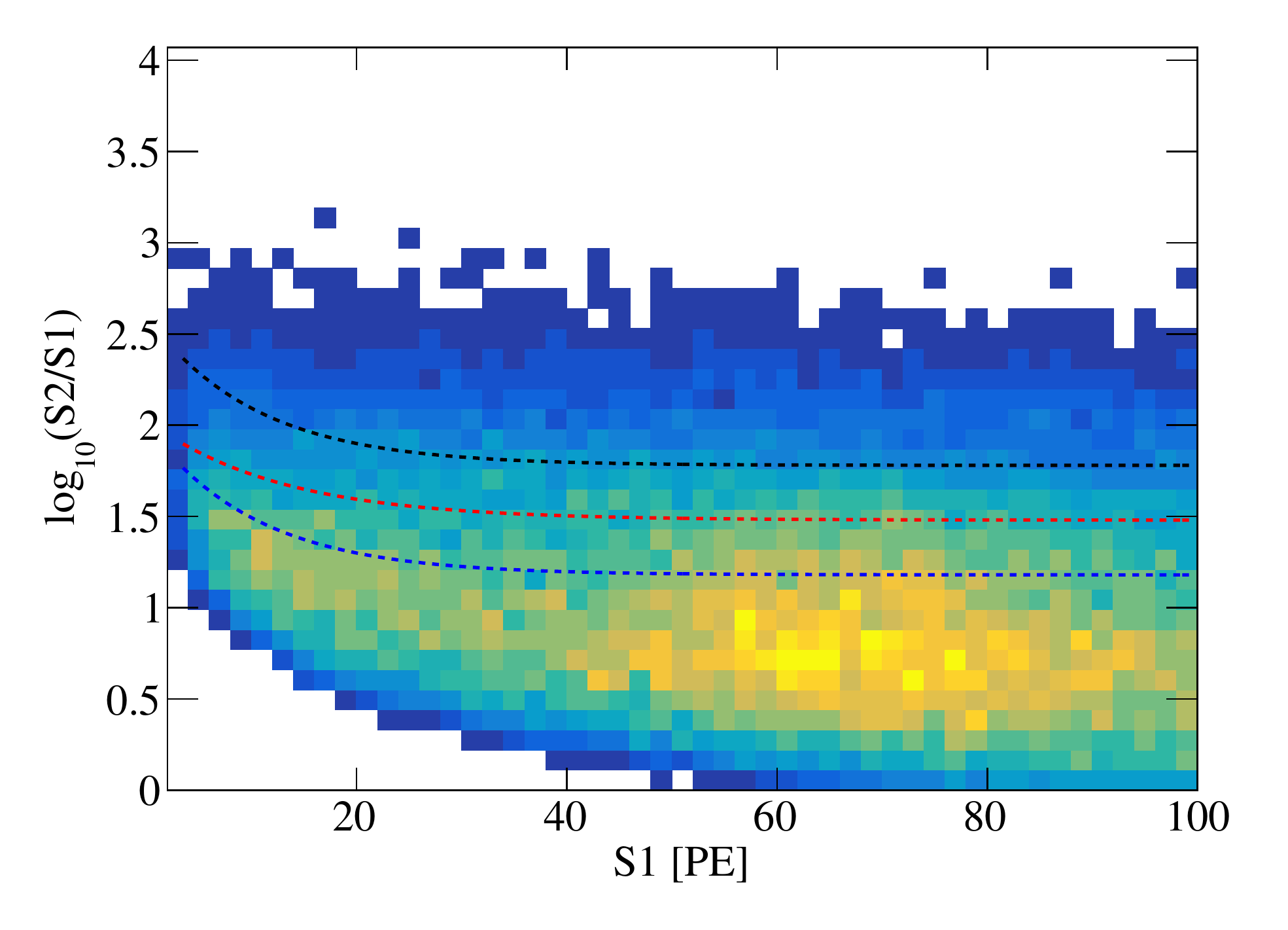}
    \caption{$S1\in (3, 45)$PE }
  \end{subfigure}
  \caption{(a) and (b): Comparison between the $R^{2}$ distribution of
    surface events (dots) and model predictions (lines) for Run 9
    (blue), Run 10 (magenta) and Run 11 (green). (c) The
    $\log_{10}(S2/S1)$ vs. $S1$ band for events outside the FV. The
    median of the ER band (black dashed line) and the NR band (red
    dashed line) and $-4\sigma$ of the ER and (blue dashed line) are
    overlaid.}
  \label{fig:wall_dis}
\end{figure}

The $\beta$-decay of daughter $^{210}$Pb (T$_{1/2}$=22.2~y) on the
PTFE surface is observed, presumably due to the $^{220}$Rn
plate-out. These events have a characteristic suppressed $S2$, likely
caused by the charge loss onto the PTFE wall during the drift.  We
also observe a temporal increase of the surface background (events
that are reconstructed very close to, or outside, the PTFE wall). A
data-driven surface background model~\cite{Zhang:2019evc} is developed
to estimate the surface background in the present analysis.  Events
with $S1>50$~PE are used to model the radial distribution of surface
events related to $S2$, serving as a shape template in ($R^2,
S2$). The background is then normalized by the DM data outside the
FV. A comparison between the scaled template and data along $R^{2}$ is
shown in Fig.~\ref{fig:wall_dis}. The number of events below the
$-4\sigma$ line of the ER band in the data (model) with
$S1\in(50,100)$ and $R^2\in (0,720)$ cm$^{2}$ is 20 (17.4) in Run 9,
28 (31.6) in Run 10, and 187 (161.8) in Run 11, with the same quality
cuts for DM events applied. The model uncertainty is estimated to be
$20\%$ due to the resolution of 5~mm in the position
reconstruction. The expected number of surface events in the DM search
region is $2.9\pm0.6$, $3.6\pm{0.8}$, and $15.4\pm{3.3}$ events in
Runs 9, 10 and 11, respectively.

\section{Final candidates from dark matter search data}
\label{sec:candidate}
Due to blind analysis, the selection cuts for final candidates are set
without considering the real data for Run 11. The signal window to
search for DM candidates and the fiducial radius are optimized by
requiring the best DM sensitivity at the mass of 40 GeV/$c^2$, with a
below-NR-median (BNM) signal acceptance within which the background is
evaluated with a cut-and-count approach. For $S1$, we inherit the
range of $[3, 45]$~PE as in the previous analysis, as the sensitivity
flattens for upper cuts from 45 to 70 PE. As was done previously, $S2$
is selected between 100 (raw) and 10000~PE, together with the 99.99\%
NR acceptance line and an additional 99.9\% ER acceptance cut to
eliminate a few events with unphysically large sizes of $S2$. All runs
share the same selection cuts on the fiducial radius, i.e.,
$R^2<720~\rm{cm}^2$. The range of the drift time is determined to be
$(18, 310)$~$\mu$s in Run 9, and $(50, 350)$~$\mu$s in Runs 10 and 11
(lower cut is higher than that in Ref.~\cite{Cui:2017nnn}), based on
the vertical distribution of events with $S1$ between 50 and 70~PE.
The xenon mass within the FV is estimated to be $328.9\pm9.9$~kg in
Run 9 and $328.6\pm9.9$~kg in Runs 10 and 11, where the uncertainties
are estimated using a 5-mm resolution in the position
reconstruction. The final exposures used in this analysis are 26.2
ton$\cdot$day in Run 9, 25.3 ton$\cdot$day in Run 10, and 80.3
ton$\cdot$day in Run 11.

The number of events in the DM search data passing the cuts is
summarized in Tab.~\ref{tab:evtnum}. In total, 1222 candidates are
obtained in the three runs. A post-unblinding event-by-event waveform
check is then performed, in which two spurious events in Run 11 are
identified (detailed waveforms are shown in
Appendix~\ref{sec:app_a}). One event is a double $S2$ event, with a
second small $S2$ being split into a few $S1$s in our clustering
algorithm which are, therefore, not properly registered. The other
event has a small $S1$ formed by three coincidental PMT hits, but two
of the hits are due to coherent noise pickup. The final number of
candidates is 1220. The sequential reduction of events after various
cuts is summarized in Table~\ref{tab:evtnum}.

\begin{table}[htb]
\centering
\begin{tabular}{cccc}
\hline\hline
Cut & Run 9 & Run 10 & Run 11\\
\hline
All triggers & 24502402 & 18369083 & 49885025\\
Single S2 cut & 9806452 & 6731811 & 20896629\\
Quality cut & 331996 & 543393 & 2708838\\
DM search window & 76036 & 74829 & 257111\\
FV cut & 392 & 145 & 710\\
BDT cut & 384 & 143 & 695\\
Post-unblinding cuts & 384 & 143 & 693 \\
\hline\hline
\end{tabular}
\caption{Number of events in Runs 9, 10, and 11 after successive
  selection cuts.}
\label{tab:evtnum}
\end{table}

\begin{sidewaysfigure}[p]
  \centering
  \begin{subfigure}{0.32\textwidth}
    \includegraphics[width=1.0\textwidth]{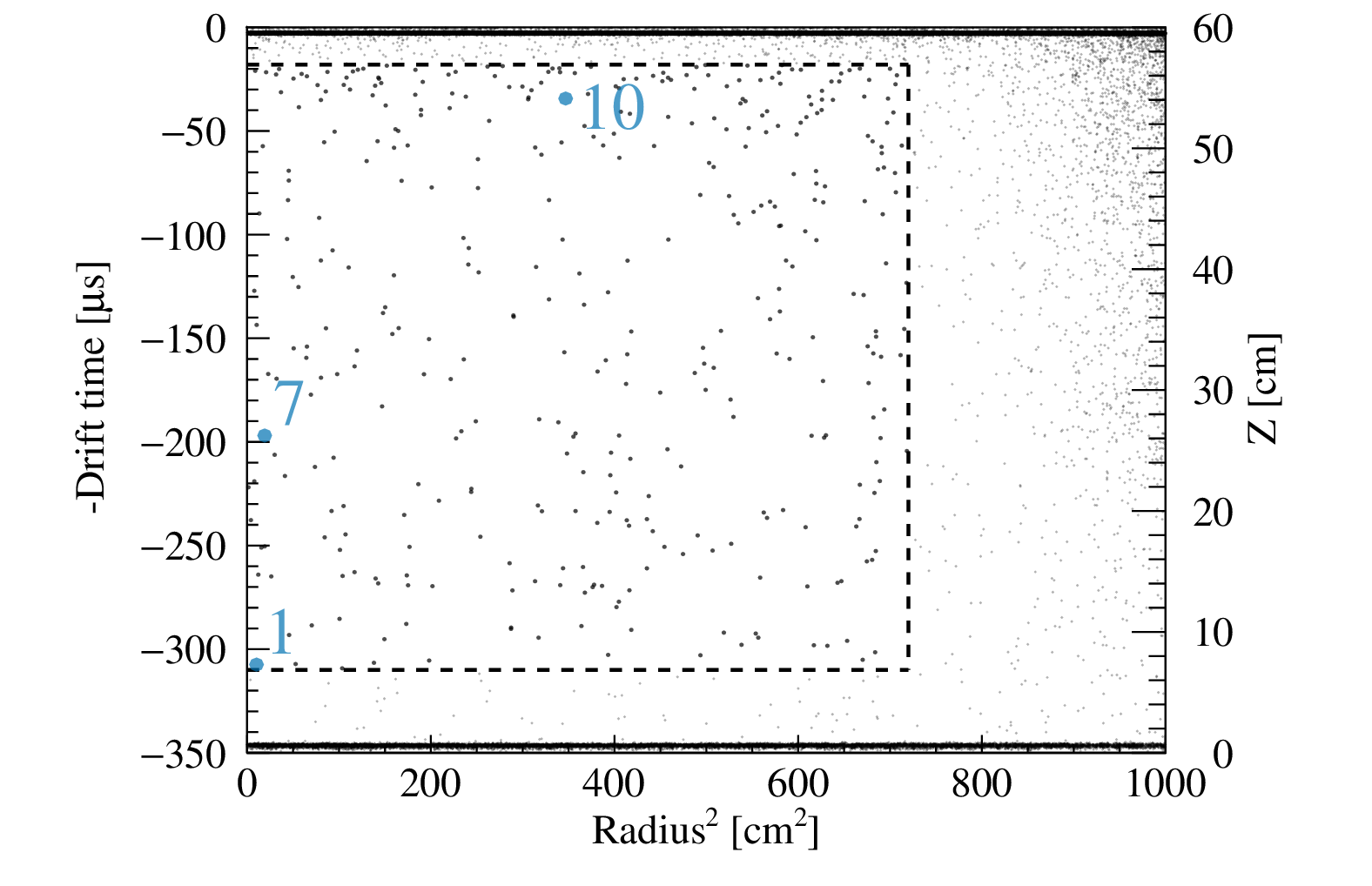}
    \caption{$R^2$-$z$ in Run 9}
    \label{fig:r2_z_run9}
  \end{subfigure}
  \begin{subfigure}{0.32\textwidth}
    \includegraphics[width=1.0\textwidth]{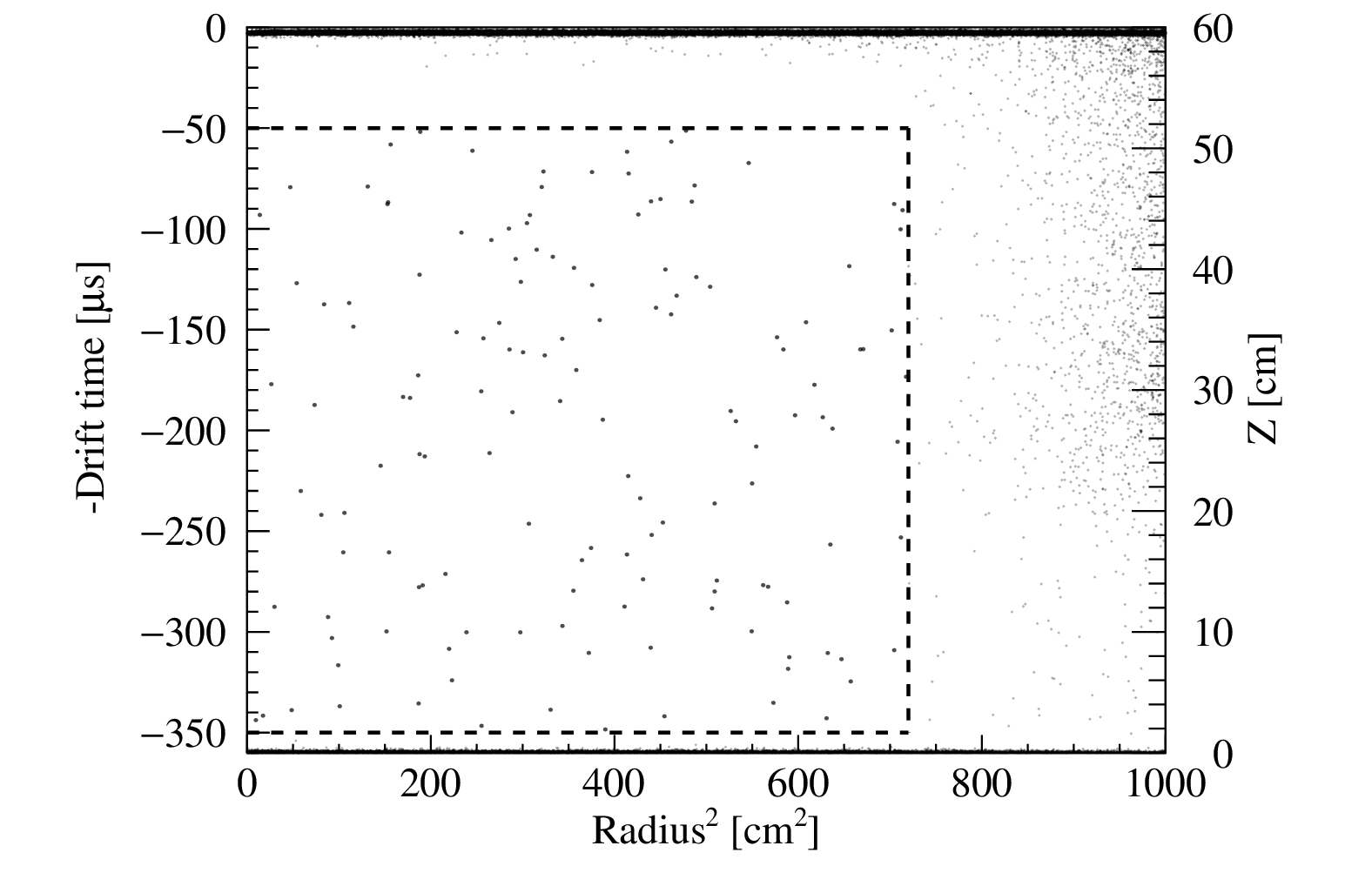}
    \caption{$R^2$-$z$ in Run 10}
    \label{fig:r2_z_run10}
  \end{subfigure}
  \begin{subfigure}{0.32\textwidth}
    \includegraphics[width=1.0\textwidth]{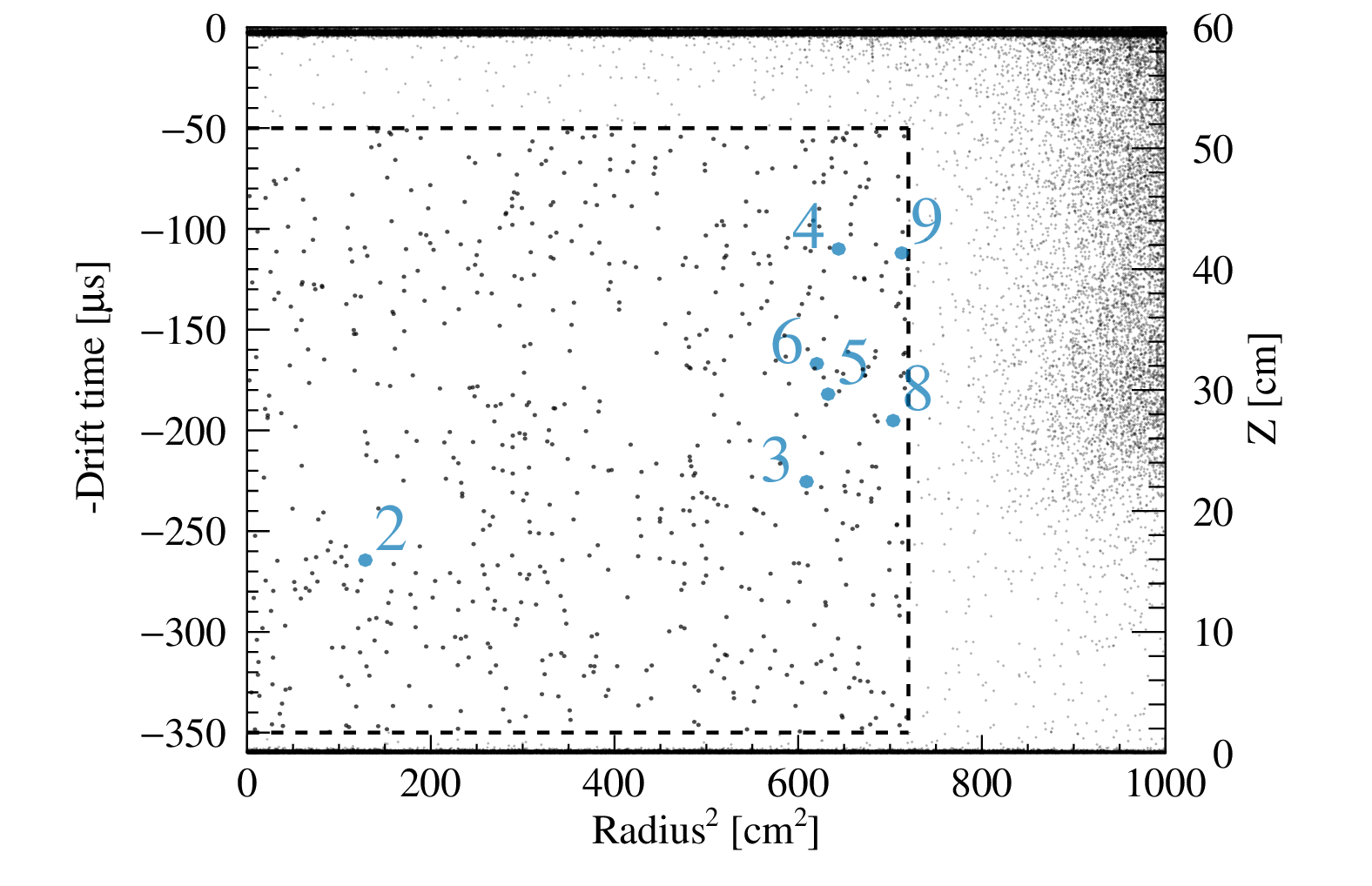}
    \caption{$R^2$-$z$ in Run 11}
    \label{fig:r2_z_run11}
  \end{subfigure}
  \begin{subfigure}{0.32\textwidth}
    \includegraphics[width=1.0\textwidth]{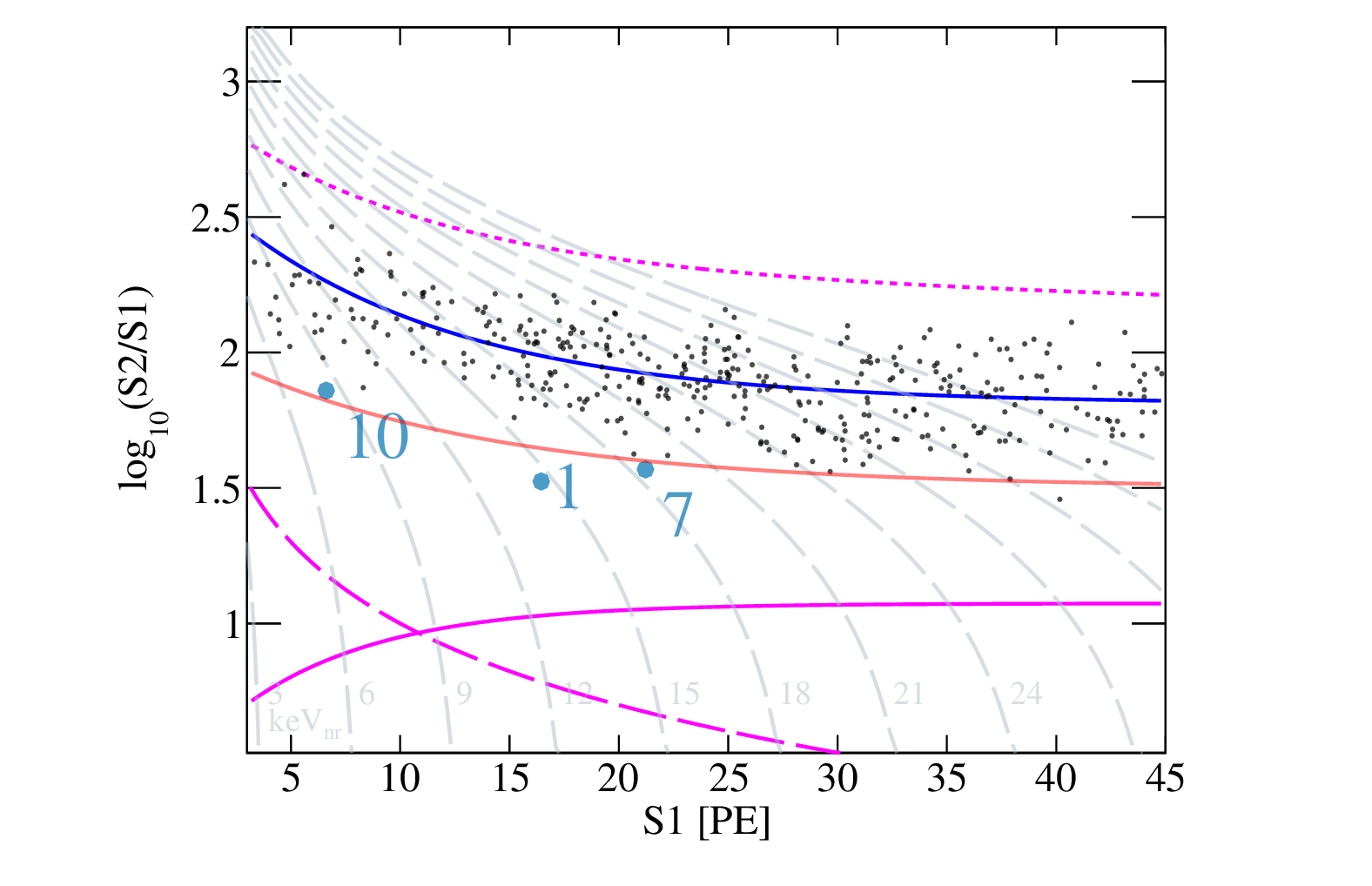}
    \caption{$\log_{10}(S2/S1)$-$S1$ in Run 9}
    \label{fig:band_run9}
  \end{subfigure}
  \begin{subfigure}{0.32\textwidth}
    \includegraphics[width=1.0\textwidth]{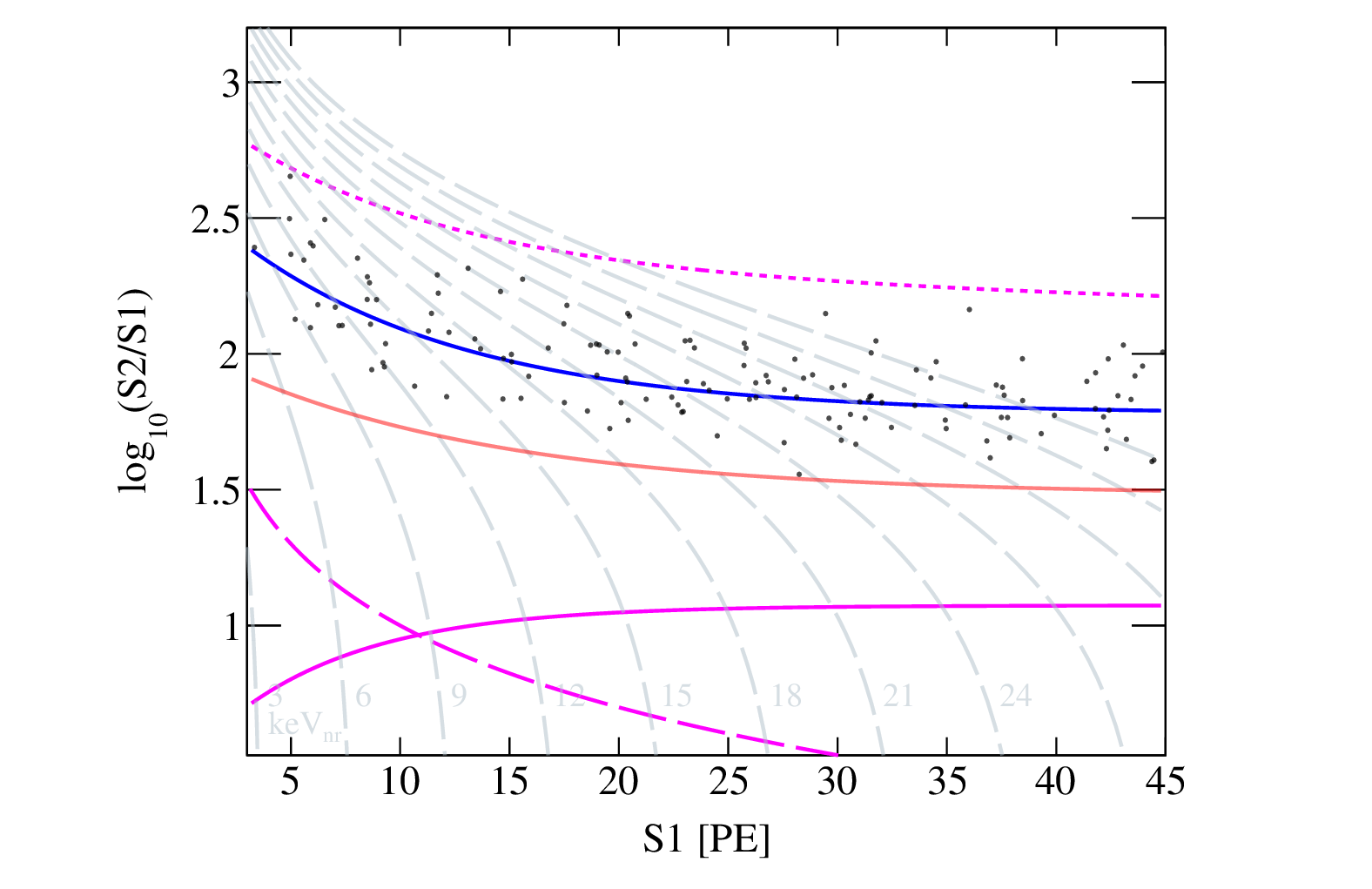}
    \caption{$\log_{10}(S2/S1)$-$S1$ in Run 10}
    \label{fig:band_run10}
  \end{subfigure}  
      \begin{subfigure}{0.32\textwidth}
    \includegraphics[width=1.0\textwidth]{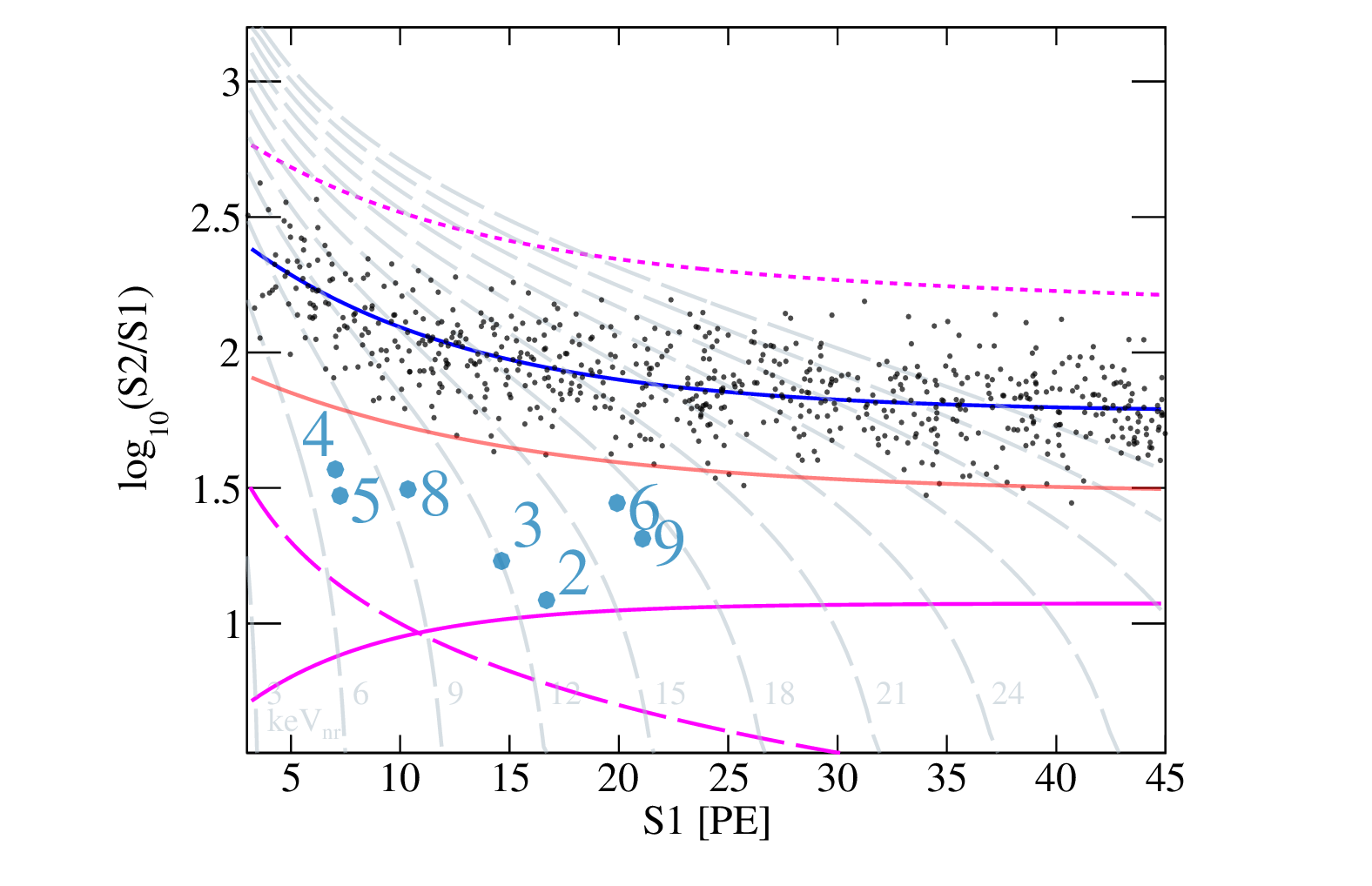}
    \caption{$\log_{10}(S2/S1)$-$S1$ in Run 11}
    \label{fig:band_run11}
  \end{subfigure}
  \caption{The spatial and signal distributions of events within $S1$
    and $S2$ range cuts. The events outside the FV are also presented
    for Run 9(a), Run 10(b) and Run 11(c). Only the final candidates
    in the DM search data are shown on the signal distributions for
    Run 9(d), Run 10(e) and Run 11(f). The ten most likely DM
    candidates are labeled. References of the ER band median (blue
    solid line) and NR band median (pink solid line) are shown in the
    signal distributions. The magenta lines are the boundaries of the
    acceptance window. The solid and dotted magenta lines are the
    99.99\% NR and 99.9\% ER acceptance cuts, respectively. The dashed
    magenta line is the $S2=100$ PE boundary. The dashed grey curves
    represent the equal-energy curves in nuclear recoil energy
    (keV$_{\rm NR}$).}
  \label{fig:vertex_dis}
\end{sidewaysfigure}

The spatial distribution of events inside and outside the FV (in the
same $S1$ and $S2$ selection region) is shown in
Fig.~\ref{fig:vertex_dis}. In Run 11, more events are clustered to the
wall, consistent with the increase of the surface background.

The distributions of the candidate events in $\log_{10}(S2/S1)$
vs. $S1$ for the three runs are also shown in
Fig.~\ref{fig:vertex_dis}, with NR median lines shown for
reference. The number of BNM candidates in Runs 9, 10, and 11 are 4,
0, and 34, respectively. Although the statistical interpretation of
the data is given in Sec.~\ref{sec:limit}, we discuss some general
features here. One of the BNM events in Run 9 was the same in the
previous analysis~\cite{Tan:2016zwf}, with $S1\sim$40~PE and
$R^2\sim330$~cm$^2$. Another three appear reasonably close to the
center of the TPC, which was above the NR-median in the previous
analysis, but appear as BNM after the improved uniformity
correction. The majority of the BNMs in Run 11 are consistent with the
surface background and the ER background. For example, if we reduce
the maximum radius cut to $R^2<600$~cm$^2$ in Run 11, the BNMs
decrease to 14, with 11 of them quite close to the NR median. A
comparison between the observed candidates and the expected background
is given in Table~\ref{tab:background_budget}, and the best fit
background values (see Sec.~\ref{sec:limit}) are also given in the
table. From a simple cut-and-count point of view, no significant
excess is found above the background.

\begin{table}[htb]
  \centering
  \begin{tabular}{ccccccc}
    \hline\hline
         & ER & Accidental & Neutron & Surface & Total fitted& Total observed\\
    \hline
    Run 9         & 381.5 & 2.20 & 0.77 & 2.13 & $387\pm23$ & 384\\
    Below NR median & 2.7 & 0.46 & 0.37 & 2.12  & $5.6\pm 0.5$  & 4\\
    \hline
    Run 10        & 141.7 & 1.08 & 0.48 & 2.66 & $145.9\pm16$ & 143\\
    Below NR median &  1.7 & 0.24 & 0.22 & 2.65 & $4.8\pm 0.6$ & 0\\
    \hline
       Run 11, span 1 & 216.5 & 1.04 & 0.60 & 6.24 & $224\pm 22$ & 224\\
       Below NR median & 4.2 & 0.32 & 0.32 & 6.22 & $11.1\pm 1.1$ &13\\
     \hline
       Run 11, span 2 & 448.2 & 1.60 & 0.92 & 9.58& $460\pm35$ & 469\\
      Below NR median & 8.26 & 0.50 & 0.50 & 9.54 & $18.8\pm 1.7$ & 21 \\
    \hline
    Total        & 1187.9 & 5.9 & 2.77 & 20.6 & $1217\pm60$ & 1220\\
    Below NR median&  16.8 & 1.52 & 1.42 &  20.5  & $40.3\pm3.1$ & 38
    \\
    \hline\hline
  \end{tabular}
  \caption{The best fit total and below-NR-median background events in
    Run 9, Run 10 and Run 11 in the FV with the signal model
    $m_{\chi}=400$ GeV/$c^2$. The BNM backgrounds are estimated with
    the PDFs. The nuisance parameters can be found in
    Tab.~\ref{tab:best_fit_nuisance}, the uncertainties of which are
    propagated into the total fitted event uncertainties. Numbers of
    observed events are shown in the last column.}
  \label{tab:background_budget}
\end{table}

\section{Fitting method and results}
\label{sec:limit}
The statistical interpretation of the data is carried out using a
profile likelihood ratio (PLR) approach, very similar to the treatment
in Refs.~\cite{Aprile:2011hx,Tan:2016zwf,Cui:2017nnn}. For the ER
background, except for $^{127}$Xe and tritium, the others are mostly flat
within the region of interest. We combine them into a single ``flat ER
background'' to avoid degeneracy in the likelihood fit. The unbinned
likelihood function is constructed as

\begin{multline}
\label{eq:likelihood}
\mathcal{L}_{\rm pandax} = \left\{\prod_{n=1}^{\textrm{nset}} \left[
    {\rm Poiss}(\mathcal{N}_{\text{obs}}^n|\mathcal{N}_{\text{fit}}^n)
    \times
    \prod_{i=1}^{\mathcal{N}_{\text{obs}}^n} (l_{s}^{n,i}+\sum_b {l_{b}^{n,i}})\right]\right\}\times\\
\left[G(\delta_s, \sigma_s) \prod_{b}G(\delta_b, \sigma_b)\right],
\end{multline}

with
\begin{align}
  &\mathcal{N}_{\text{fit}}^{n}=N_s^n(1+\delta_s) + \sum_bN_b^n(1+\delta_{b}),\\
  \label{eq:frac_likelihood_sig}  &l_{s}^{n,i}=\frac{N_s^n(1+\delta_s)P_{\rm DM}^n(S1^i,S2^i,r^i,z^i)}{\mathcal{N}_{\text{fit}}^{n}},\\
  \label{eq:frac_likelihood_bkg}  &l_{b}^{n,i}=\frac{N_{b}^n(1+\delta_{b})P_{b}^n(S1^i,S2^i,r^i,z^i)}{\mathcal{N}_{\text{fit}}^{n}},\\
  &G(\delta, \sigma)=\frac{1}{\sqrt{2\pi}\sigma}\exp\left(-\frac{\delta^{2}}{2\sigma^2}\right).
\end{align}
 
Instead of simply dividing the data into three runs, we separated the
data into 14, 4, and 6 sets in Runs 9, 10, and 11 (so $n$ runs up to
24), respectively, according to different operating conditions, such
as the drift/extraction fields and electron lifetime, which affect the
expected signal distributions. For each set, the number of observed
events is $\mathcal{N}_{\text{obs}}^n$; $N_s^n$ and $N_b^n$ are the
number of signal and backgrounds events, respectively. In this
analysis, $N_s^n$ is related to the DM-nucleon cross-section
$\sigma_{\chi n}$ by the incoming flux (standard halo), the number of
target xenon nuclei, and the Helms form
factor~\cite{Savage:2006qr}. The nuisance normalization parameters
$\delta_s$ and $\delta_b$ are constrained by the uncertainties
$\sigma_s$ and $\sigma_b$, respectively, by a Gaussian penalty
function $G(\delta,\sigma)$. $\sigma_s$ is set to be $20\%$ to capture
the global uncertainties in the DM flux, target mass, and detector
efficiency, and $\sigma_b$ is obtained from
Table~\ref{tab:er_background}. For $^{127}$Xe, accidental, and
neutrino backgrounds, in all data sets we assume a common $\delta_b$
to reflect the correlated systematic uncertainty. On the other hand,
the flat ER and surface backgrounds have independent values of
$\delta_b$ to reflect the set-to-set changes. The tritium background
is left to float in the fit (no corresponding penalty).

\begin{figure}[htb]
  \centering
\includegraphics[width=0.45\textwidth]{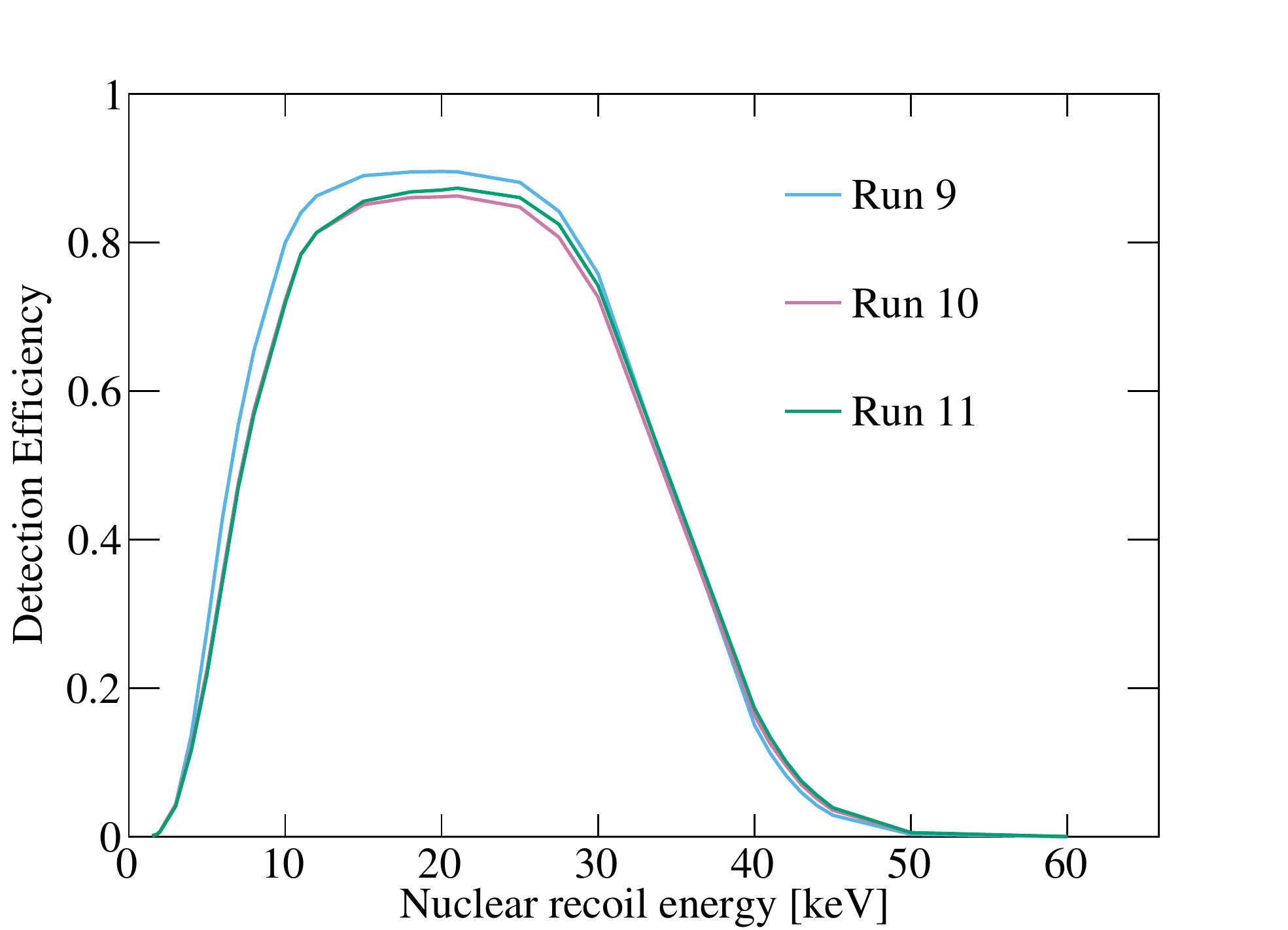}
\caption{The detection efficiencies as functions of the NR energy for
  Run 9 (blue), Run 10 (magenta), and Run 11 (green).}
  \label{fig:det_eff_nr}
\end{figure}

The PDFs for signals and backgrounds, $P_s^n$ and $P_{b}^n$, are
extended to four dimensions ($S1$, $S2$, $r$, $z$). Except for the
surface background, the signal distributions of DM and other
backgrounds are treated to be independent from their spatial
distributions. The spatial distributions of neutron and $^{127}$Xe
backgrounds are extracted from Geant4-based simulations, and that for
the accidental background is obtained from random isolated-$S1$-$S2$
pairs. The four-dimensional distribution of the surface background is
produced with the data-driven surface model~\cite{Zhang:2019evc},
within which $S1$, $S2$, and $r$ are correlated (see
Fig.~\ref{fig:wall_dis}), and $z$ is independent. Spatial
distributions of all other backgrounds and DM signals are uniform. The
ER background PDF in $S1$ and $S2$ follows the NEST2-based modeling in
Sec.~\ref{sec:calibration_model}. The DM signal PDF is obtained by
assuming a standard halo model and an NR energy spectrum of the
spin-independent (SI) elastic DM-nucleus scattering used in previous
analyses~\cite{Tan:2016diz, Tan:2016zwf, Cui:2017nnn}, together with
the updated NR model mentioned earlier. The selection efficiency is
embedded in the PDF, by generating events weighted by the overall
efficiency (Eqn.~\ref{eq:eff}). The detection efficiency for NR events
as a function of the recoil energy is illustrated in
Fig.~\ref{fig:det_eff_nr}.

Standard fitting is performed on the combined data to minimize the PLR
test statistic ($q_{\sigma}$) at different DM masses $m_{\chi}$. The
best fit of DM events for $m_{\chi}>200$ GeV/$c^2$ is almost the same
as the best fit of nuisance parameters in
Tab.~\ref{tab:best_fit_nuisance}. As an example, for $m_{\chi}=400$
GeV/$c^2$, the best fit of $\sigma_{\chi n}$ is $4.4\times 10^{-46}$
cm$^{2}$, corresponding to a detected signal number of 5.7. Based
on the background-only toy Monte Carlo tests, the best fit corresponds
to a $p$-value of 0.17, which is equivalent to a significance of 0.96
$\sigma$, consistent with no significant excess above the background.
\begin{table}[hbt]
  \centering
  \begin{tabular}{cc}
    \hline \hline
    & $m_{\chi}=400$ GeV/$c^2$ \\
    \hline
    $\delta_{^{3\rm H}}$
&$-0.03\pm0.27$\\	
	$\delta_{\rm {flat\;ER,\;Run\; 9}}$ &  $-0.08\pm0.07$ \\
	$\delta_{\rm flat\; ER,\; Run\; 10}$ & $0.02\pm0.13$\\
	$\delta_{\rm flat\; ER,\; Run\; 11,\; span\; 1}$ &
	$0.08\pm0.11$ \\
	$\delta_{\rm flat\; ER,\; Run\; 11,\; span\; 2}$ &
	$0.10\pm 0.08$ \\
	$\delta_{^{127}\rm Xe}$ & $0.00\pm 0.13$ \\
	$\delta_{\rm Accidental}$  & $0.02\pm 0.29$\\
	$\delta_{\rm Neutron}$ & $-0.04\pm0.49$ \\
	$\delta_{\rm wall\; Run\; 9 \;and\; 10}$  &
	$-0.26\pm0.20$ \\
	$\delta_{\rm wall\; Run\; 11}$& $0.03\pm0.16$\\
    \hline
  \end{tabular}
  \caption{The best fit nuisance parameters for $m_{\chi}=400$ GeV/$c^{2}$.}
  \label{tab:best_fit_nuisance}
\end{table}

\begin{figure}[hbt]
  \centering
  \includegraphics[width=.6\textwidth]{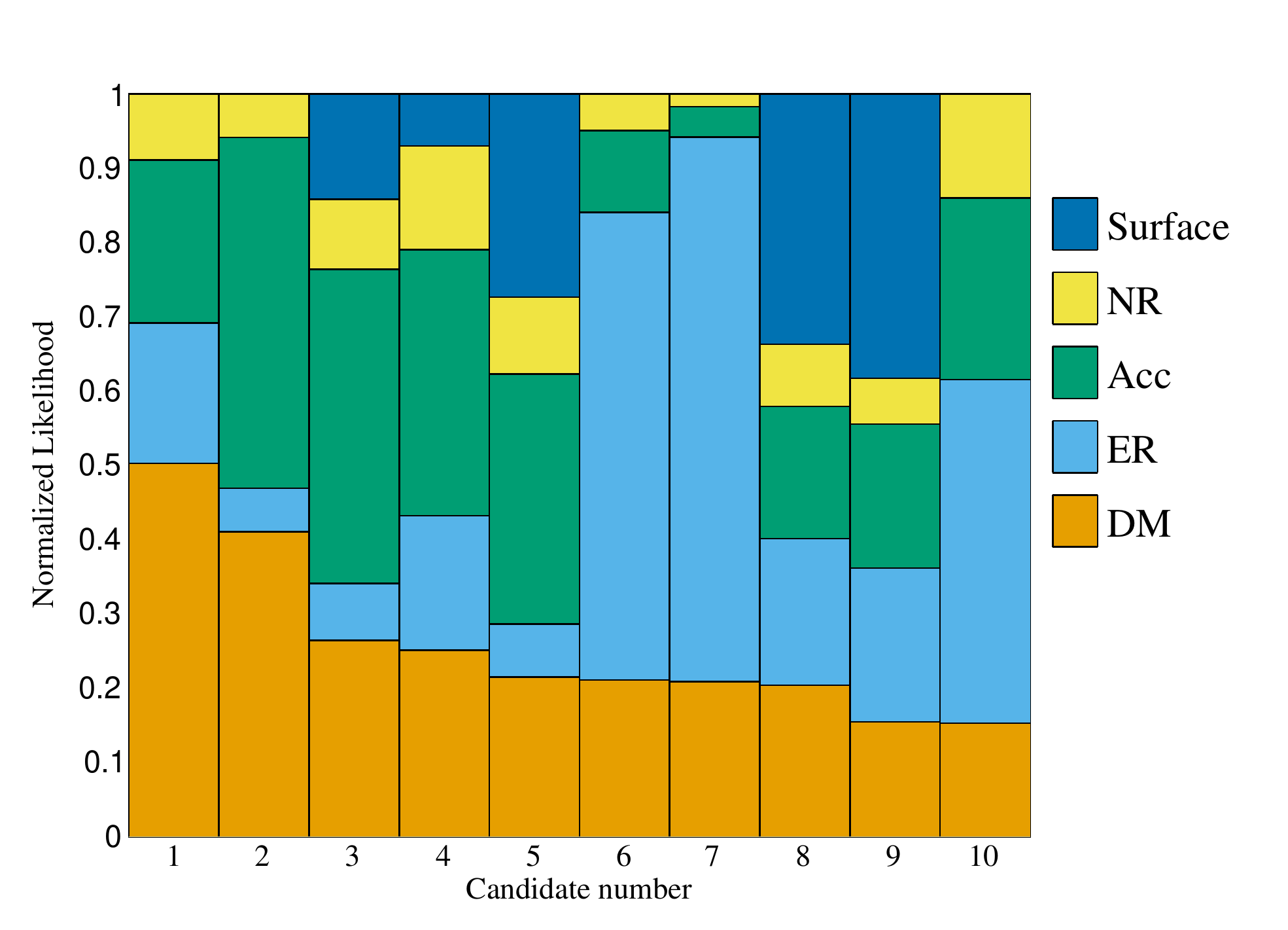}
  \caption{The normalized likelihoods of the most likely DM events for
    $m_{\chi}=400$ GeV/$c^{2}$.}
  \label{fig:evt_dm_likelihood}
\end{figure}

We also examine the likelihoods (Eqns.~\ref{eq:frac_likelihood_sig}
and~\ref{eq:frac_likelihood_bkg}) of individual events using
backgrounds and DM PDFs. For the top ten DM-like events with
$m_{\chi}=400$~GeV/$c^{2}$ labeled in Fig.~\ref{fig:vertex_dis}, the
ratios of the likelihoods of different hypotheses for each event are
presented in Fig.~\ref{fig:evt_dm_likelihood}. This confirms our
observations that out of the 38 BNM events, most of them are likely to
be a surface background or an ER background. Events 1 and 2 have the
highest probability of being either a DM or accidental background.

\begin{figure}[htb]
  \centering
\includegraphics[width=0.6\textwidth]{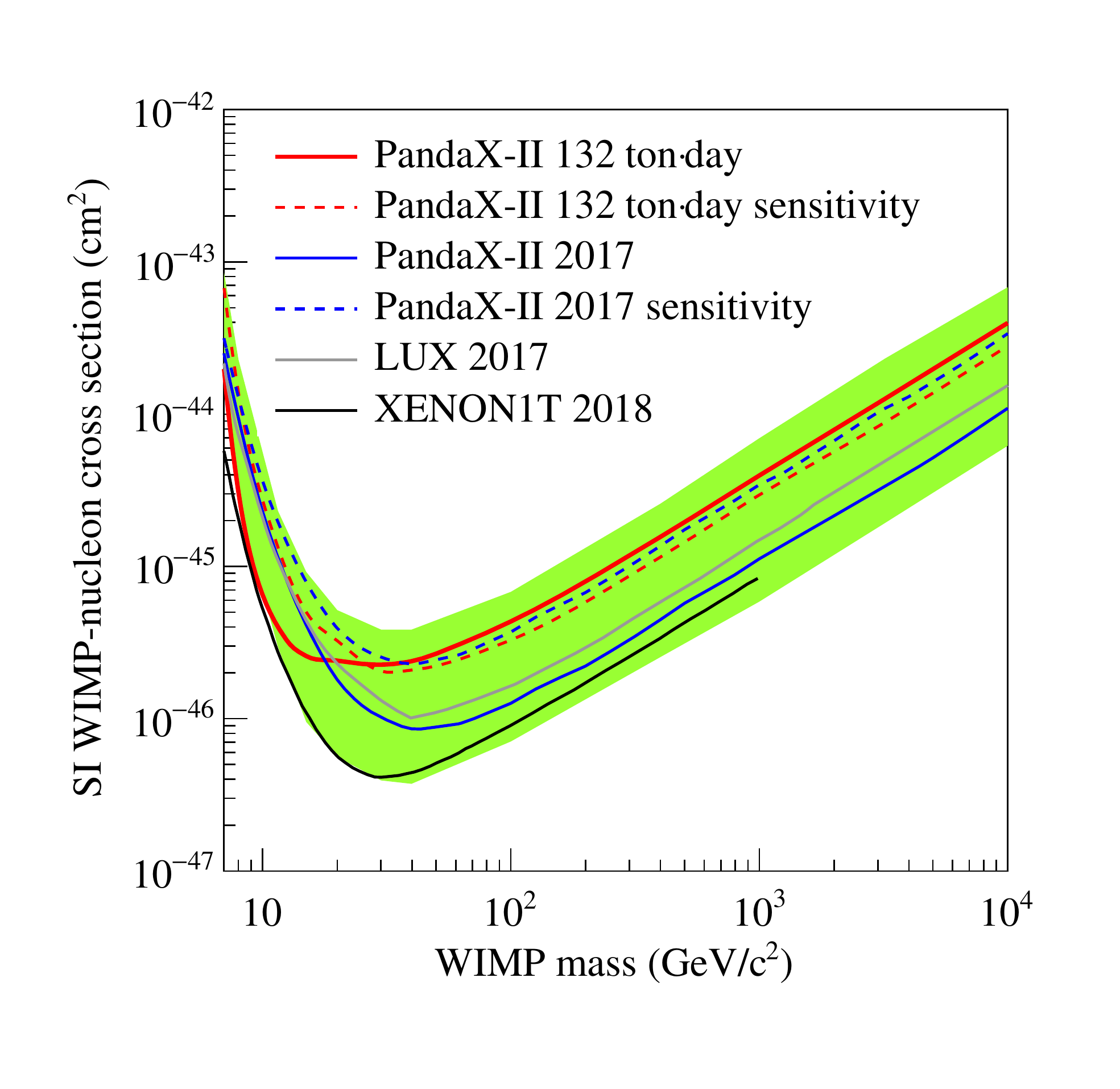}
\caption{The $90\%$ confidence level upper limits vs. $m_{\chi}$ for
  the spin-independent WIMP-nucleon elastic cross sections from the
  full exposure data of PandaX-II, overlaid with that from PandaX-II
  2017~\cite{Cui:2017nnn}, LUX 2017~\cite{Akerib:2016vxi}, and XENON1T
  2018~\cite{Aprile:2018dbl}. The green band represents the
  $\pm1\sigma$ sensitivity band. Below 8 GeV/c$^2$, the median
  sensitivity of this analysis is slightly weaker than that in 2017,
  as our nuclear recoil model is re-calibrated, leading to a lower
  signal efficiency for low mass WIMPs.}
  \label{fig:limits}
\end{figure}

Based on the above, we choose to report the upper limit of the cross
section of this search. The standard CL$_{s+b}$
approach~\cite{Junk:1999kv} is adopted, for which we performed a
two-dimensional scan in ($m_{\chi}$,$n_{\chi n}$). On each grid, a
large number of toy Monte Carlo simulations with similar statistics
are generated and fitted with the signal hypothesis, with the
resulting distribution of $q_{\sigma,MC}$ compared to the observed
$q_{\sigma,data}$ to define the 90\% confidence level of the
exclusion. The results below 10 GeV/$c^2$ are power constrained at
$-1\sigma$ of the sensitivity band~\cite{Cowan:2011an}, which is
obtained by generating 90\% exclusion lines using background-only
Monte Carlo simulation data sets, with the same PLR procedures. The
results are shown in Fig.~\ref{fig:limits}. The minimum excluded
$\sigma_{\chi n}$ is $2.2\times10^{-46}$~cm$^2$ at $m_{\chi}$ of 30
GeV/$c^2$, corresponding to a detected DM number of 1.7. At higher
masses, the limit is set at $2.5\times10^{-46}$
($1.6\times10^{-45}$)~cm$^2$ for a WIMP mass of 40 (400)~GeV/$c^2$,
and the corresponding number of detected DM signal events is 13.7
(20.1). The limit curve is weakened from that in
Ref.~\cite{Cui:2017nnn}, in which a downward fluctuation of the background
was observed and the limit was power-constrained to -1$\sigma$. The
turning of the limit curve around 15~GeV/c$^2$ is due to the fact that
the most "DM-like" events have $S1>10$ PE (see
Fig.~\ref{fig:vertex_dis}); thus, their DM-likelihoods increase with
increasing WIMP mass after roughly 15~GeV/c$^2$ or so.

\section{Conclusions and Outlook}
\label{sec:conclusion}
In summary, we report the WIMP search results with the 132
ton$\cdot$day full exposure data of the PandaX-II experiment, which
include a combination of data corresponding to 401 live-days with
several running conditions. Several major improvements have been made
in the data correction, selection, signal modeling, and data fitting
in this analysis. No significant excess of events is identified above
the background. A $90\%$ upper limit is set on the SI elastic
DM-nucleon cross section with the lowest excluded value of
$2.2\times10^{-46}$~cm$^2$ at a WIMP mass of 30 GeV/$c^2$.

The long duration of the PandaX-II operation, the systematic studies
performed, and the analysis techniques are all crucial for the
development of the next generation of PandaX programs, i.e.,
PandaX-4T~\cite{Zhang:2018xdp}. With the four-ton scale of a sensitive liquid
xenon target in a lower-background detector, the PandaX-4T experiment
is under preparation in the second phase of CJPL (CJPL-II). Together
with worldwide multi-ton scale experiments~\cite{Aprile:2015uzo,
  Mount:2017qzi}, the sensitivity of the DM search will be advanced by
more than one order of magnitude in the near future.

\section*{Acknowledgement}
  This project is supported in part by the Double First Class Plan of
  the Shanghai Jiao Tong University, grants from National Science
  Foundation of China (Nos. 11435008, 11455001, 11525522, 11775141 and
  11755001), a grant from the Ministry of Science and Technology of
  China (No. 2016YFA0400301) and a grant from China Postdoctoral
  Science Foundation (2018M640036). We thank the Office of Science and
  Technology, Shanghai Municipal Government (No. 11DZ2260700,
  No. 16DZ2260200, No. 18JC1410200) and the Key Laboratory for
  Particle Physics, Astrophysics and Cosmology, Ministry of Education,
  for important support. We also thank the sponsorship from the
  Chinese Academy of Sciences Center for Excellence in Particle
  Physics (CCEPP), the Hongwen Foundation in Hong Kong, and Tencent
  Foundation in China. Finally, we thank the CJPL administration and
  the Yalong River Hydropower Development Company Ltd. for the
  indispensable logistical support and other help.
\bibliographystyle{unsrt}
\bibliography{refs}

\appendix
\section{Events removed by post-unblinding cuts}
\label{sec:app_a}
\begin{figure}[hbt]
  \centering
  \includegraphics[width=.8\textwidth]{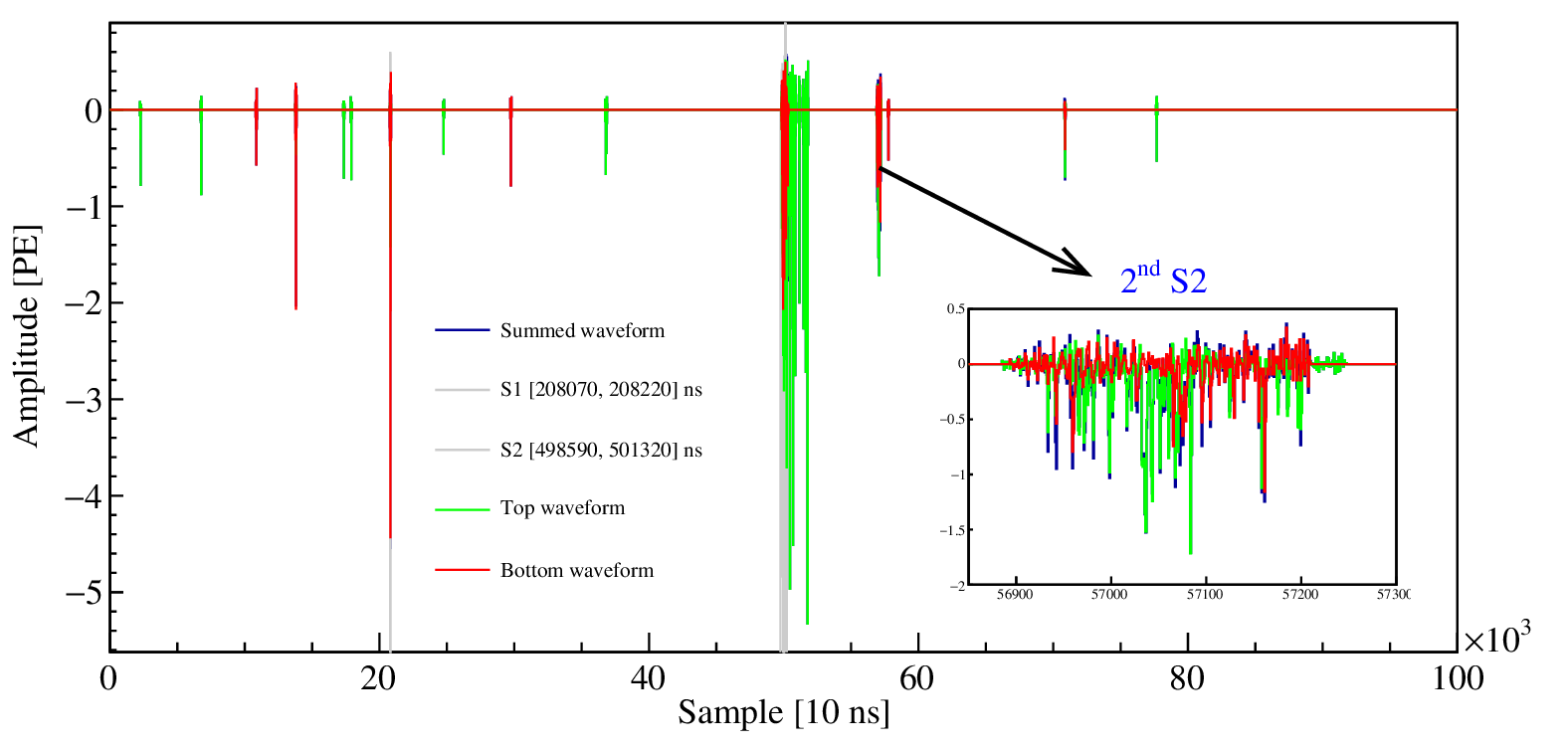}
  \caption{The full waveform of event 167193 in run 20922. The second
    small $S2$ was split into a few $S1$s in our clustering algorithm,
    so that it was incorrectly recognized as a single scattering
    events.}
  \label{fig:wf_small_s2}
\end{figure}

\begin{figure}[hbt]
  \centering
  \includegraphics[width=.8\textwidth]{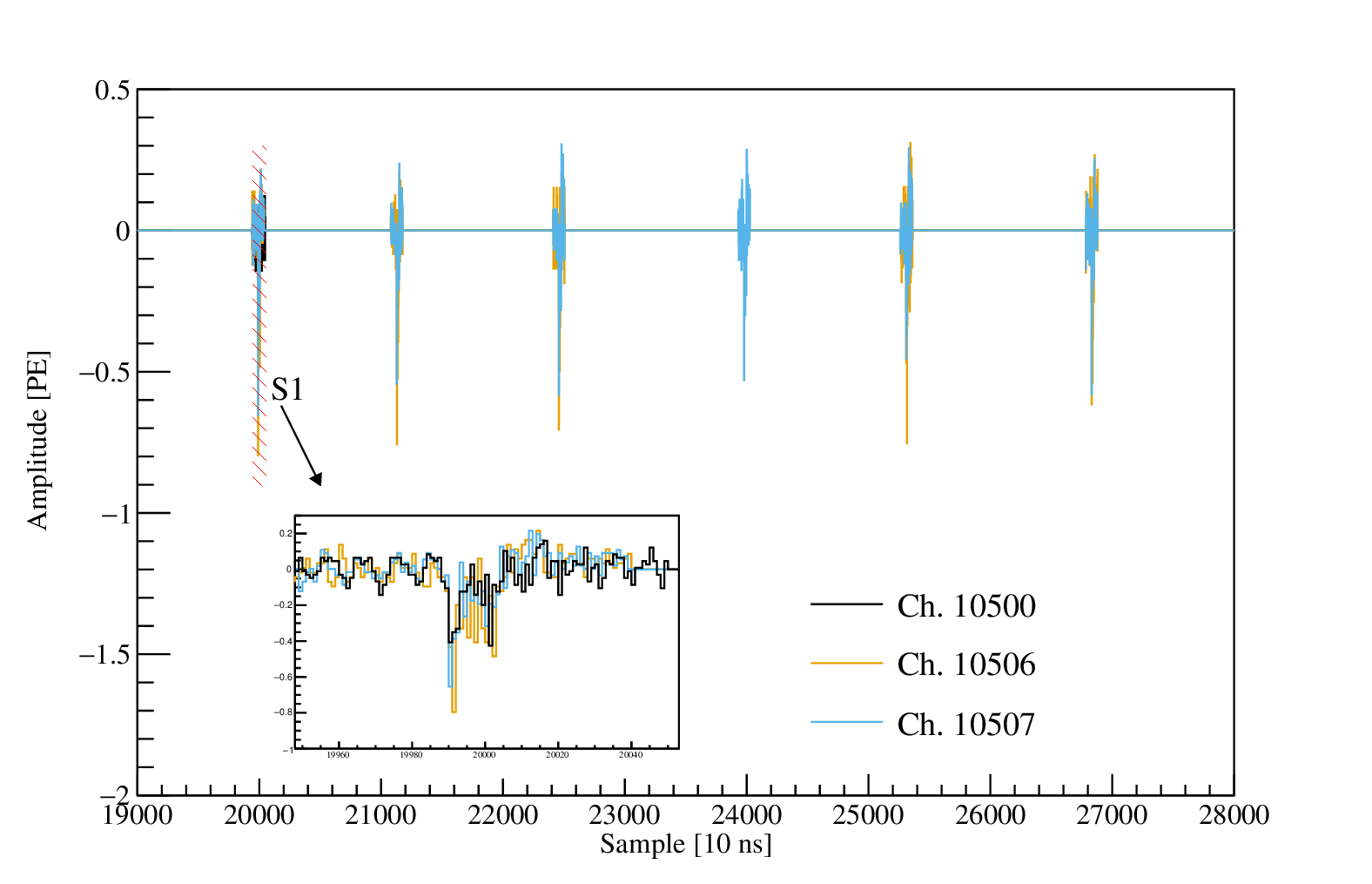}
  \caption{The partial waveform of event 112727 in run 22940. Two of
    the three hits in the reconstructed $S1$ are due to the coherent
    noise pickup in channel 10506 and 10507.}
  \label{fig:wf_cross_talk}
\end{figure}

\section{Horizontal distribution of the events in the $S2$ and $S1$ range cuts}
\label{sec:app_b}
\begin{figure}[hbt]
  \centering
  \begin{subfigure}{0.32\textwidth}
 \includegraphics[width=1.0\textwidth]{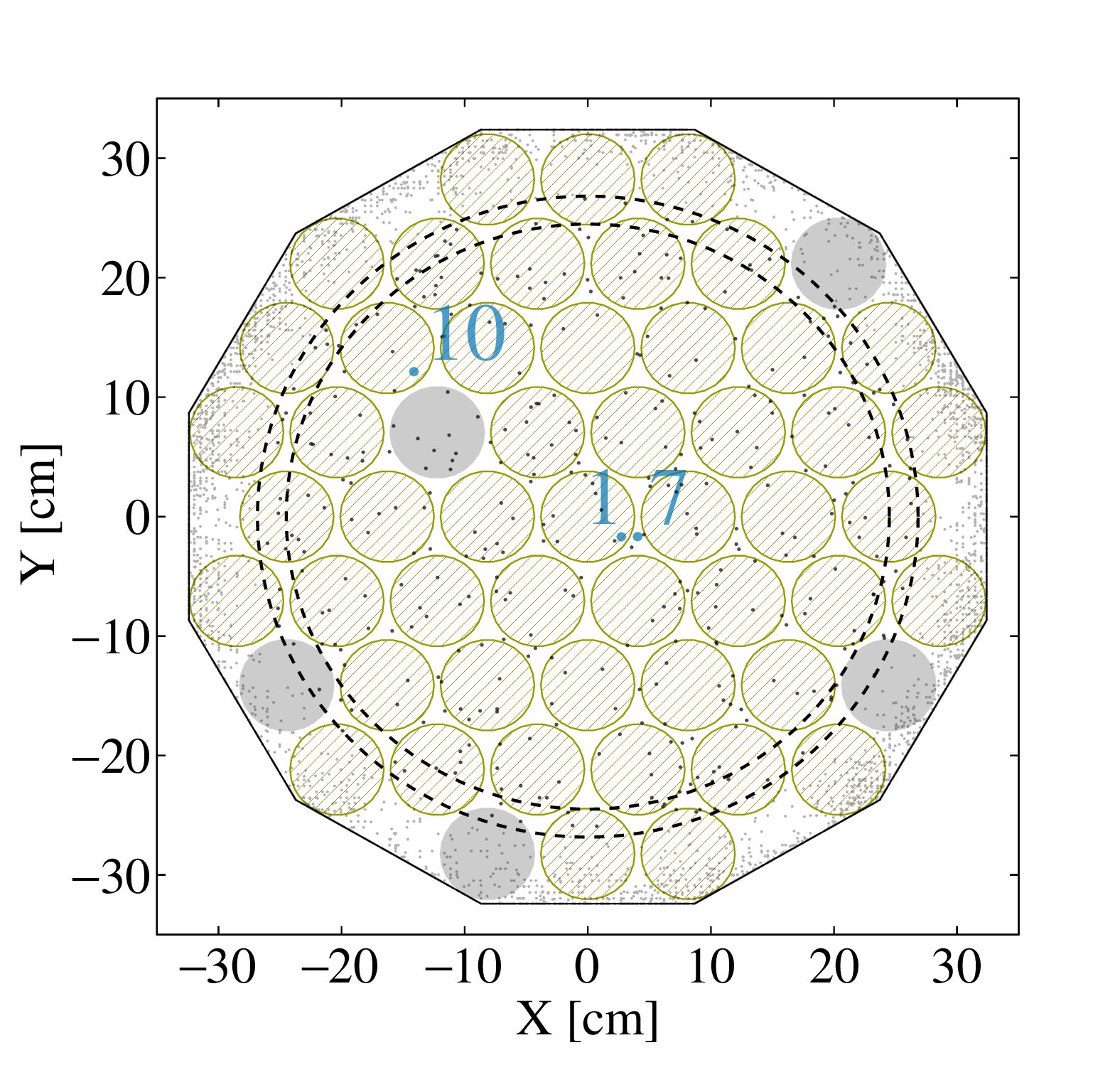}
  \caption{$x$-$y$ in Run 9}
   \label{fig:x_y_run9}
 \end{subfigure}
    \begin{subfigure}{0.32\textwidth}
 \includegraphics[width=1.0\textwidth]{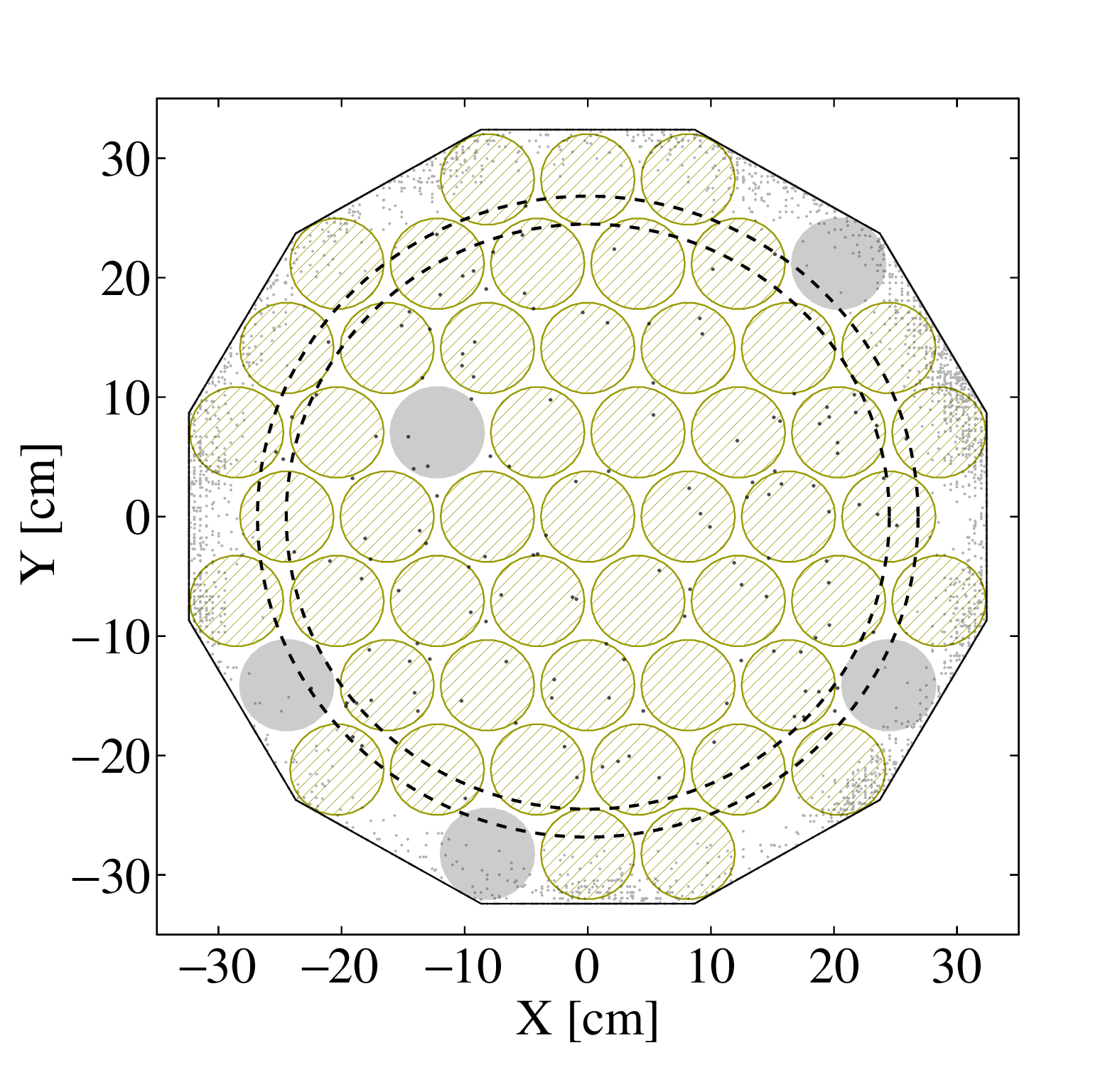}
  \caption{$x$-$y$ in Run 10}
   \label{fig:x_y_run10}
 \end{subfigure}
   \begin{subfigure}{0.32\textwidth}
 \includegraphics[width=1.0\textwidth]{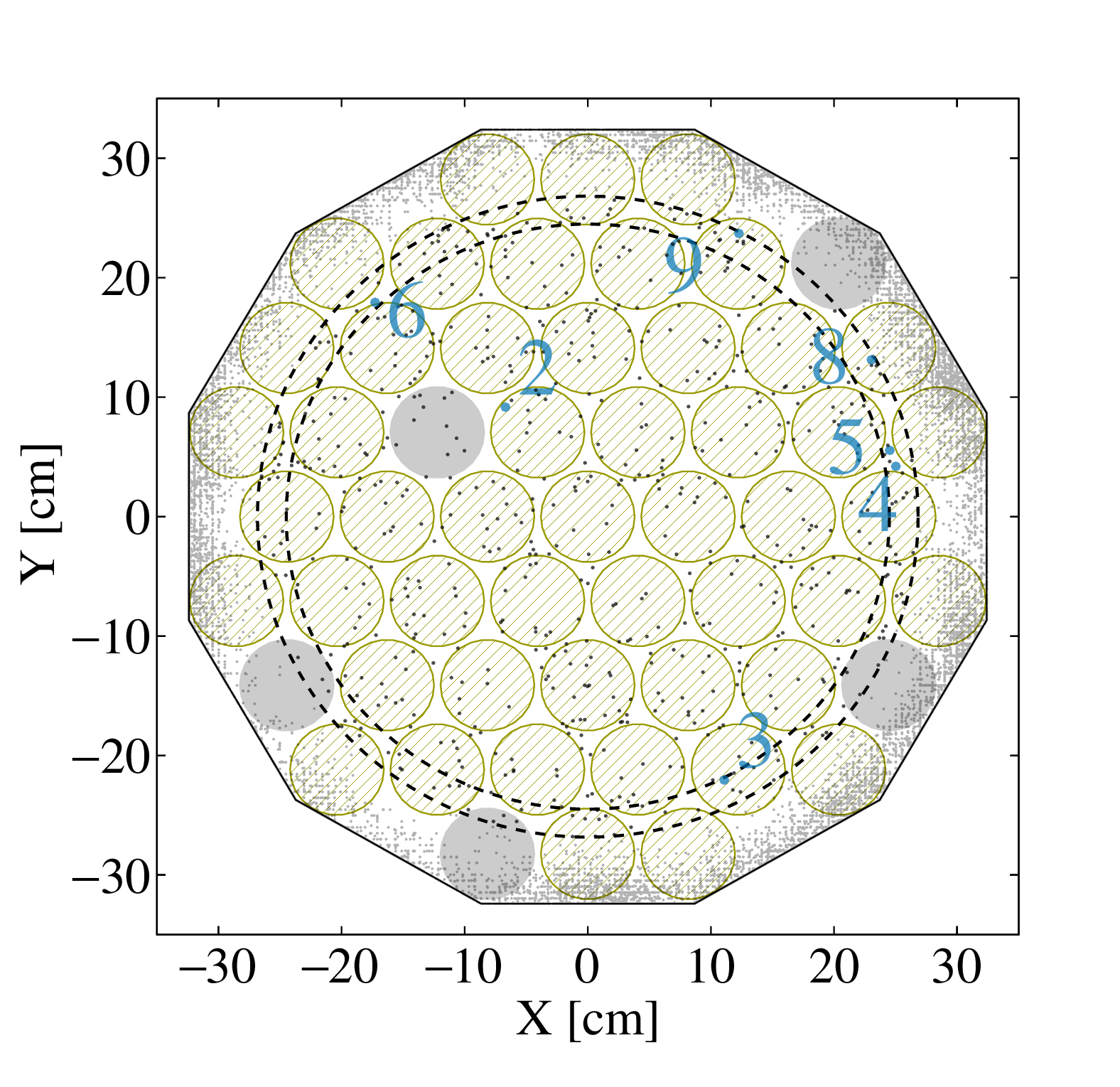}
  \caption{$x$-$y$ in Run 11}
   \label{fig:x_y_run11}
 \end{subfigure}
 \caption{$x$ vs. $y$ distribution of the events in the $S2$ and $S1$
   range cut of DM search runs. The drifting time cut,
   $(18, 310) \mu$s in Run 9 (a) and $(50, 350) \mu$s in Run 10 (b)
   and 11 (c) is applied to all events. The top ten DM-like candidates
   are labeled. The dashed lines mark the $R^{2}=720$ cm$^{2}$ and
   $R^2=600$ cm$^{2}$. The dodecagon is the boundary of the
   detector. The yellow (gray) circles represent the normal
   (inhibited) PMTs of the top array.}
  \label{fig:xycandidates}
\end{figure}

\end{document}